\documentclass[twocolumn, twocolappendix]{aastex631}

\usepackage{xspace}
\usepackage[encapsulated]{CJK}

\newcommand\grizli{\textsc{Gri$z$li}}
\newcommand\eazy{\textsc{EA$z$Y}}
\newcommand\se{\textsc{Source Extractor}}
\newcommand\zspec{$z_{\rm spec}$\xspace}
\newcommand\zphot{$z_{\rm phot}$\xspace}
\newcommand*{\JWST}{\emph{JWST}\xspace}
\newcommand*{\HST}{\emph{HST}\xspace}
\newcommand*{\Spitzer}{\emph{Spitzer}\xspace}

\usepackage{footmisc}
\usepackage{amsmath}
\usepackage[flushleft]{threeparttable}

\graphicspath{{./}{figures/}}
\shorttitle{The UNCOVER Photometric Catalog}
\shortauthors{J. R. Weaver et al.}

\begin{document}

\title{The UNCOVER Survey:\\
A first-look HST+JWST catalog of 60,\,000 galaxies near Abell~2744 and beyond}

\correspondingauthor{John R. Weaver}
\email{john.weaver.astro@gmail.com}

\author[0000-0003-1614-196X]{John R. Weaver}
\affiliation{Department of Astronomy, University of Massachusetts, Amherst, MA 01003, USA}
\author[0000-0002-7031-2865]{Sam E. Cutler}
\affiliation{Department of Astronomy, University of Massachusetts, Amherst, MA 01003, USA}
\author[0000-0002-9651-5716]{Richard Pan}
\affiliation{Department of Physics and Astronomy, Tufts University, 574 Boston Ave., Medford, MA 02155, USA}
\author[0000-0001-7160-3632]{Katherine E. Whitaker}
\affiliation{Department of Astronomy, University of Massachusetts, Amherst, MA 01003, USA}
\affiliation{Cosmic Dawn Center (DAWN), Denmark}
\author[0000-0002-2057-5376]{Ivo Labb\'{e}}
\affiliation{Centre for Astrophysics and Supercomputing, Swinburne University of Technology, Melbourne, VIC 3122, Australia}
\author[0000-0002-0108-4176]{Sedona H. Price}
\affiliation{Department of Physics and Astronomy and PITT PACC, University of Pittsburgh, Pittsburgh, PA 15260, USA}
\author[0000-0001-5063-8254]{Rachel Bezanson}
\affiliation{Department of Physics and Astronomy and PITT PACC, University of Pittsburgh, Pittsburgh, PA 15260, USA}
\author[0000-0003-2680-005X]{Gabriel Brammer}
\affiliation{Cosmic Dawn Center (DAWN), Denmark}
\affiliation{Niels Bohr Institute, University of Copenhagen, Jagtvej 128, K{\o}benhavn N, DK-2200, Denmark}
\author[0000-0001-9002-3502]{Danilo Marchesini}
\affiliation{Department of Physics and Astronomy, Tufts University, 574 Boston Ave., Medford, MA 02155, USA}
\author[0000-0001-6755-1315]{Joel Leja}
\affiliation{Department of Astronomy \& Astrophysics, The Pennsylvania State University, University Park, PA 16802, USA}
\affiliation{Institute for Computational \& Data Sciences, The Pennsylvania State University, University Park, PA 16802, USA}
\affiliation{Institute for Gravitation and the Cosmos, The Pennsylvania State University, University Park, PA 16802, USA}
\author[0000-0001-9269-5046]{Bingjie Wang (\begin{CJK*}{UTF8}{gbsn}王冰洁\ignorespacesafterend\end{CJK*})}
\affiliation{Department of Astronomy \& Astrophysics, The Pennsylvania State University, University Park, PA 16802, USA}
\affiliation{Institute for Computational \& Data Sciences, The Pennsylvania State University, University Park, PA 16802, USA}
\affiliation{Institute for Gravitation and the Cosmos, The Pennsylvania State University, University Park, PA 16802, USA}
\author[0000-0001-6278-032X]{Lukas J. Furtak}
\affiliation{Physics Department, Ben-Gurion University of the Negev, P.O. Box 653, Beer-Sheva 8410501, Israel}
\author[0000-0002-0350-4488]{Adi Zitrin}
\affiliation{Physics Department, Ben-Gurion University of the Negev, P.O. Box 653, Beer-Sheva 8410501, Israel}

\author[0000-0002-7570-0824]{Hakim Atek}
\affiliation{Institut d'Astrophysique de Paris, CNRS, Sorbonne Universit\'e, 98bis Boulevard Arago, 75014, Paris, France}
\author{Iryna Chemerynska}
\affiliation{Institut d'Astrophysique de Paris, CNRS, Sorbonne Universit\'e, 98bis Boulevard Arago, 75014, Paris, France}
\author[0000-0001-7410-7669]{Dan Coe}
\affiliation{Space Telescope Science Institute (STScI), 3700 San Martin Drive, Baltimore, MD 21218, USA}
\affiliation{Association of Universities for Research in Astronomy (AURA), Inc. for the European Space Agency (ESA)}
\affiliation{Center for Astrophysical Sciences, Department of Physics and Astronomy, The Johns Hopkins University, 3400 N Charles St. Baltimore, MD 21218, USA} 
\author[0000-0001-8460-1564]{Pratika Dayal}
\affiliation{Kapteyn Astronomical Institute, University of Groningen, P.O. Box 800, 9700 AV Groningen, The Netherlands}
\author[0000-0002-8282-9888]{Pieter van Dokkum}
\affiliation{Department of Astronomy, Yale University, New Haven, CT 06511, USA}
\author[0000-0002-1109-1919]{Robert Feldmann}
\affiliation{Institute for Computational Science, University of Zurich, Winterhurerstrasse 190, CH-8057 Zurich, Switzerland}
\author[0000-0003-4264-3381]{Natascha M. Förster Schreiber}
\affiliation{Max-Planck-Institut f{\"u}r extraterrestrische Physik, Gie{\ss}enbachstra{\ss}e 1, 85748 Garching, Germany}

\author[0000-0002-8871-3026]{Marijn Franx}
\affiliation{Leiden Observatory, Leiden University, P.O.Box 9513, NL-2300 AA Leiden, The Netherlands}

\author[0000-0001-7201-5066]{Seiji Fujimoto}\altaffiliation{NASA Hubble Fellow}
\affiliation{
Department of Astronomy, The University of Texas at Austin, Austin, TX 78712, USA
}
\author[0000-0001-7440-8832]{Yoshinobu Fudamoto}
\affiliation{Waseda Research Institute for Science and Engineering, Faculty of Science and Engineering, Waseda University, 3-4-1 Okubo, Shinjuku, Tokyo 169-8555, Japan}
\affiliation{National Astronomical Observatory of Japan, 2-21-1, Osawa, Mitaka, Tokyo, Japan}

\author[0000-0002-3254-9044]{Karl Glazebrook}
\affiliation{Centre for Astrophysics and Supercomputing, Swinburne University of Technology, Melbourne, VIC 3122, Australia}
\author[0000-0002-2380-9801]{Anna de Graaff}
\affiliation{Max-Planck-Institut f{\"u}r Astronomie, K{\"o}nigstuhl 17, D-69117, Heidelberg, Germany}
\author{Jenny E. Greene}
\affiliation{Department of Astrophysical Sciences, 4 Ivy Lane, Princeton University, Princeton, NJ 08544, USA}
\author[0000-0002-0000-2394]{St\'ephanie Juneau}
\affiliation{NSF’s National Optical-Infrared Astronomy Research Laboratory, 950 N. Cherry Avenue, Tucson, AZ 85719, USA}
\author[0000-0002-3838-8093]{Susan Kassin}
\affiliation{Space Telescope Science Institute (STScI), 3700 San Martin Drive, Baltimore, MD 21218, USA}
\author[0000-0002-7613-9872]{Mariska Kriek}
\affiliation{Leiden Observatory, Leiden University, P.O.Box 9513, NL-2300 AA Leiden, The Netherlands}
\author[0000-0002-3475-7648]{Gourav Khullar}
\affiliation{Department of Physics and Astronomy and PITT PACC, University of Pittsburgh, Pittsburgh, PA 15260, USA}
\author[0000-0003-0695-4414]{Michael V. Maseda}
\affiliation{Department of Astronomy, University of Wisconsin-Madison, 475 N. Charter St., Madison, WI 53706 USA}
\author[0000-0002-8530-9765]{Lamiya A. Mowla}
\affiliation{Dunlap Institute for Astronomy and Astrophysics, 50 St. George Street, Toronto, Ontario, M5S 3H4, Canada}
\author[0000-0002-9330-9108]{Adam Muzzin}
\affiliation{Department of Physics and Astronomy, York University, 4700 Keele Street, Toronto, Ontario, ON MJ3 1P3, Canada}
\author[0000-0003-2804-0648]{Themiya Nanayakkara}
\affiliation{Centre for Astrophysics and Supercomputing, Swinburne University of Technology, Melbourne, VIC 3122, Australia}
\author[0000-0002-7524-374X]{Erica J. Nelson}
\affiliation{Department for Astrophysical and Planetary Science, University of Colorado, Boulder, CO 80309, USA}
\author[0000-0001-5851-6649]{Pascal A. Oesch}
\affiliation{Department of Astronomy, University of Geneva, Chemin Pegasi 51, 1290 Versoix, Switzerland}
\affiliation{Cosmic Dawn Center (DAWN), Niels Bohr Institute, University of Copenhagen, Jagtvej 128, K{\o}benhavn N, DK-2200, Denmark}
\author[0000-0003-4196-0617]{Camilla Pacifici}
\affiliation{Space Telescope Science Institute (STScI), 3700 San Martin Drive, Baltimore, MD 21218, USA}

\author[0000-0001-7503-8482]{Casey Papovich}
\affiliation{Department of Physics and Astronomy, Texas A\&M University, College Station, TX, 77843-4242 USA}
\affiliation{George P. and Cynthia Woods Mitchell Institute for Fundamental Physics and Astronomy, Texas A\&M University, College Station, TX, 77843-4242 USA}
\author[0000-0003-4075-7393]{David J. Setton}
\affiliation{Department of Physics and Astronomy and PITT PACC, University of Pittsburgh, Pittsburgh, PA 15260, USA}
\author[0000-0003-3509-4855]{Alice E. Shapley}
\affiliation{Physics \& Astronomy Department, University of California: Los Angeles, 430 Portola Plaza, Los Angeles, CA 90095, USA}
\author{Heath V. Shipley}
\affiliation{Department of Physics, Texas State University, San Marcos, TX 78666, USA}
\author[0000-0001-8034-7802]{Renske Smit}
\affiliation{Astrophysics Research Institute, Liverpool John Moores University, 146 Brownlow Hill, Liverpool L3 5RF, UK}
\author[0000-0001-7768-5309]{Mauro Stefanon}
\affiliation{Departament d'Astronomia i Astrofisica, Universitat de Valencia, C. Dr. Moliner 50, E-46100 Burjassot, Valencia, Spain}
\affiliation{Unidad Asociada CSIC ``Grupo de Astrofisica Extragalactica y Cosmologi'' (Instituto de Fisica de Cantabria - Universitat de Valencia)}
\author[0000-0002-5522-9107]{Edward N.\ Taylor}
\affiliation{Centre for Astrophysics and Supercomputing, Swinburne University of Technology, Melbourne, VIC 3122, Australia}
\author[0000-0001-8928-4465]{Andrea Weibel}
\affiliation{Department of Astronomy, University of Geneva, Chemin Pegasi 51, 1290 Versoix, Switzerland}
\author[0000-0003-2919-7495]{Christina C. Williams}
\affiliation{NSF’s National Optical-Infrared Astronomy Research Laboratory, 950 N. Cherry Avenue, Tucson, AZ 85719, USA}
\affiliation{Steward Observatory, University of Arizona, 933 North Cherry Avenue, Tucson, AZ 85721, USA}

\begin{abstract}

In November 2022, the \textit{James Webb Space Telescope} (\JWST) returned deep near-infrared images of Abell~2744 -- a powerful lensing cluster capable of magnifying distant, incipient galaxies beyond it. Together with the existing \textit{Hubble Space Telescope} (\HST) imaging, this publicly available dataset opens a fundamentally new discovery space to understand the remaining mysteries of the formation and evolution of galaxies across cosmic time. In this work, we detect and measure some 60\,000 objects across the 49\,arcmin$^2$ \JWST footprint down to a $5\,\sigma$ limiting magnitude of $\sim$30\,mag in 0.32\arcsec{} apertures. Photometry is performed using circular apertures on images matched to the point spread function of the reddest NIRCam broad band, F444W, and cleaned of bright cluster galaxies and the related intra-cluster light. To give an impression of the photometric performance, we measure photometric redshifts and achieve a $\sigma_{\rm NMAD}\approx0.03$ based on known, but relatively small, spectroscopic samples. With this paper, we publicly release our \HST and \JWST PSF-matched photometric catalog with optimally assigned aperture sizes for easy use, along with single aperture catalogs, photometric redshifts, rest-frame colors, and individual magnification estimates. These catalogs will set the stage for efficient and deep spectroscopic follow-up of some of the first \JWST-selected samples in Summer 2023. 

\end{abstract}

%% Keywords should appear after the \end{abstract} command. 
%% The AAS Journals now uses Unified Astronomy Thesaurus concepts:
%% https://astrothesaurus.org
%% You will be asked to selected these concepts during the submission process
%% but this old "keyword" functionality is maintained in case authors want
%% to include these concepts in their preprints.

%% From the front matter, we move on to the body of the paper.
%% Sections are demarcated by \section and \subsection, respectively.
%% Observe the use of the LaTeX \label
%% command after the \subsection to give a symbolic KEY to the
%% subsection for cross-referencing in a \ref command.
%% You can use LaTeX's \ref and \label commands to keep track of
%% cross-references to sections, equations, tables, and figures.
%% That way, if you change the order of any elements, LaTeX will
%% automatically renumber them.
%%
%% We recommend that authors also use the natbib \citep
%% and \citet commands to identify citations.  The citations are
%% tied to the reference list via symbolic KEYs. The KEY corresponds
%% to the KEY in the \bibitem in the reference list below. 

\section{Introduction} \label{sec:intro}

% quick summary of photometry
% history of abell2744
% advantages and challenges of cluster surveys
% future work

The vast distance scales of our Universe relative to the human timescale implicitly means that there are very few astrophysical processes we can observe changing in real time.  Dynamical processes in galaxies transpire over timescales of millions to billions of years.  Thus to understand the formation and evolution of galaxies across cosmic time necessitates the study of statistically representative snapshots.  Observational campaigns are forced to make decisions in survey design, generally prioritizing either depth \citep[e.g.,][]{Williams1996, Giavalisco2004, Beckwith2006, Bouwens2011,Illingworth2013, Illingworth2016, Lotz2017}, volume \citep[e.g.,][]{Scoville2007, Jarvis2013, Aihara2018, Abbott2018}, or a mix therein \citep[e.g.,][]{Grogin2011, Koekemoer2011}, in order to assemble unbiased galaxy populations.  All of these surveys share a common theme: they require a synthesis of the panchromatic flux measurements for detected sources into a photometric catalog as the first step towards modeling the stellar populations. Such photometric catalogs serve as the foundation of any galaxy survey, necessary for identifying robust galaxy samples and for enabling the vast majority of subsequent science investigations.

% Deep field vs clusters, introduce the FF and Abell 2744
% Advantages and challenges of cluster surveys

The deepest surveys of our Universe to date come from single ultra-deep pointings with the Hubble Space Telescope (\HST), either of ``blank fields'' \citep[i.e., relatively dark lines of sight through our own Milky Way galaxy;][]{Williams1996, Beckwith2006, Bouwens2011, Illingworth2013} or by targeting known clusters of galaxies at intermediate redshifts 
\citep{Lotz2017, Coe2019, Salmon2020, Sharon2020}. One particular advantage of targeting galaxy clusters is the added boost from strong gravitational lensing; the richest clusters magnify background galaxies by factors of a few up to dozen typically, depending on the size and position of the background galaxy with respect to the lens \citep[e.g.,][]{Coe2019}. Strong lens clusters unveil some of the most distant \citep[e.g.,][]{Zheng2012, Coe2013, Zitrin2014, Strait2021, Bradley2022, Furtak2022a, Hsiao2022, Roberts-Borsani2022, Williams2022, Adams2023, Atek2023} and lowest-mass \citep[e.g.,][]{Livermore2017, Bouwens2017, Bouwens2022, Atek2018, Bhatawdekar2019, Kikuchihara2020, Furtak2021} galaxies known, even single candidate stars in some cases \citep{Welch2022}. However, this boost from cosmic telescopes comes with a cost in terms of contamination from  cluster galaxies/intracluster light and the complex and non-linear distortions to galaxy morphologies \citep[e.g.,][]{shipley:18, Bhatawdekar2019, Pagul2021, Fox2022}.  In addition, the source-plane area that is being probed behind a lens is smaller. As such, there is a trade-off between detecting a higher number of galaxies that are boosted in flux but for a smaller area probed relative to an unlensed field; an effect known as the magnification bias \citep[e.g.,][]{Broadhurst1995}. However, since the luminosity function of high-redshift galaxies is steep enough \citep[e.g.,][]{Finkelstein2015, Mason2015}, the net effect is a gain in the number density of detections.

Despite these challenges, galaxy clusters afford our best opportunity to push to the most extreme depths and thus to the frontiers of galaxy formation. 
Campaigns such as the Director's Discretionary Hubble Frontier Fields program (HFF) have amassed a rich archival data set of \HST imaging \citep{Lotz2017, Steinhardt2020} that has set the stage for \JWST imaging and spectroscopic programs \citep[e.g.,][]{Willott2017, Treu2022, bezanson:22, Windhorst2023}. One cluster is particularly compelling to study: in addition to a spectacular central core \citep{Lotz2017, shipley:18, Pagul2021, Kokorev2022}, Abell~2744 contains prominent lensing features within two additional massive cluster sub-structures in the north and north-west \citep{Furtak2022}. Abell~2744 thus contains an unusually large area of high magnification when combining the various structures. Several early \JWST programs target Abell~2744; here we combine publicly available \HST and \JWST photometry from the 
JWST-GO-2561, JWST-DD-ERS-1324, and JWST-DD-2756 programs.

In this paper, we present the space-based photometric catalog for the UNCOVER survey \citep{bezanson:22} as part of Data Release 2 (DR2)\footnote{\noindent Note that an earlier set of mosaics, photometric catalog, and manuscript were provided as part of DR1; these too are available on our website (\url{https://jwst-uncover.github.io/DR1.html}\label{foot:dr1}) and Zenodo (\url{https://doi.org/10.5281/zenodo.8199803}).}, including derived photometric redshifts and magnification corrections from updated lensing models originally presented in \citet{Furtak2022}. Catalogs are available online.\footnote{\noindent Catalogs are available on our website (\url{https://jwst-uncover.github.io/DR2.html}) and Zenodo (\url{https://doi.org/10.5281/zenodo.8199802})\label{foot:catalog}}. The software used to produce these catalogs, \texttt{aperpy}, is generally applicable to any \JWST/NIRCam data and is freely available.\footnote{\noindent \texttt{aperpy} is available though Github (\url{https://github.com/astrowhit/aperpy}) and Zenodo (\url{https://doi.org/10.5281/zenodo.8280270}).\label{foot:aperpy}}

In Section~\ref{sec:data}, we present an overview of the data processing, including the reduction and astrometric correction of images from \HST and \JWST.  Section 3 describes two approaches for handling the added complexity of intracluster light in the Abell~2744 cluster. Source detection and photometry are described in Section 4, including a description of the methodology adopted to homogenize the point spread function (PSF), the measurement of total fluxes from aperture photometry, and a quantification of representative errors. We present the photometric catalog properties in Section 5, including depths, galaxy number counts, and comparisons to other surveys.  Finally, we summarize our findings in Section 6. An appendix contains additional relevant information regarding the stability of the \JWST point spread function in time and across the detector.

 All magnitudes in this paper are expressed in the AB system \citep{Oke1974}, for which a flux $f_\nu$ in $10\,\times\,$nJy ($10^{-28}$~erg~cm$^{-2}\,$s$^{-1}\,$Hz$^{-1}$) corresponds to AB$_\nu=28.9-2.5\,\log_{10}(f_\nu/\mu{\rm Jy})$. When computing physical properties such as rest-frame fluxes, we adopt a standard $\Lambda$CDM cosmology with $\mathrm{H_0=70\,km\,s^{-1}\,Mpc^{-1}}$, $\Omega_M=0.3$, and $\Omega_{\Lambda}=0.7$.

\section{Data} \label{sec:data}

\begin{figure}
    \centering
    \includegraphics[width=0.5\textwidth]{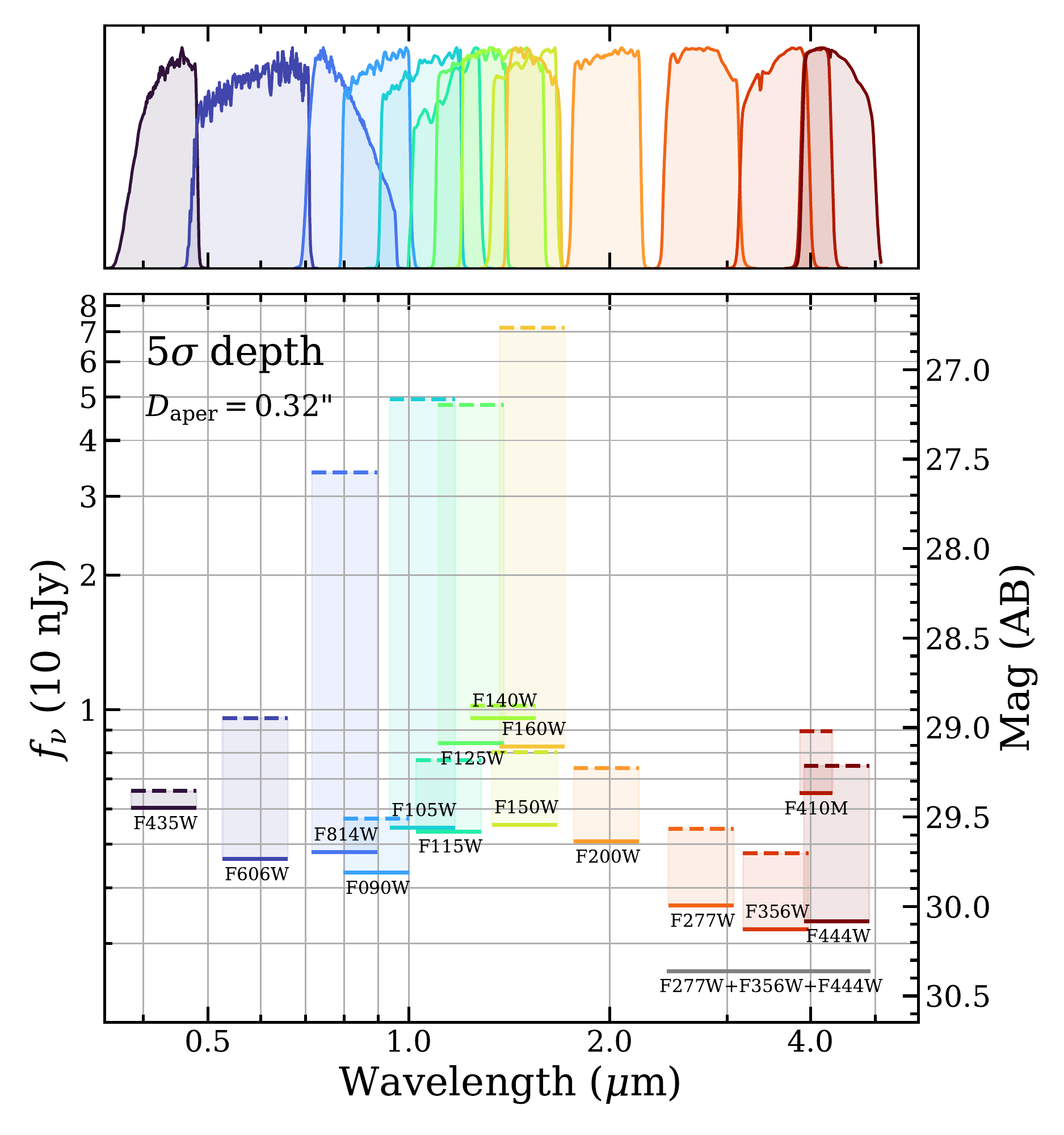}
    \caption{Effective catalog depths over the Abell~2744 \JWST footprint for the 15 available photometric bands and their transmission curves. The effective depth of our LW detection image is also included. Depths are quoted in 0.32\arcsec{} diameter apertures and correspond to the area-weighted 50$^{th}$ (median) and 10$^{th}$ percentiles (dashed and solid lines, respectively). Areas in \HST imaging without \JWST coverage are not considered. See the text for details.}
    \label{fig:depths}
\end{figure}

\setlength{\tabcolsep}{4pt}
\renewcommand{\arraystretch}{1.2}

\begin{table}[]
    \centering
    \begin{tabular}{c|ccc|cccc}
    \hline
        Filter & \multicolumn{3}{c}{Depth ($5\,\sigma$ AB)}  & \multicolumn{4}{c}{Area (arcmin$^2$)} \\
            &  10$^{th}$ & 50$^{th}$ & 90$^{th}$ & 10$^{th}$ & 50$^{th}$ & 90$^{th}$ & Total \\
            \hline
    \hline
LW DET & 30.36 & 29.85 & 29.01 &  5.78 & 27.10  & 44.56 & 48.91 \\
\hline
F435W & 29.45 & 29.35 & 28.30 &  0.90 & 4.30  & 15.85 & 18.52 \\
F606W & 29.73 & 28.95 & 27.42 &  0.95 & 19.97  & 33.11 & 37.54 \\
F814W & 29.70 & 27.58 & 27.17 &  2.24 & 19.28  & 28.05 & 32.98 \\
F090W & 29.81 & 29.51 & 29.30 &  1.46 & 6.76  & 11.65 & 12.84 \\
F105W & 29.56 & 27.17 & 27.06 &  4.17 & 12.86  & 17.95 & 21.11 \\
F115W & 29.58 & 29.18 & 28.15 &  6.18 & 27.28  & 44.53 & 48.25 \\
F125W & 29.09 & 27.20 & 27.09 &  4.32 & 12.24  & 17.48 & 20.97 \\
F140W & 28.95 & 28.88 & 28.54 &  1.66 & 4.04  & 4.84 & 5.62 \\
F150W & 29.54 & 29.14 & 28.23 &  6.07 & 27.89  & 45.00 & 48.63 \\
F160W & 29.11 & 26.77 & 26.53 &  3.35 & 12.00  & 18.89 & 21.03 \\
F200W & 29.64 & 29.23 & 28.47 &  5.93 & 27.00  & 44.10 & 47.82 \\
F277W & 29.99 & 29.57 & 28.76 &  5.53 & 26.25  & 44.11 & 48.58 \\
F356W & 30.13 & 29.70 & 28.88 &  5.74 & 26.25  & 44.20 & 48.82 \\
F410M & 29.37 & 29.02 & 28.58 &  3.45 & 15.84  & 25.90 & 28.66 \\
F444W & 30.08 & 29.21 & 28.24 &  5.60 & 26.31  & 44.37 & 48.25 \\
    \end{tabular}
    \caption{Effective catalog depths, quoted in 0.32\arcsec{} diameter apertures and correspond to the area-weighted 10$^{th}$, 50$^{th}$ (median), and 90$^{th}$ percentiles. Total areas reflect the union of the LW detection footprint with the coverage available for each band.}
    \label{tab:depths}
\end{table}

\subsection{James Webb Space Telescope} \label{sec:data:jwst}

The photometric catalogs presented herein include all public \JWST/NIRCam imaging of Abell~2744 available to date: the Ultradeep NIRSpec and NIRCam ObserVations before the Epoch of Reionization (UNCOVER) Treasury survey \citep[PIs Labb\'{e} \& Bezanson, JWST-GO-2561;][]{bezanson:22}, the Early Release Science (ERS) GLASS-\JWST program \citep[PI: Treu, JWST-DD-ERS-1324;][]{Treu2022}, and a Directors Discretionary (DD) program (JWST-DD-2756, PI: Chen).  
As described in \citet{bezanson:22}, our dataset combines the deep NIRCam imaging with 4-6 hour exposures in 7 filters (F115W, F150W, F200W,
F277W, F356W, F410M, and F444W) from UNCOVER, with the ultra-deep imaging with 9-14 hour exposures from GLASS-ERS in 7 filters (F090W, F115W, F150W, F200W,
F277W, F356W, and F444W). The GLASS-ERS NIRCam pointing is taken in parallel and thus offset to the cluster outskirts, thereby extending the combined science area. Additionally, the DDT program includes two epochs of NIRCam imaging in 6 filters (F115W, F150W, F200W, F277W, F356W, and F444W), totaling $\sim1$ hour per filter. All together, images in 8 unique \JWST filters from both short-wavelength (SW) and long-wavelength (LW) channels are combined to extend the coverage of Abell~2744 to include the nearby cluster sub-structures. Although the UNCOVER NIRISS parallel imaging has been reduced and released in \citet{bezanson:22}, photometry and cataloging of the parallel is planned for future work.

Next, we summarize the key steps of the image reduction, referring the reader to Section 3.1 of \citet{bezanson:22} for further details. Imaging mosaics are produced from the flux-calibrated NIRCam exposures released in Stage 2b of the \JWST calibration pipeline (v1.8.4) and combined with calibration set jwst\_1039.pmap. The exposures are then processed, aligned, and co-added using the GRIsm redshift and LIne analysis software for space-based spectroscopy (\grizli{}\footnote{\noindent \url{https://github.com/gbrammer/grizli}}, 1.8.16.dev12; \citealt{Brammer2019, Kokorev2022}).  The pipeline has been optimized to handle known \JWST artifacts \citep{Rigby2022}. Our flat-field calibration image is custom made from on-sky commissioning data (COM-1063), updating the official calibration files to correct for smoothly varying large-scale structure in the flats and to further optimize pixel-to-pixel variations. The data reduction pipeline next subtracts a large scale sky background, performs an astrometric alignment (see Section~\ref{sec:data:specz}), identifies and removes hot pixels, and drizzles the images to a common pixel grid of 0.02\arcsec{} per pixel for SW bands and 0.04\arcsec{} per pixel for LW bands using \textsc{astrodrizzle} \citep{Gonzaga12}.

\subsection{Hubble Space Telescope} \label{sec:data:hst}

A wide range of imaging of the Abell~2744 cluster and surrounding area exists within the public \HST archive. 
 Briefly, we summarize the programs relevant herein.
Program HST-GO-11689 (PI: Dupke) and HST-GO-13386 (PI: Rodney) include deep HST/ACS imaging in the cluster center in 3 filters (F435W, F606W, and F814W), with program HST-DD-13495 \citep[PI: Lotz;][]{Lotz2017} acquiring complementary deep \HST/WFC3 observations in 4 filters (F105W, F125W, F140W, and F160W). While each of the above programs are (deep) individual pointings limited by the ACS and WFC3 field of view, respectively, the data was later expanded by a factor of 4 with shallower imaging in 2 ACS filters (F606W and F814W) and 3 WFC3 filters (F105W, F125W, and F160W) by the BUFFALO survey \citep[Program HST-GO-15117, PIs: Steinhardt \& Jauzac;][]{Steinhardt2020}. Most recently, the deep optical coverage was further expanded by Program JWST-DD-17231 (PI: Treu). A summary of the instruments, filters, program IDs, and orbit depths can be found in Table 3 in \citet{bezanson:22}.  Taken together, they contribute 7 unique \HST filters. These images are reduced following the same procedure as described in Section~\ref{sec:data:jwst} onto the same 0.04\arcsec{} pixel grid as the \JWST images. 

\subsection{Astrometry} \label{sec:data:specz}
\begin{figure}
    \centering
    \includegraphics[width=0.5\textwidth]{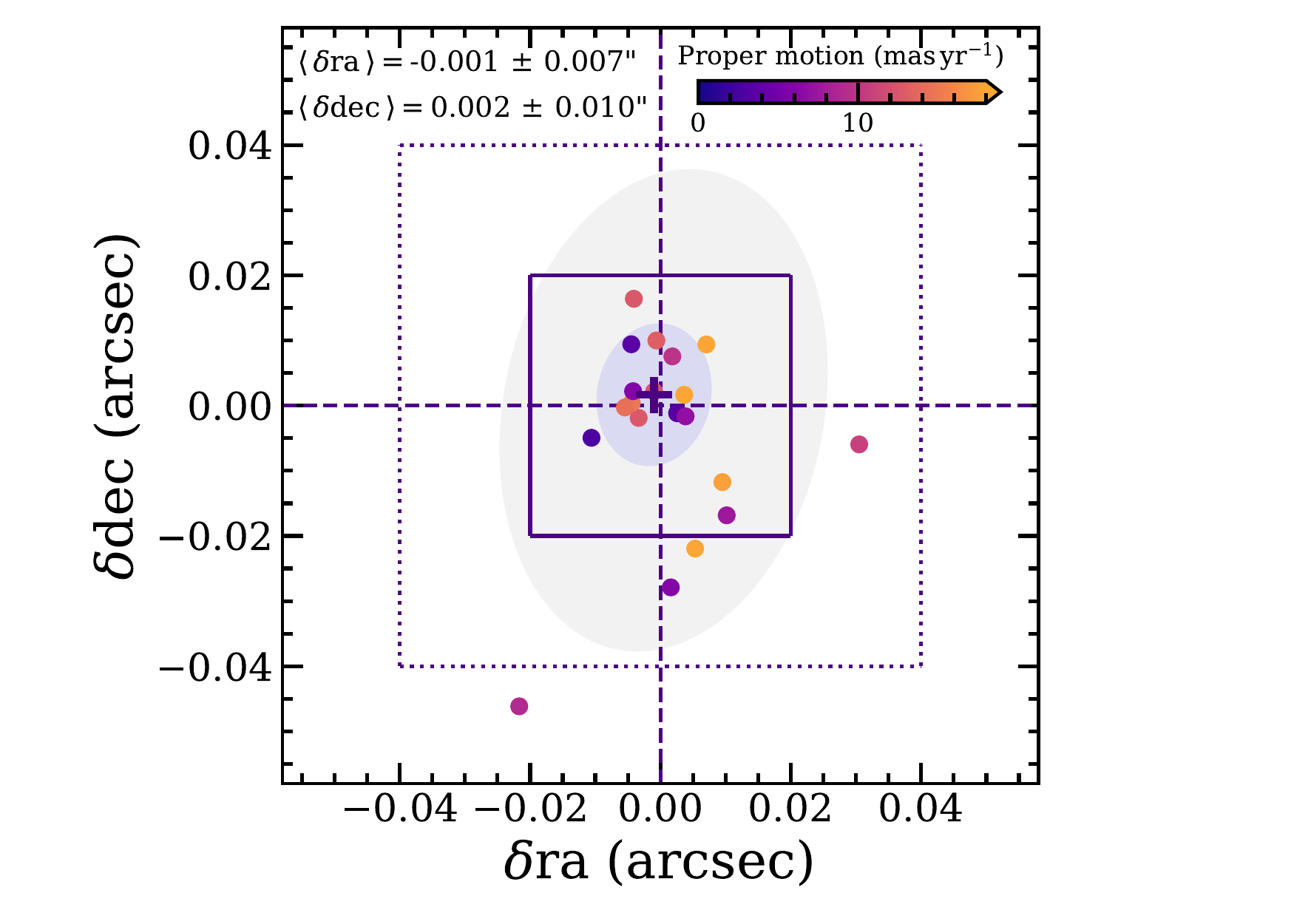}
    \caption{Astrometric performance of the imaging dataset based on positions of bright stars in \HST/F160W compared to GAIA DR3. One and four pixel areas are shown by the solid and dashed squares, respectively. Filled purple and grey elliptical contours enclose roughly 50\% and 90\% of bright stars, respectively. The median deviation (purple cross) and the standard deviation of the absolute median deviation are quoted for each axis corresponding to the innermost 50\% of stars.}
    \label{fig:astrometry}
\end{figure}

Astrometric registration of the images is performed by \grizli{} using F444W. We adopt star positions from GAIA DR3 \citep{GaiaCollaboration2016, GaiaCollaboration2022}. Using their well known proper motions, the positions of GAIA stars observed in 2015 are projected to November 2022, the epoch during which the \JWST imaging was acquired. The remaining images are then registered consistently to the F444W filter. In order to independently test the accuracy of the astrometry, we opt to compare to stars in the F160W filter instead of F444W, where saturation and central star clipping is less of an issue. We perform an additional correction to the proper motions to shift to the median epoch of the wider F160W BUFFALO \HST imaging in July 2019 \citep{Steinhardt2020}. Figure~\ref{fig:astrometry} demonstrates our achieved precision of $\approx0.008$\arcsec{} or a fifth of a pixel, measured by the standard deviation of the median absolute deviation for the innermost 50\% of sources (purple shaded region). The median bias, based on the same sources, is also small at $\approx0.002$\arcsec{}, or 5\% of a single pixel. 

\subsection{Spectroscopic redshifts} \label{sec:data:specz}
Spectroscopic redshifts (hereafter \zspec) over our survey footprint are taken from a compilation by \citet{Kokorev2022}. We find 518 secure entries with \zspec values with confidence flags 3 or 4 within 0.3\arcsec{} of our sources. Generally, a flag of 3 or 4 note secure \zspec{} from a single strong emission line, multiple emission lines, or easily distinguished continuum features; note that flagging definitions can vary somewhat between subsets of the compilation. 156 are found in the NASA/IPAC Extragalactic Database,\footnote{\noindent \url{https://ned.ipac.caltech.edu/}} 340 are cluster members with \zspec{} from \cite{Richard2021} with VLT/MUSE, 22 are grism redshifts from GLASS \citep{Treu2015}. The corresponding values are stored in the catalog in the \texttt{z\_spec} column. Please note that we do not include any \zspec{} from the UNCOVER follow-up program (see Price et al., in prep.).

\section{Removal of Sky, Intra-Cluster Light, and Bright Cluster Galaxies}
In order to achieve the science objectives, we need to first mitigate the contamination of foreground light from the many bright cluster galaxies (bCGs) and the powerful intra-cluster light (ICL). Otherwise, the photometry of distant sources seen through this foreground light will be inaccurate, potentially mischaracterized, or missing altogether the rare high-$z$ galaxies magnified by the strong gravitational lensing of these cluster members. We note that throughout this paper we adopt the term bCG, which is not synonymous with the traditional brightest cluster galaxy (BCG).

\begin{figure*}
    \centering
    \includegraphics[width=\textwidth]{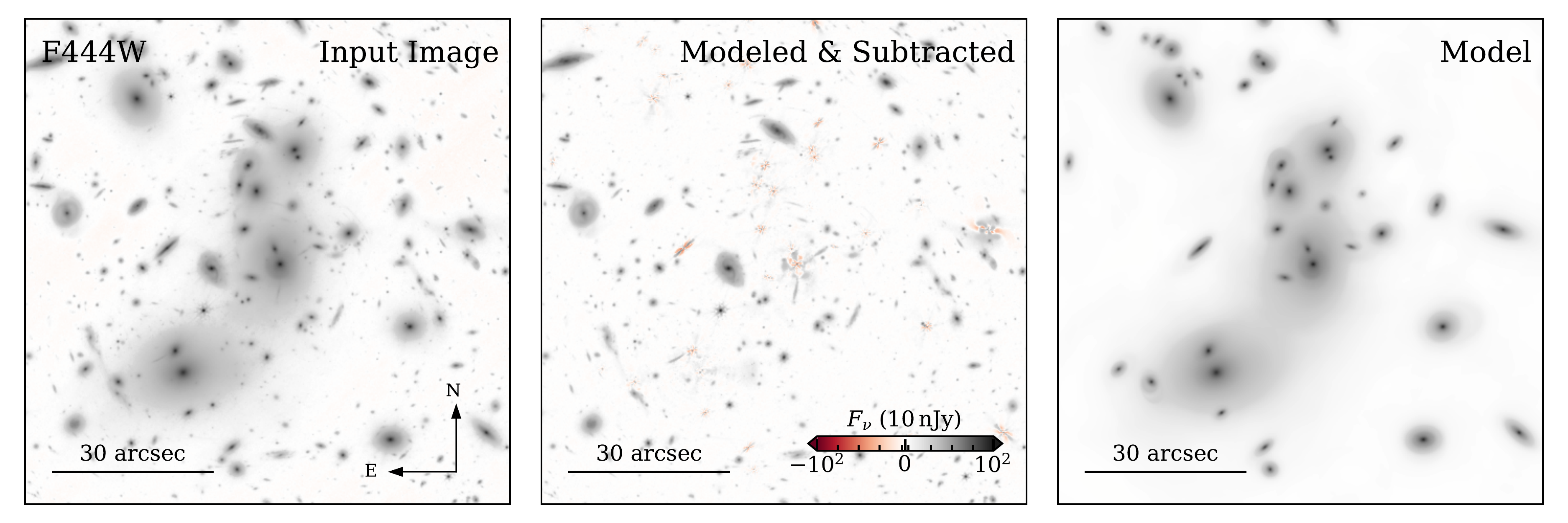}
    \includegraphics[width=\textwidth]{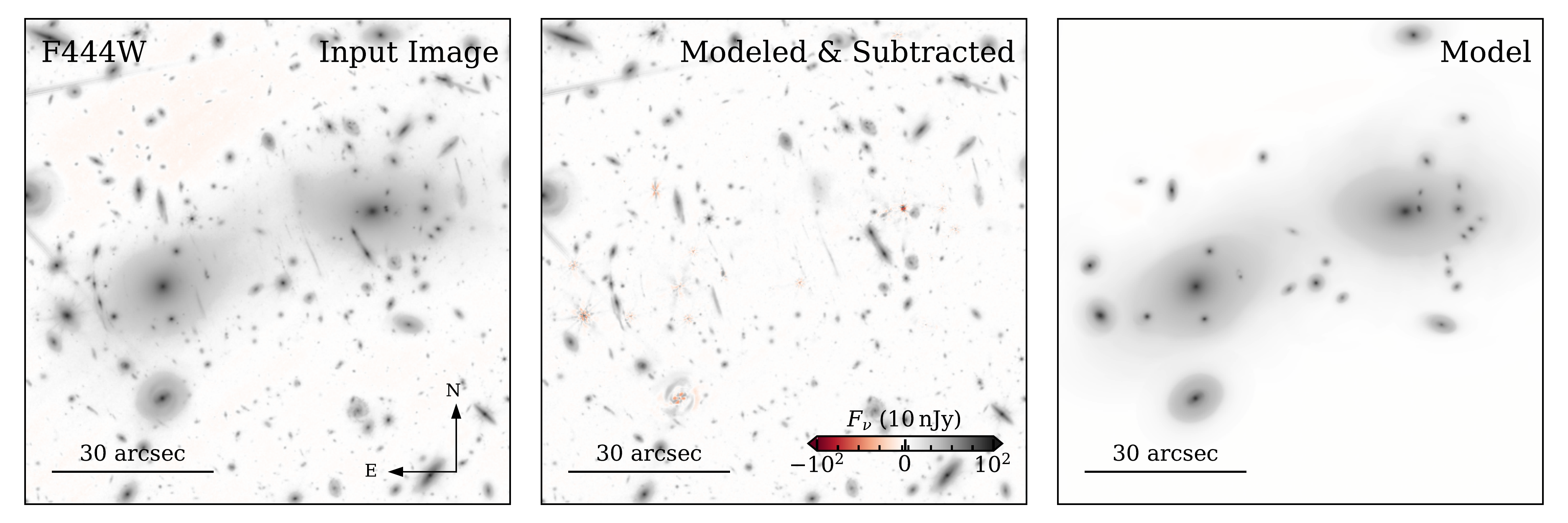}
    \caption{Zoom-in around the two primary cluster cores of Abell~2744 in F444W. The bright cluster galaxies and intra-cluster light visible in the input image (left) is removed by subtracting fitted models (middle), with the models themselves on the right.}
    \label{fig:icl_demo}
\end{figure*}

\subsection{Subtracting fitted models}
\label{sec:bcgmodel}

For robust and tested bCG and ICL subtraction, we adopt the method described in \citet{ferrarese:06} and implemented by \citet{shipley:18} in the HFF-DeepSpace (HFFDS) photometric catalogs of six lensing clusters, including Abell~2744.

The bCGs that contribute significantly to the total cluster luminosity are first identified from the HFFDS catalogs; we refer the reader to \citet{shipley:18} for a more detailed description of the selection process. We further expand our selection to accommodate the wider footprint of the present data set \citep[see Figure 2 of][]{bezanson:22}. We note that fewer bCGs are subtracted from the F410M mosaic, which has a smaller footprint. To minimize computation time, we generate ``cropped mosaics'' using the IRAF \texttt{IMCOPY} task. The boundaries of these mosaics are defined by the outermost isophotal cluster radii.

We use \se{} to create a crude mask of all background sources (excluding cluster members). This is done by using the parameters: \texttt{DETECT\_THRESH} = 1.2, \texttt{DEBLEND\_NTHRESH} = 10, \texttt{DEBLEND\_MINCONT} = 0.01, which identifies more isolated sources. We repeat this detection on the masked image, providing a more accurate and precise mask, especially near tightly clustered galaxies. These two masks are combined to isolate the cluster galaxies and ultimately smoothed with a Gaussian kernel to account for nearby, poorly modeled light.

Using this mask to isolate each cluster member, we run \texttt{ELLIPSE} to measure and extract the isophotal parameters out to an arbitrary radius. This is then given to \texttt{BMODEL} to create the galaxy model. This galaxy model is subtracted from the cropped mosaic. This process is repeated for each cluster member, yielding an initial residual image. 

Although these initial models provide a good approximation of the bCGs and ICL, we adopt an iterative approach, repeating the modeling process on residual mosaics to improve the models and subsequent residuals. To do this, we construct intermediate images whereby all but one galaxy is subtracted. We model and subtract the best-fit model, and repeat the process for the rest of the sample. This allows us to better model the galaxy without any contamination from neighbors. We see a convergence after 10 rounds after which the models do not visibly change (see \citealt{shipley:18} for details). For crowded regions with multiple bCGs we allow for \se{} to recreate and improve the original mask as a replacement for the primary mask. The improvement is most noticeable for galaxies whose masks are strongly affected by multiple nearby bCGs.

The final model images are produced by averaging over the ten individual model images from each iteration. As found by \citealt{shipley:18}, further noise reduction of the averaged model image can be achieved by rejecting the lowest 4 and highest 2 values on a pixel-by-pixel basis. We use the IRAF task \texttt{IMCOMBINE} with the following parameters: \texttt{combine} = ``average'', \texttt{reject} = ``minmax'', \texttt{nlow} = ``4'', and \texttt{nhigh} = ``2'' \citep[see Section 3.1.3 of][for details]{shipley:18}. Finally, we subtract the average galaxy model from the cropped mosaic to produce the final residual mosaic. Figure~\ref{fig:icl_demo} shows the effect of this careful bCG and ICL subtraction relative to the original mosaic near the primary cluster cores.

With this final subtracted mosaic, we revert our initial cropping by using the IRAF \texttt{IMCOPY} task. We copy our subtracted mosaic onto the original mosaic. We then perform a sky subtraction to remove excess light near the edges of the galaxy models using a Gaussian interpolation. The background is measured using the \se{} \texttt{AUTO} setting with the following parameters: mesh size of 192 for SW bands (0.02\arcsec{} scale) and 96 for LW bands (0.04\arcsec{} scale), limiting magnitude of 15, and a maximum threshold of 0.01. The background subtraction does not significantly change the residual mosaic, but near the borders where the differences are well-defined due to our initial cropping this step smooths the previously defined edges.

As shown in Figure~\ref{fig:icl_demo}, there are residuals left near the bCG centers due to real structures not described by the smooth elliptical models. Given that the positive residuals are liable to be detected as sources and that photometry within their vicinity will be unreliable, we opt to mask objects detected near the known bCGs (see Section~\ref{sec:flags} for details). Furthermore, there is some stray ICL not immediately associated with known bCGs that is bright enough to not be removed by the final background subtraction. While photometry of objects detected near this stray residual ICL will be biased, the regions affected make up less than 1\% of the total area.

\section{Source Detection and Photometry} 
In this section we discuss the construction of our catalogs. Briefly, the sub-sections include the following: source detection, PSF homogenization, aperture photometry, corrections to photometry accounting for magnification via strong gravitational lensing, identification of star candidates, and a general recommended `use' flag. Catalogs corresponding to Data Release 2 (DR2) are available online.\footref{foot:catalog} The software used to produce these catalogs, \texttt{aperpy}, is generally applicable to any \JWST/NIRCam data and is freely available.\footref{foot:aperpy} We report flux densities and their uncertainties in $F_\nu$ units of 10\,nJy corresponding to an AB magnitude zeropoint of 28.9.

\label{sec:catalog}
% mention data release + zeropoint
\subsection{Source Detection} \label{sec:detection}
\begin{figure*}
    \centering
    \includegraphics[width=\textwidth]{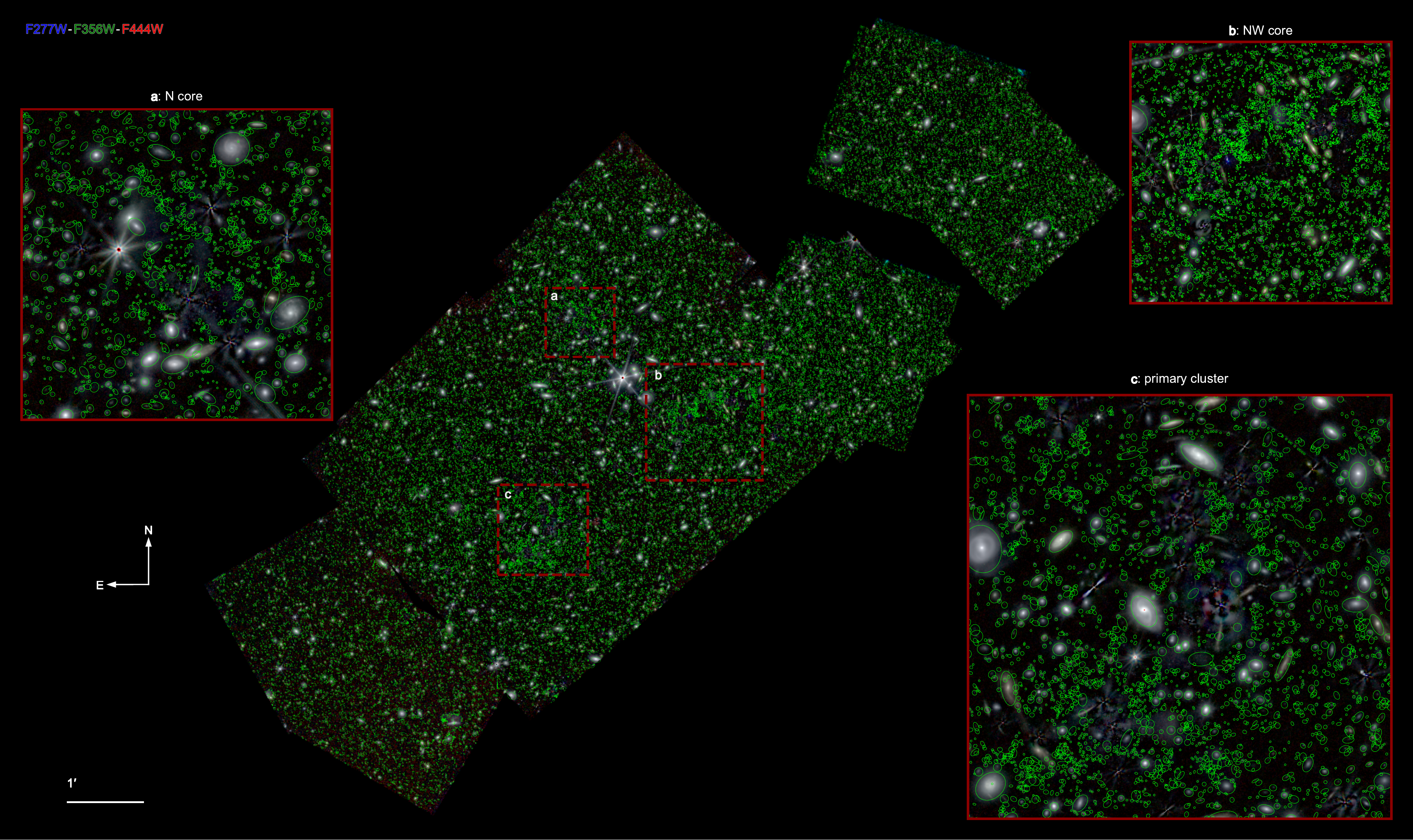}
    \caption{Color composite image of the \JWST footprint of Abell~2744, with three cluster cores highlighted. Bright cluster galaxies and intra-cluster light have been subtracted. Apertures consistent with our super photometric catalog are shown in green for reliable objects (see Section~\ref{sec:flags}).}
    \label{fig:detection_mosaic}
\end{figure*}

\begin{figure*}
    \centering
    \includegraphics[width=\textwidth]{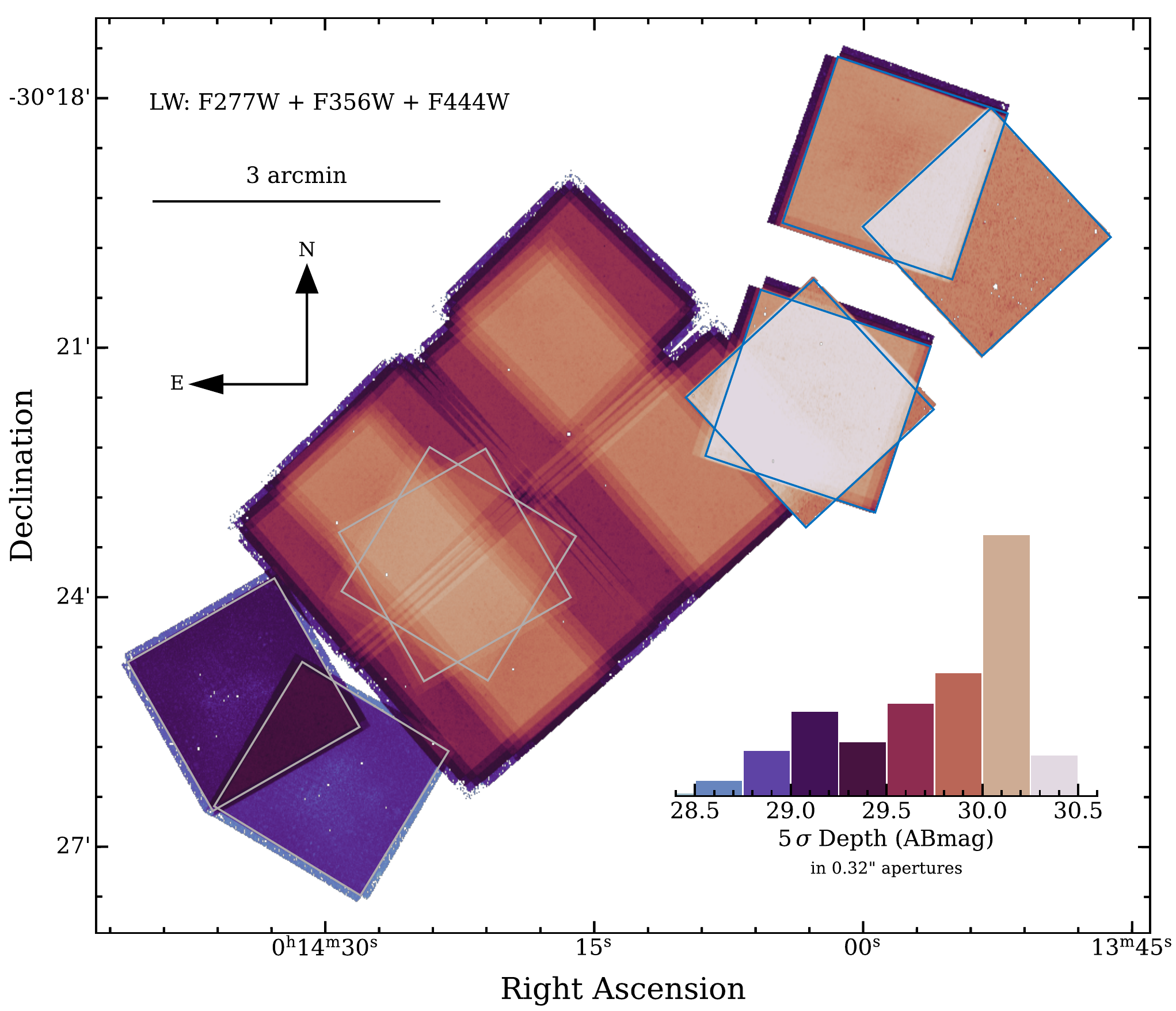}
    \caption{Schematic of depth variation across the noise-equalized LW coadd detection image aggregated from the DDT (grey), GLASS (blue), and UNCOVER programs. Combining F277W, F356W, and F444W, the 5\,$\sigma$ depths in our LW detection band measured in 0.32\arcsec apertures span $28.5-30.0$\,mag. Poisson contributions of the brightest objects feature prominently in our \JWST weight maps. Although provided in our mosaic release, the UNCOVER NIRISS parallel imaging is not yet cataloged and so is not shown. }
    \label{fig:detection_depths}
\end{figure*}

Sources are detected on a sky-subtracted noise-equalized (i.e. inverse variance weighted) co-added image based on our deepest \JWST imaging in the three LW broadband filters: F277W, F356W, and F444W. Detection is performed with \texttt{SEP} \citep{Barbary2016}, adopting the configuration listed in Table~\ref{tab:sep_config}. While aperture photometry is performed on point spread function (PSF) matched images (Section~\ref{sec:psfmatching}), we combine LW images at their native resolution to maximize sensitivity to identify faint sources.  We detect 61\,648 sources across the 49\,arcmin$^{2}$ area. Figure~\ref{fig:detection_mosaic} shows 
a RGB image of the bCG-subtracted detection co-add 
with ellipses marking 
all non-flagged (see Section~\ref{sec:flags}) sources detected from the bCG-subtracted LW co-add.
The effective catalog depth in 0.32\arcsec{} apertures of the noise-equalized LW bCG-subtracted co-add detection image is shown in Figure~\ref{fig:detection_depths} (see Section~\ref{subsec:depthcurve} for further details).

\begin{table}[h]

\centering
\renewcommand{\arraystretch}{0.9}
\begin{threeparttable}
\caption{\textsc{SEP} parameters used for source detection on the noise-equalized F277W+F356W+F444W co-added image. Not supplying a weight map implicitly tells \texttt{SEP} to use \texttt{THRESH\_TYPE} = \texttt{ABSOLUTE}, suitable for noise-equalized detection images.}

\begin{tabular}{ll}
 \hline \hline
Name & Value \\
 \hline
\texttt{KERNEL} & \texttt{3.5\,pixel FWHM Gaussian} \\
\texttt{MINAREA} & \texttt{3\,pixels} \\
\texttt{THRESH} & \texttt{1.2\,$\sigma$} \\
\texttt{DEBLEND\_NTHRESH} & \texttt{32} \\
\texttt{DEBLEND\_CONT} & \texttt{0.0001} \\
\texttt{CLEAN} & \texttt{N} \\

 \hline
\end{tabular}
\label{tab:sep_config}
\end{threeparttable}
\end{table}

\subsection{Point Spread Function Matching} \label{sec:psfmatching}
Before extracting aperture photometry, the PSF of each image is matched (or ``homogenized''). This approach allows for consistent photometric measurements within the same aperture size across all bands, which leads to a better recovery of source colors, \zphot{}, and physical parameters. We adopt our longest wavelength NIRCam band, F444W, as our target PSF. This choice is motivated by F444W being our broadest NIRCam PSF, meaning that the corresponding images are matched to the lowest resolution. Additionally, F444W will probe the reddest rest-frame light (e.g. $1-2\mu$\,m stellar bulk) at $z\gtrsim1$, making it an ideal band with which to derive total fluxes from our aperture photometry. Preserving the original F444W image properties will maximize consistency within the photometry.

Following methodology described in \citet{Skelton2014} and \citet{Whitaker2019}, we generate empirical PSFs in all \HST and \JWST bands using stars identified within the FOV. Point sources are known to inhabit a locus within a size-magnitude plane where size is approximated here by the ratio of flux in 0.16\arcsec{} to 0.32\arcsec{} diameter apertures and magnitude within 0.32\arcsec{} diameter apertures. Instead of a selection box, we fit the slope based on a first pass selection of stars. Sources with aperture magnitudes fainter than 24\,AB or outside the locus by 2.8$\sigma$ are rejected, with the total number of candidates in each filter varying from 14 (F090W) to 120 (F814W). The candidates are extracted in stamps, re-centered using cubic interpolation based on their center of mass evaluated in a window around the initial stamp center, and then normalized to unity within 16 pixels (0.64\arcsec{} diameter). Final PSFs for each filter are constructed by averaging the centered and normalized stamps, discarding any pixel whose flux is outside 3$\sigma$ of the flux distribution at that position. Lastly, we renormalize each empirical PSF such that its energy enclosed within a 4\arcsec{} diameter aligns with typical calibration levels.\footnote{\noindent\HST/ACS: \url{https://www.stsci.edu/hst/instrumentation/acs/data-analysis/aperture-corrections},\\ \HST/WFC3: \url{https://www.stsci.edu/hst/instrumentation/wfc3/data-analysis/photometric-calibration/ir-encircled-energy},\\ \JWST/NIRCam: \url{https://jwst-docs.stsci.edu/jwst-near-infrared-camera/nircam-performance/nircam-point-spread-functions}}

We choose to build empirical PSFs for \JWST/NIRCam as opposed to using the simulated PSFs from WebbPSF \citep{Perrin2012, Perrin2014} as we find that all PSFs provided by WebbPSF are $1-2\%$ narrower than real stars in our mosaics even after accounting for broadening introduced during image reduction. Further details can be found in Appendix~\ref{app:nircam_psfs}.

We note that our relatively simple technique does not consider the inhomogenous position angles (PAs) of the various observations in each mosaic. For example, due to a guide star failure 1 of the 4 visits of UNCOVER was observed at a slightly different PA whose effect is visible in the mosaics. This is in addition to the significantly different PAs of the other two programs in the field. Additionally, while the expected spatial and temporal variation of the PSF as measured in circular apertures appears negligible based on predictions from WebbPSF (see Appendix~\ref{app:psf_stability}), this has not yet been measured empirically. A more sophisticated treatment of the PSF including spatial and rotational variation will be explored in future work.

\begin{figure}
    \centering
    \includegraphics[width=0.48\textwidth]{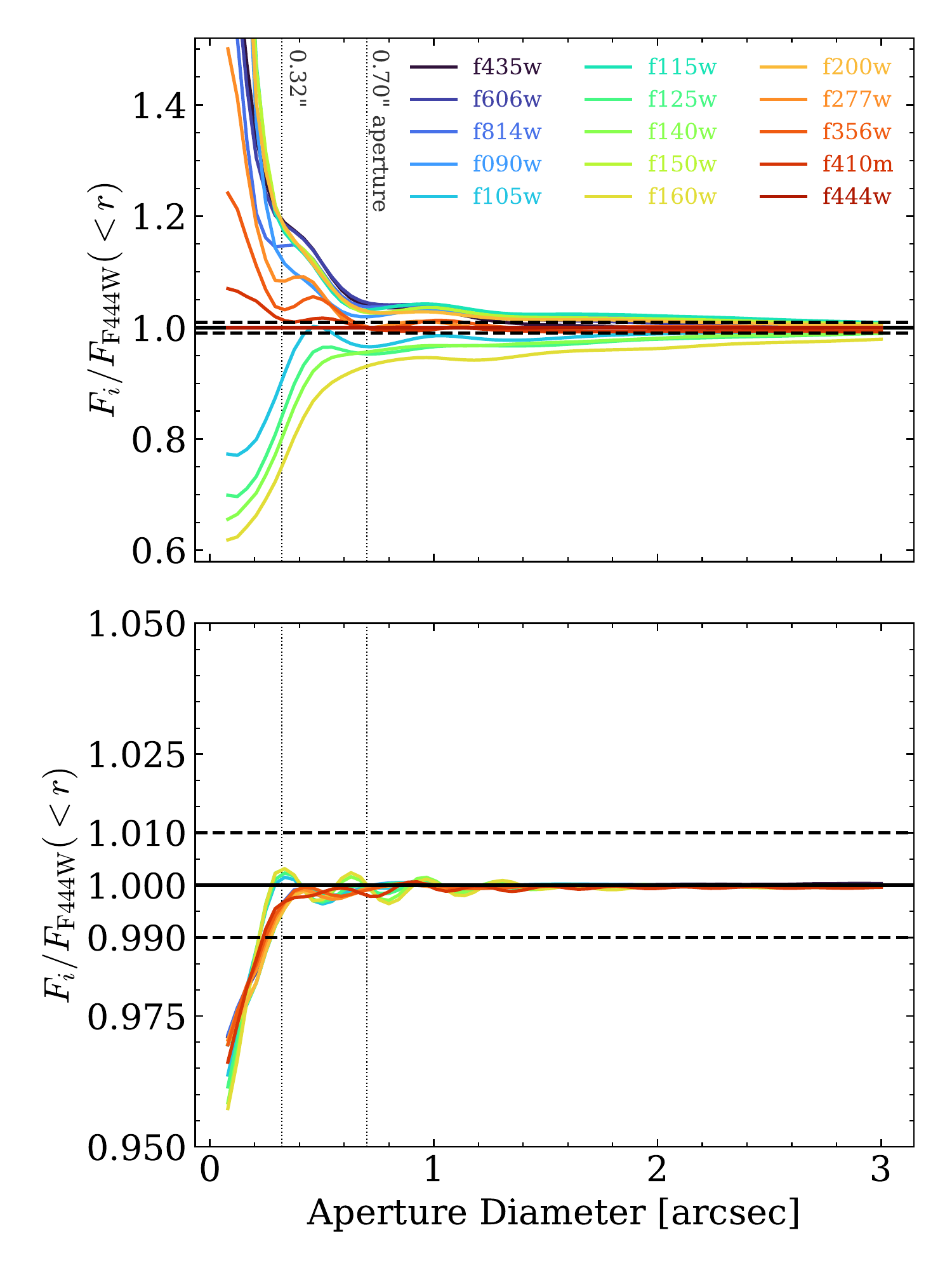}
    \caption{PSF growth curves for each filter before (top) and after (bottom) matching to F444W. After matching, all filters have deviations below the 1\% level at the smallest aperture diameter used (0.32\arcsec{}). Growth curves are shown relative to the F444W growth curve; a value of 1 indicates perfect matching with F444W. Dashed lines indicate the $\pm1\%$ deviations from exact matching (solid black line). Dotted lines indicate the location of 0.32\arcsec{} and 0.70\arcsec{} aperture diameters.}
    \label{fig:psf_matching}
\end{figure}

Kernels are produced using \textsc{Pypher} \citep{Boucaud2016}. \textsc{Pypher} generates PSF-matching kernels using an algorithm based on Wiener filtering \citep{Wiener1949} with a tunable regularization parameter, which we set to $3\times10^{-3}$. Matching is done in $3\times$ oversampled space, after which the kernels are rescaled to the original pixel scale. The lower value of the regularization parameter helps avoid high-frequency noise in the kernels, while still maintaining $<1$\% deviations at all aperture diameters of interest. All filters are matched to the reference filter (F444W). Figure \ref{fig:psf_matching} shows the PSF growth curves of every filter relative to the F444W growth curve, both before (top) and after (bottom) convolving with a matching kernel. Matched PSFs have almost identical growth curves to the reference filter, with deviations below the 1\% level. The \HST NIR filters (F105W, F125W, F140W, F160W) have the most variation, but this is only significant for aperture diameters smaller than 0.32\arcsec{}.  This discrepancy is unavoidable given the systematic differences in the shapes of the PSF between \HST and \JWST, where the latter has significantly more sub-structure and ``snowflake''-like patterns that complicate the PSF matching. This substructure should not significantly impact our photometric measurements, as it is azimuthally averaged out within the circular apertures.

Alternatively, we have similar success matching PSFs with \textsc{Photutils} \citep{photutils}, which uses ratios of Fourier transforms to generate a matching kernel. \textsc{Photutils} also requires the selection of one of several window functions, used to filter high-frequency noise from the Fourier ratios. The best window function for a given reference PSF and the ideal values for the tuning parameters used to scale the window function are not straightforward and usually require testing different combinations. While \textsc{Photutils} has similar PSF-matching success to \textsc{Pypher}, it is less convenient due to the additional parameters that need tuning. Finally, we find that PSF-matching methods that utilize linear combinations of shapelets (e.g. Gauss-Hermite or Gauss-Laguerre polynomials) to generate kernels \citep[e.g.,][]{Skelton2014} should be avoided when matching to \JWST filters. The intrinsic symmetry of these functions cannot effectively match the rotational asymmetry of the diffraction spikes and extended structure of \JWST PSFs. However, shapelet-based algorithms perform the best for PSF-matching to an \HST reference filter (e.g., F160W).

\subsection{Aperture Photometry} \label{sec:photometry}
Photometry is measured in 0.32\arcsec{}, 0.48\arcsec{}, 0.70\arcsec{}, 1.00\arcsec{}, and 1.40\arcsec{} diameter circular apertures with \textsc{SEP}. These ``color'' aperture measurements are then corrected to total fluxes by scaling them by the ratio of the flux estimated from elliptical Kron-like apertures \citep{Kron1980} to the flux measured in consistently sized color apertures. To ensure that all detected objects have robust Kron measurements, photometry is extracted on a F444W-matched inverse variance weighted F277W+f356W+F444W co-added image (consistent with the construction of our detection image) in elliptical apertures whose semi-major and semi-minor axes are grown by a factor of $2.5\times$ the Kron ``radius'' (a unitless factor), with unity as the minimum allowed factor. Then, for sources whose circularized Kron radius is less than the circular aperture radius, the circular apertures are used instead of the Kron-scaled ellipse. 
The resulting measurements from this procedure are commonly referred to as ``AUTO'' flux densities. Note that unlike \textsc{Source Extractor}, \texttt{SEP} does not mask neighboring objects by default and so may produce catastrophically large Kron radii in certain cases. As of v1.1, \texttt{SEP} now provides functionality to mask neighboring sources which in our case dramatically reduces the number of such failures. We leverage this functionality.

We additionally correct each measurement for light missed at large radii (especially important for \JWST) by dividing the flux measurement of each source by the fraction of the total light from the F444W PSF curve of growth within each respective circularized Kron radius. We stress that this correction for the \JWST bands in particular is on the order of $10-20$\%, significantly larger than for \HST due to the large fraction of light characteristically scattered to large radii in \JWST PSFs.

\begin{figure*}
    \centering
    \includegraphics[width=\textwidth]{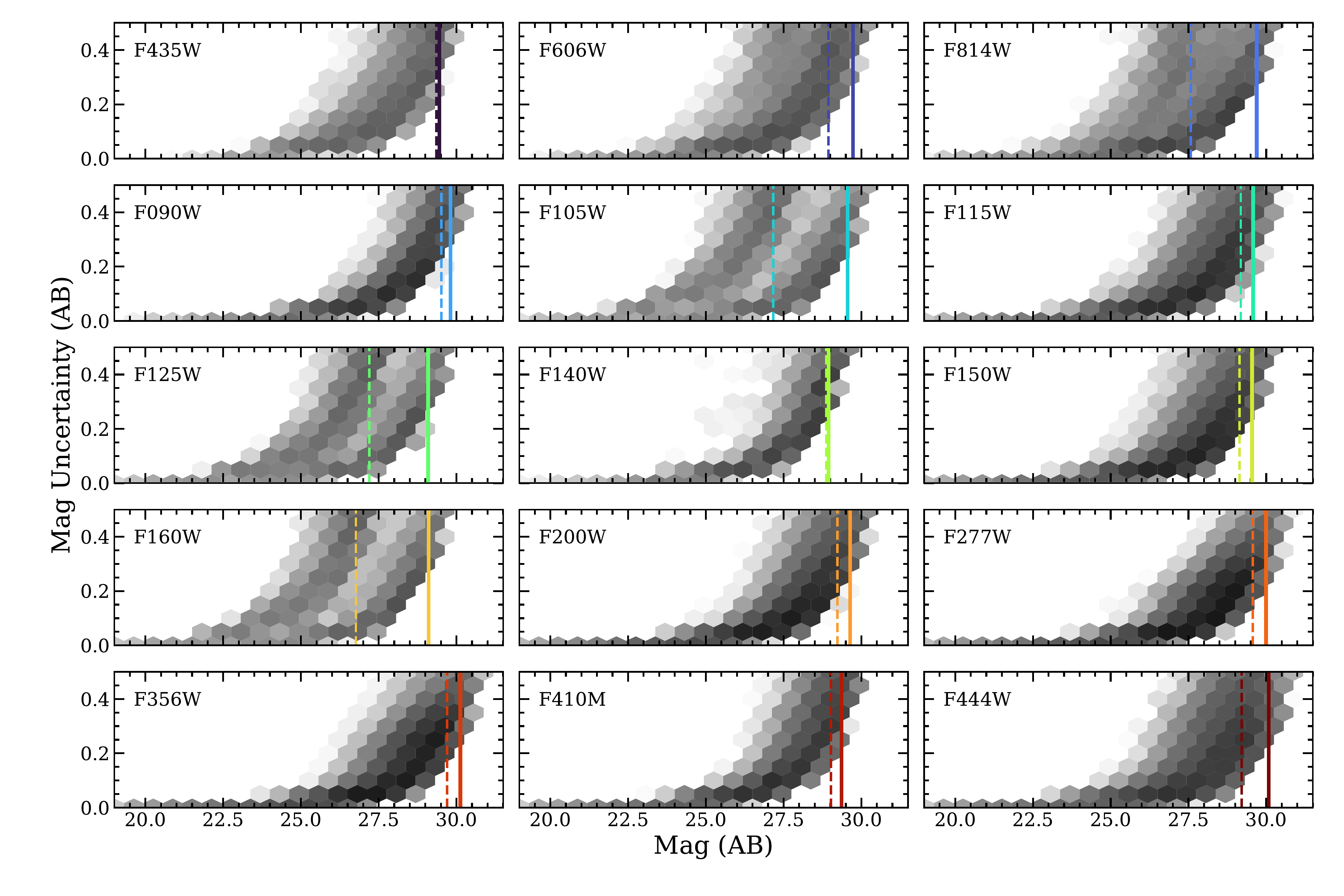}
    \caption{Photometric uncertainty as a function of magnitude shown by log$_{10}$-scaled 2D histograms for total fluxes derived in 0.32\arcsec{} diameter apertures. Catalog depths for each filter measured in the same aperture size corresponding to the 50$^{th}$ and 10$^{th}$ percentile areas are indicated by the dashed and solid lines, respectively.}
    \label{fig:unc_corr}
\end{figure*}

Photometric uncertainties are derived by means of an independent estimate of the background noise. For each band we place 10\,000 circular apertures in regions outside detected sources, within our detection footprint, and with good coverage in that band. We opt to use the segmentation maps to mark detected sources, making the placement of the apertures dependent not only on the footprint of the detection image, but also its union with the footprint for that particular filter. For a given filter, apertures are placed on the corresponding noise-equalize image to account for the variation in depth across the field (e.g., see Figure~\ref{fig:detection_depths}). Outlier measurements greater than 5$\sigma$ are removed, and the width of the flux distribution is estimated by the standard deviation, and is unitless. To obtain physical noise estimates consistent with our photometry, the width is multiplied by the effective local noise around each source estimated by the inverse of the square root of the median weight computed within a $9\times9$ pixel box. Total flux errors are then computed by multiplying the resulting per-object noise by the ratio of total to aperture flux measured on the F444W-matched F277W+F356W+F444W co-add. 

Figure~\ref{fig:unc_corr} illustrates the growth of photometric uncertainty with magnitude. While the growth of photometric uncertainty in some \HST bands show dual loci corresponding to the shallower BUFFALO and deeper HFF observations, that of \JWST bands follow many loci which are blended together due to the multi-modal depths produced by the overlapping DDT, UNCOVER, and GLASS programs (see also Section~\ref{subsec:depthcurve}).

Given the extreme depths of our detection images and the crowded nature of galaxy clusters, 17\,695 objects (29\%) are flagged as potentially being blended (\texttt{FLAG\_KRON}). Estimates of Kron radii are known to fail in such cases, and so for these sources we do not correct to total flux using their likely corrupted Kron-aperture fluxes. Instead we simply correct for the missing light beyond the color aperture using the F444W curve of growth, assuming point-source morphology. In most cases this will underestimate the total flux and some physical parameter estimates (e.g., stellar mass), while leaving their redshifts and colors robust. There also exist an additional 5\,561 sources (9\%) small enough that their circularized Kron diameters are less than 0.32\arcsec{} and so their total flux is robustly estimated by applying the same point-like correction to total using the F444W curve of growth. Together these two sub-samples make up the 23\,256 objects (38\%) flagged as \texttt{USE\_CIRCLE}.

While providing photometry in five different aperture sizes is useful for comparisons and cross-checks, it is liable to produce awkward workflows and confusion. To improve the accessibility of these early \JWST catalogs, we additionally build a `super' catalog following \citet{Labbe2003} who use the isophotal areas for each object based on the detection image (i.e. the number of pixels in their segment). In short, unblended objects are assigned photometry corresponding to the smallest aperture that does not exceed the size of an aperture that would enclose the isophotal area. For blended objects, however, their true isophotal area is uncertain and so to be conservative we shrink the equivalent isophotal aperture diameter by 20\%. This factor is dependent on the particular data set and is determined experimentally, see Section~5.2 of \citeauthor{Labbe2003} for details. In rare cases where most suitable aperture does not provide reliable photometry, such as a large aperture containing a masked pixel, the next largest usable aperture is chosen. Overall, 48\,360 sources (78\%) of the super catalog are assigned photometry based on 0.32\arcsec{} apertures, with 5\,648 and 3\,267 sources being assigned photometry based on 0.48\arcsec{} and 0.70\arcsec{} apertures, respectively; 4\,373 sources are assigned either 1.0\arcsec{} or 1.4\arcsec{} apertures. We encourage users to default to this super catalog for most science applications.

Photometry is corrected for line-of-sight attenuation through the Galaxy, adopting the dust maps of \citet{Schlafly2011} and the attenuation law of \citet{Fitzpatrick2007}. Given the small footprint, we opt to apply the median ${\rm E(B-V)}=0.01$. The dust column density in the direction of Abell~2744 is favorably low, resulting in attenuation corrections in each band on the order of 1\% or less. We report flux densities and their uncertainties in $F_\nu$ units of 10\,nJy corresponding to an AB magnitude zeropoint of 28.9.

\subsection{Source Magnification}
Abell~2744 is one of the most powerful lensing clusters known \citep[e.g.][]{Merten2011, Jauzac2015}. According to the most recent estimates of all three cluster cores by \citet{Furtak2022}, Abell~2744 magnifies most objects in our survey footprint by at least $\mu=2$, with objects seen nearest to the cluster cores magnified by $\mu\sim10-100$ (see Figure~5 in \citeauthor{Furtak2022}). Magnifications and shear parameters for sources in each catalog are computed from the latest version (\texttt{v1.1}\footnote{\noindent Publicly available at \url{https://jwst-uncover.github.io/DR2.html\#LensingMaps}}) of the \citet{Furtak2022} analytic strong lensing mass model assuming \zphot{} derived from \eazy{} in Section~\ref{sec:photoz} (and from \zspec{} where available). The \texttt{v1.1} of the lens model includes an additional spectroscopic redshift from \citet{Bergamini2023} and two additional multiple image systems compared to \texttt{v1.0} presented in \citet{Furtak2022}. The model is thus constrained by 141 multiple images belonging to 48 sources and achieves a lens plane RMS of $\Delta_{\mathrm{RMS}}=0.51\arcsec$. Future work presented in B. Wang et al. (submitted) will provide updated magnification estimates using \zphot{} from \textsc{Prospector-$\beta$} \citep{Johnson2021, Wang2023}. Note that while the magnifications are provided in the catalogs, measurements (e.g. fluxes) are \textit{not} corrected for magnification.

\subsection{Identifying Stars}
\label{sec:stars}

\begin{figure*}
    \centering
    \includegraphics[width=0.95\textwidth]{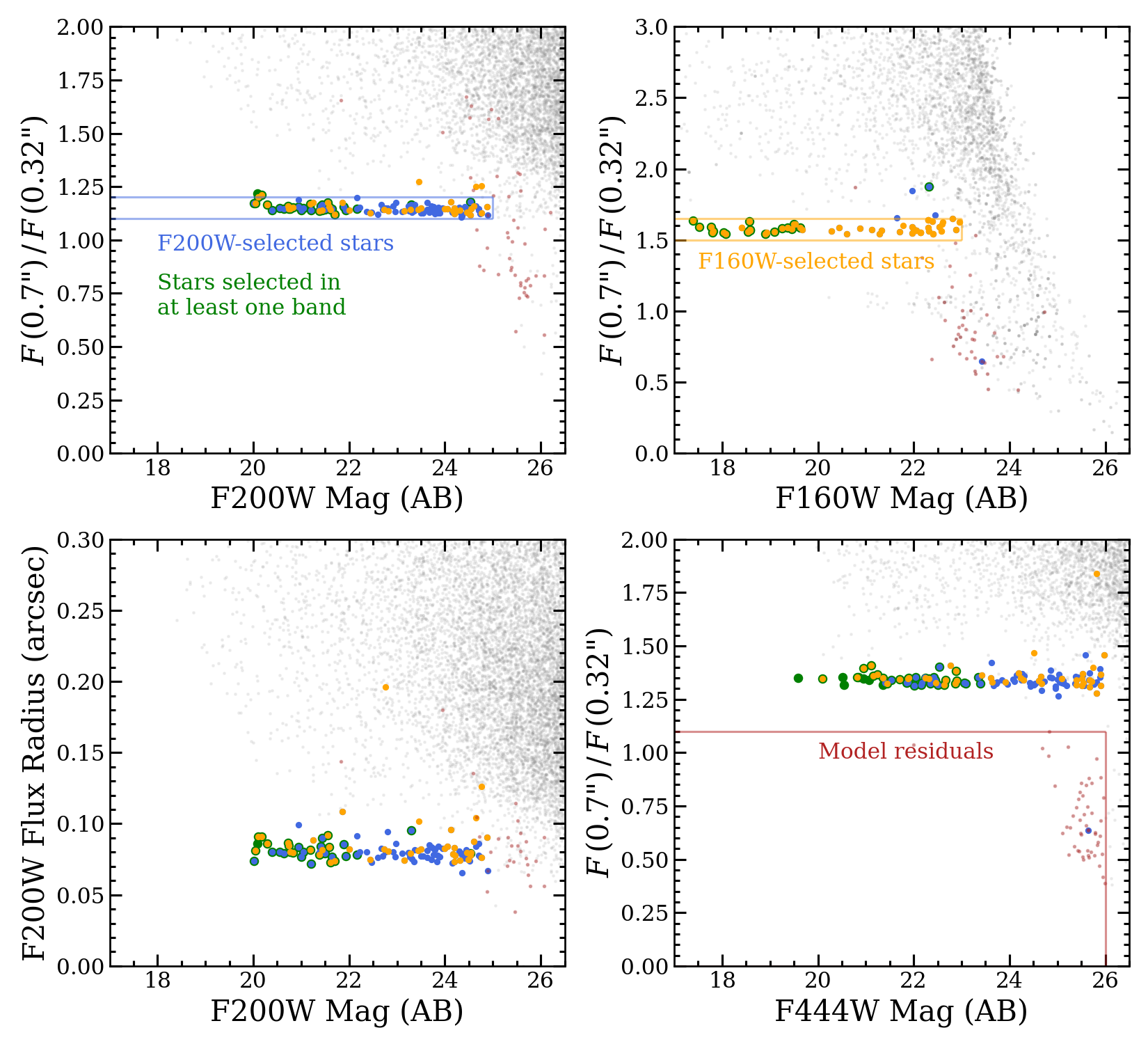}
    \caption{The selection of stars and spurious objects (bad pixels or model residuals) are determined from the flux ratio in large to small apertures above a limiting magnitude threshold. Stars are identified from the \JWST/F200W image (blue) at positions corresponding to the catalog, and/or directly from the \HST/F160W image (yellow) to account for high proper motion. We also include any star that contributed to PSF models in any band (green) that has a match to an object in the catalog. Some artifacts such as bCG model residuals can also be similarly flagged (red).
    }
    \label{fig:badflag}
\end{figure*}

% why this is now a challenge
Constructing reliable photometric catalogs of extragalactic sources requires identifying foreground stars and other spurious sources that would otherwise contaminate galaxy samples. While no one identifier is complete and strictly pure, stars in \HST phototmetric surveys have been traditionally identified by the fact that they are typically bright and unresolved point sources, noting that not all point sources are stars (e.g. quasars). However, due to the staggering efficiency of \JWST/NIRCam, stars of similar brightness often saturate the detector pixels making their identification surprisingly difficult.

% our method
To overcome this new obstacle, we first identify stars using traditional methods as described in \citet{Skelton2014}, with fluxes measured on the original, non PSF-matched mosaics. Figure~\ref{fig:badflag} presents stars selected in F200W (blue) having a flux ratio in 0.70\arcsec{} over 0.32\arcsec{} apertures brighter than 25.0\,AB. While this selection alone accounts for most of the stars, there are a handful that are better identified from \HST F160W where their profiles are less affected by saturation. However, despite measuring stellar sizes on the native resolution \HST images, we found that the width of the stellar locus was surprisingly large. Further inspection revealed that a number of stars have sufficiently high proper motions that both their \JWST-derived centroids miss their \HST era positions. As a workaround, we have added an additional flag that identifies stars detected on the F160W image itself having a flux ratio in 0.70\arcsec{} over 0.32\arcsec{} apertures between 1.5 and 1.65, and are brighter than 23 AB in 0.7\arcsec{} diameter apertures. For completeness, we also flag any object that contributed to a PSF model (see Section~\ref{sec:psfmatching}) that has a match in our catalog (green). A total of 113 ($<1\%$) stars are flagged by at least one of these criteria. Note that bright stars that are saturated in our LW detection image are not flagged as stars because they are not detected as single objects; instead each drives multiple spurious detections near their saturated cores that we classify as artifacts (see Section~\ref{sec:flags}).

% caveats and choices
However, a point-like indicator is unsuitable for the faintest stars that are likely to intermix with the general galaxy population. Some literature studies have opted to identify stars by comparing their fit quality between galaxy and stellar templates, either alone or in combination with resolution criteria, e.g. \citet{Weaver2022}. To enable these comparisons, we use \eazy{} to quantify the goodness-of-fit of our spectral energy distributions (SEDs) to theoretical \texttt{PHOENIX} BT-Settl stellar templates \citep{Allard_2012} and include the corresponding $\chi^{2}$  estimates in our catalog. However, we \textit{do not} use $\chi^{2}$ to flag additional faint stars. Such a comparison is prone to incorrectly rejecting a non-zero number of potentially interesting and little known objects for which we have poor spectral templates, e.g. high-$z$ galaxies. Consequently, galaxy candidates fainter than $\sim25$\,mag are liable to be contaminated by difficult to identify foreground stars and so require additional scrutiny. 

\subsection{Getting Started: The Use Flag}
\label{sec:flags}

A ``use'' flag is particularly useful when familiarizing oneself with any given photometric catalog.  Ideally, one simple selection yields a clean sample of galaxies across cosmic time for further analysis.  Following \citet{Skelton2014}, the \texttt{USE\_PHOT} flag in our photometric catalog requires stars to be removed (see Section~\ref{sec:stars}) and sets a minimum signal-to-noise of 3 in a color aperture measured on the F277W+F356W+F444W co-add to ensure robust Kron radii and aperture-to-total corrections.  However, with the addition of the novel \JWST photometry, there are further selections required to sufficiently clean up the catalog as described below.  For this first generation, we opt to be more conservative at the expense of completeness to benefit more exotic parameter spaces (e.g., extreme high redshifts). 

% residuals
While our bCG modeling and subtraction enables searches for objects within the immediate vicinity of the bCGs, it also produces undesirable residual features in some cases which are detected in our catalogs. We find that these artifacts can be identified in F444W having a flux ratio less than 1.1 with magnitudes brighter than 26\,AB (see bottom right panel of Figure~\ref{fig:badflag}). Additionally, there are a number of artifacts found near the cores of bright stars that are saturated in \JWST. We identify them as being within 3\arcsec{} of the center of groups of saturated pixels at least 10 pixels in size. We also manually build a star spike mask and flag any object whose centroid is within the mask as an artifact. Several bright, saturated stars are flagged as artifacts as a by-catch; they are not flagged as stars even though they fall within the stellar locus of one or more bands in Figure~\ref{fig:badflag}. The catalog includes a column \texttt{FLAG\_ARTIFACT} to signal the 5\,112 objects (8\%) satisfying any of these selections (and others, see below).

The areas immediately surrounding the subtracted bCGs are in all cases contaminated by residual structures. We further flag all detections within a conservative 3\arcsec{} radius of a known bCG centroid computed as part of the modeling procedure. The catalog includes a column \texttt{FLAG\_NEARBCG} to signal the 3\,336 objects (5\%) satisfying this selection. These objects are also flagged under \texttt{FLAG\_ARTIFACT}, making up 65\% of the objects flagged as such. While we caution that some residual features outside this radius remain unflagged, it is not possible to perfectly flag genuine residual features without also catching real sources of interest. A more sophisticated treatment will be explored in future work.

% explain the use flag
Together, we build a single \texttt{USE\_PHOT} flag which removes undesirable objects, including known stars (\texttt{FLAG\_STAR}), sources within 3\arcsec{} of bCGs (\texttt{FLAG\_NEARBCG}) as well as objects affected by bright star spikes, spurious detections near saturated star cores, and other model residual features (\texttt{FLAG\_ARTIFACT}). While the super catalog contains zero sources with unreliable photometry (\texttt{FLAG\_NOPHOT}), this number is non-zero in the single aperture catalogs, especially in the larger ones. To ensure the reliability of total photometry based on F277W+F356W+F444W Kron radii, the ``use'' flag also excludes 956 objects (2\%) with a SNR$<3$ in the F277W+F356W+F444W color aperture, or was masked after detection (\texttt{FLAG\_LOWSNR}). By selecting all objects where \texttt{USE\_PHOT}=1, this reduces our total sample to 55\,613 galaxy candidates with reliable photometry.

\section{Catalog Properties} \label{sec:properties}
With photometry in hand, we summarize the key properties of our catalog including effective catalog depths, galaxy number counts, photometric redshifts, and a brief comparison of measured \JWST photometry to that expected from the best-fit SED template solutions. Although we provide catalogs based on five individual aperture sizes, in this section we will refer to the more broadly applicable `super' catalog.

\subsection{Photometric depths}\label{subsec:depthcurve}

\begin{figure}
    \centering
    \includegraphics[width=0.45\textwidth]{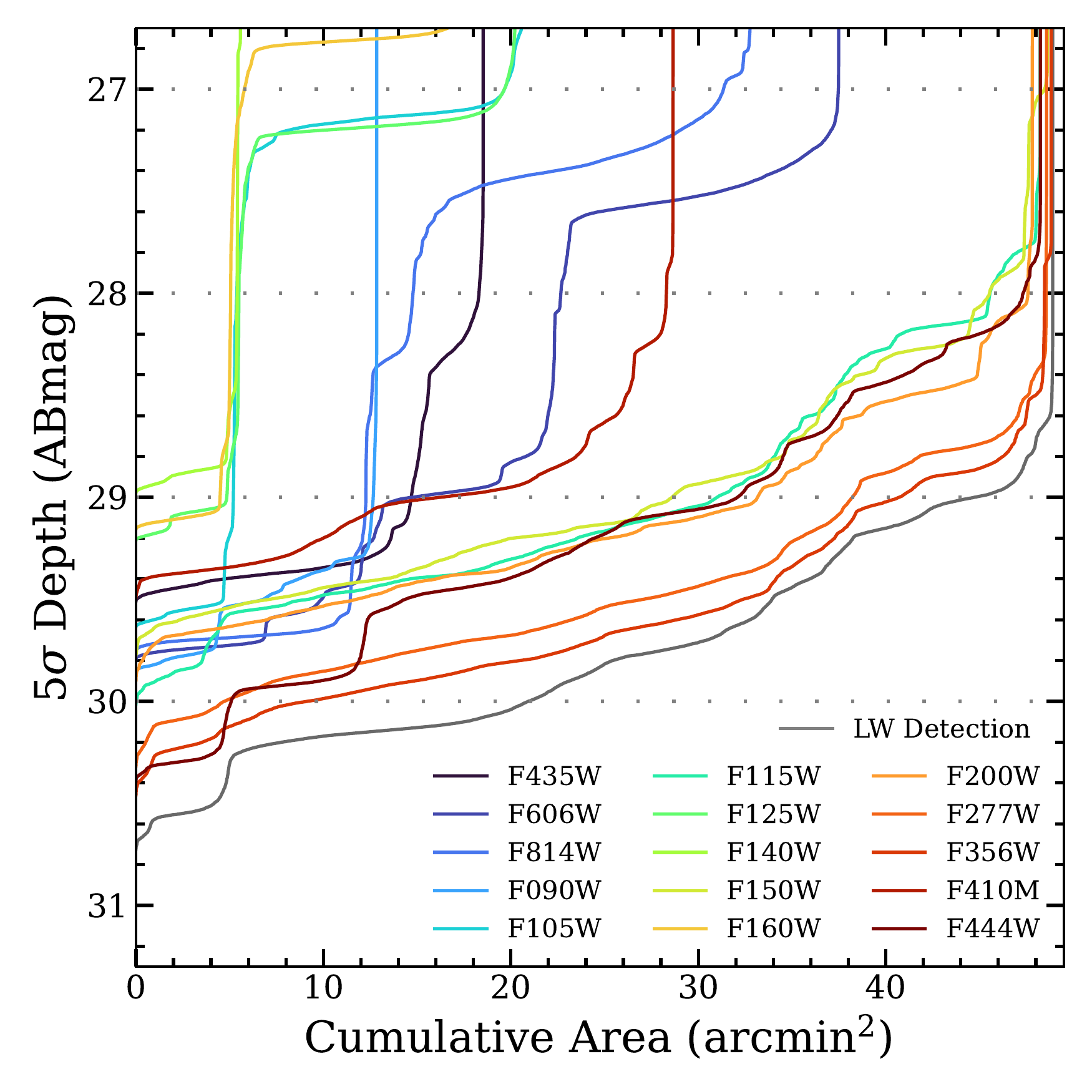}
    \caption{Depth as a function of cumulative area for each filter mosaic as well as the LW detection image. Measurements are taken from 10\,000 empty apertures of 0.32\arcsec{} diameter, per filter. Grey dotted lines mark depths at 27, 28, 29, and 30\,AB~mag.}
    \label{fig:depthcurves}
\end{figure}

As introduced in Section~\ref{sec:intro}, the Abell~2744 imaging consists of three overlapping \JWST and several overlapping \HST programs. Consequently, the depth of any of these mosaics cannot be fully described by a single value. While the photometric uncertainties are computed to account for this variation, Figure~\ref{fig:depthcurves} explicitly illustrates the effective catalog depth of each filter as a function of cumulative area corresponding to 0.32\arcsec{} apertures. The depths of some of the \HST mosaics are roughly bimodal as a result of the deep HFF observations contrasted with the wider but shallower BUFFALO coverage. Meanwhile for \JWST, the contribution of the shallow DDT observation from NIRCam's Module A is readily visible and the depths produced by including UNCOVER and GLASS create many small regions of varying depths elsewhere.

\subsection{Galaxy Number Counts}

\begin{figure*}
    \centering
    \includegraphics[width=\textwidth]{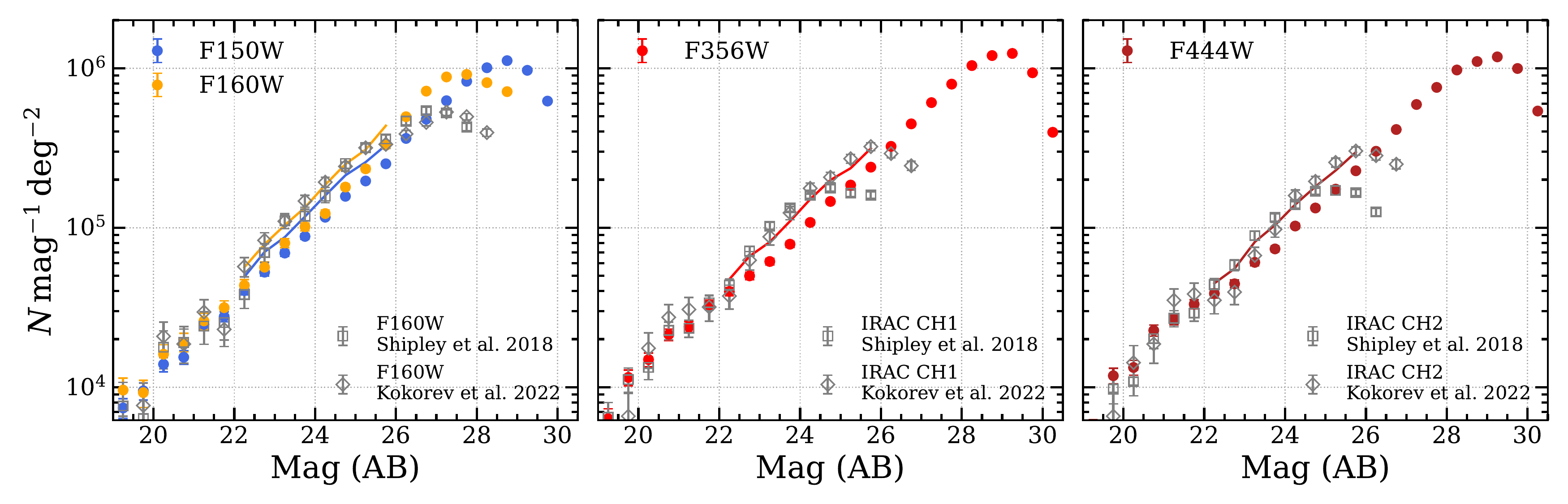}
    \caption{Galaxy number counts for sources identified from our LW detection image and measured in total magnitudes from F150W, F160W, F356W, and F444W. Stars and objects with unreliable photometry are removed. Effects of blending at intermediate brightness are shown by the colored curves where we allow Kron aperture corrections for blended sources. Literature F160W-selected counts in Abell~2744 by \citet{shipley:18} and \citet{Kokorev2022} are shown by the unfilled grey boxes and diamonds, respectively; note that bright cluster galaxies have been excluded. Errorbars denote 1\,$\sigma$ poisson uncertainties. Counts are not corrected for magnification.}
    \label{fig:number_counts}
\end{figure*}

We compute the galaxy number counts for our catalog in three \JWST/NIRCam bands (F150W, F356W, F444W) and one \HST/WFC3 band (F160W), shown in Figure~\ref{fig:number_counts}. The variation in image depth as well as source magnification make constructing number counts non-trivial, and so instead we estimate counts \textit{without} correcting for magnification. We assume the nominal total science areas listed in Table~\ref{tab:depths}. Areas for literature counts are taken from their respective papers. We remove any unreliable source with \texttt{USE\_PHOT}=0. We stress that these counts should \textit{not} be used to precisely quantify completeness nor survey depth, but are merely an accessible means by which to validate our catalogs against literature measurements. Efforts to quantify completeness are in progress (R. Pan et al., in preparation).

While these lensed counts cannot be directly compared to those from unlensed field surveys, they can be readily compared to other lensed counts also from Abell~2744 \citep{shipley:18, Kokorev2022} where F150W/F160W, F356W/Ch1, and F444W/Ch2 are sufficiently similar such that deviations above 5\% are significant (see Appendix~\ref{app:comparison}). To maximize consistency, we also remove any bCGs present in literature counts that are removed as part of this work. We find good agreement with the literature counts at bright magnitudes ($\lesssim22$\,AB) for all bands.

At intermediate magnitudes, our counts are lower due to real photometric differences. We hypothesize that the higher literature counts are caused by a combination of factors including blending from the shallower and lower resolution F160W imaging, the relatively large Kron apertures used by the literature catalogs, and the conservative circular apertures adopted for blended objects in our catalog (which are not included in the comparisons in Appendix~\ref{app:comparison}). To study this further, we allowed all objects to have their aperture fluxes corrected using the Kron ellipses despite obvious cases of blending confirmed by visual inspection. Doing so brings our number counts in agreement with the literature (solid curves in Figure~\ref{fig:number_counts}), suggesting that the literature counts are driven high by blending and hence justifying our conservative approach. In order to quantify the extent to which our fluxes are underestimated we compute aperture photometry on up a series of images containing PSF-convolved circularly symmetric galaxy light profiles with effective radii from \citet{vanderWel2014} corresponding to a $10^9\,M_\odot$ system at $z=1$, 2, 3, and 4. Variation in the light missed due to the assumed S\'ersic index and/or redshift are $\lesssim10$\%. After applying aperture corrections assuming a point-like object, we find that our smaller 0.32-0.70\arcsec{} apertures miss approximately half of the total light with the larger apertures missing $<10\%$. We note, however, that this is insufficient to elevate our number counts to match literature and so magnification notwithstanding, the true counts likely lie somewhere in between.

At faint magnitudes, our deep LW-selected catalog provides significantly more objects near the depth limit of the images out to 27\,AB in F160W relative to the literature. F150W counts go further still to $\approx29$\,AB, although they do not include the very bluest objects missed by our redder LW selection function. Furthermore, our F356W and F444W counts extend to significantly fainter populations compared to existing \Spitzer/IRAC data, making them some of the deepest ever obtained at these wavelengths. Although elementary, these observed galaxy number counts serve to demonstrate that we may confidently springboard from well-studied \HST surveys to probe orders of magnitude deeper with \JWST.

\subsection{Photometric Redshifts} \label{sec:photoz}

\begin{figure}
    \centering
    \includegraphics[width=0.45\textwidth]{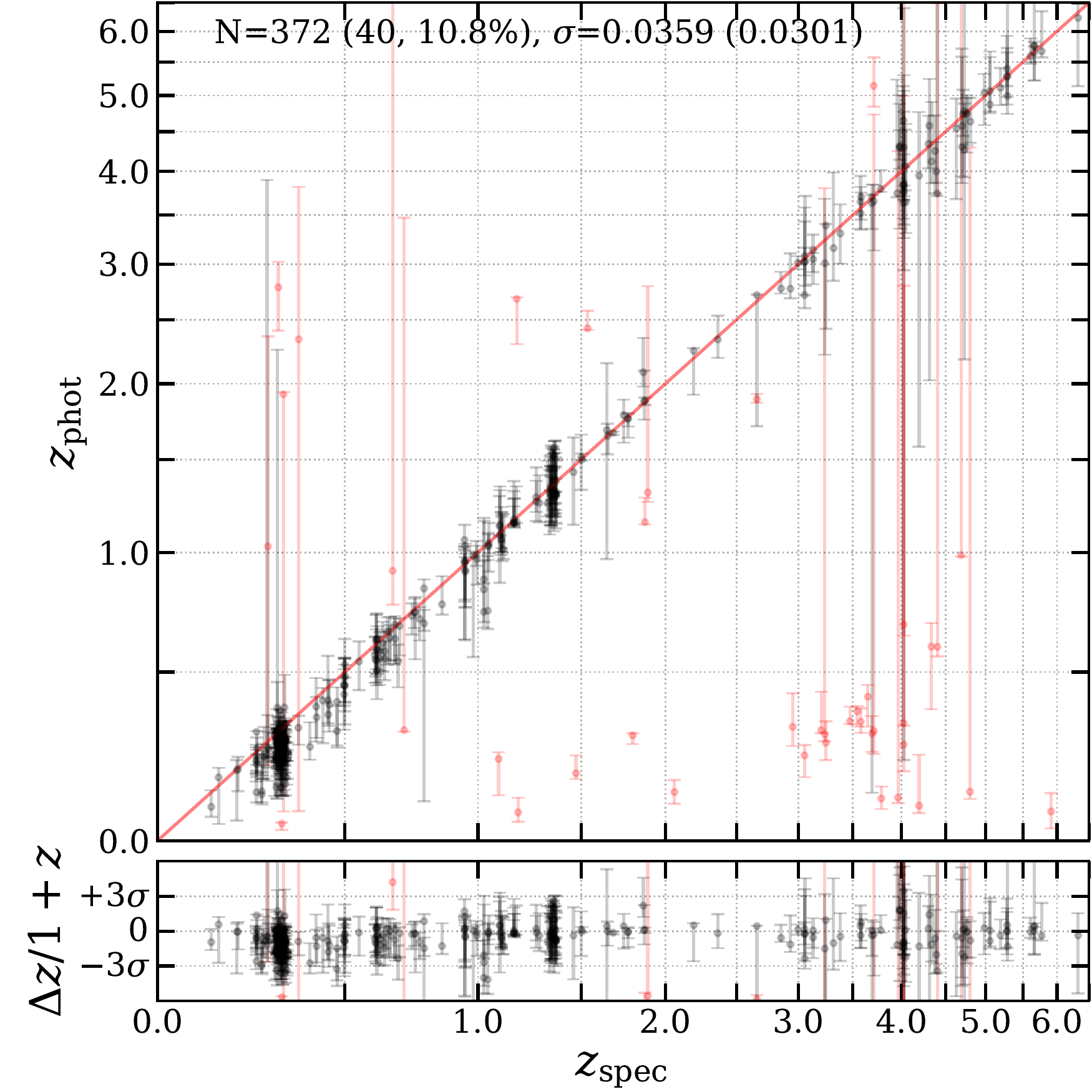}
    \caption{Performance of \zphot{} assessed by comparison with known \zspec{}, described in Section~\ref{sec:data:specz}. Outlier sources with \zphot{} wrong by more than $0.15\,\Delta z / (1+z_{\rm spec})$ are colored red. Lower panel shows residuals  in normalized units of $\sigma$ away from \zspec{}. $N=372$ sources are compared finding 40 outliers (10.8\%), an overall tightness $\sigma_{\rm NMAD}=0.0359$ and an outlier-removed $\sigma_{\rm NMAD}=0.0301$. \zphot{} are derived from super catalog SEDs fit with \eazy{} using the \texttt{SFHz\_CORR} template set.}
    \label{fig:zphot_zspec}
\end{figure}

\begin{figure}
    \centering
    \includegraphics[width=0.5\textwidth]{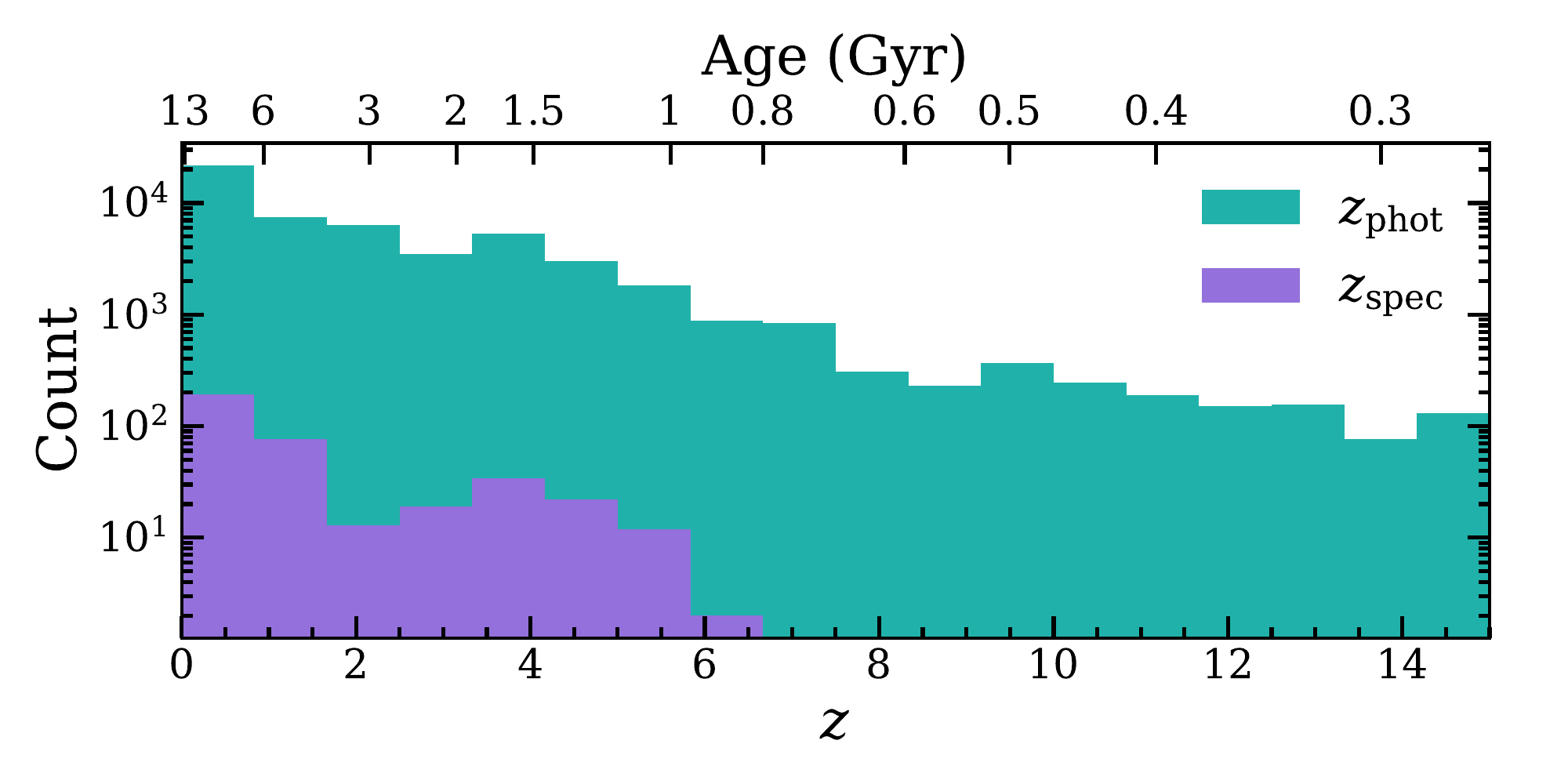}
    \caption{Distribution of \zphot{} (teal) and \zspec{} (purple) shown up to $z=15$. \zphot{} are derived from super catalog SEDs fit with \eazy{} using the \texttt{SFHz\_CORR} template set.
    }
    \label{fig:z_dist}
\end{figure}

In order to provide an impression as to photometric performance, we compute \zphot{} using \eazy{} \citep{Brammer2008}. We use all \HST and \JWST bands available for each object and set a minimum error floor of 5\%, an increase from the default 1\% to more realistically reflect the calibration uncertainties of \JWST/NIRCam. Given the considerable uncertainty as to the real high-$z$ galaxy SEDs, we forgo the usual pre-processing step of iteratively tuning zeropoints to avoid biasing our colors to those of the SED templates. Furthermore, any disagreements between our photometry and that predicted from \eazy{}'s models will be more readily visible (see Section~\ref{sec:validation}). We also turn off both magnitude and $\beta$-slope priors for similar reasons.

We compute \zphot{} for all five sets of photometry independently based on our bCG-subtracted imaging which are then combined to complement the super catalog. We additionally compute separate \zphot{} from two SED template sets: the default \texttt{FSPS\_FULL} set used frequently in the literature, and the newer \texttt{SFHZ\_CORR} set which features $z$-dependent priors on allowable star-formation histories and an observed $z=8.5$ emission line galaxy spectrum from \citet{Carnall2023} to provide realistic line ratios -- two considerations especially important for identifying robust high-$z$ galaxies candidates. We set $z=20$ as an upper limit.

One popular way of judging photometric accuracy is to compare \zphot{} and \zspec{}, where available. Of the N\,=\,518 spectroscopic sources (see Section~\ref{sec:data:specz}), we remove N\,=\,146 of them including 10 failed \eazy{} fits, and 136 unreliable objects include bright bCGs subtracted in our mosaics, stars, and artifacts (\texttt{USE\_PHOT}\,$=0$). \zphot{} performance is then assessed using the normalized median absolute deviation \citep[NMAD,][]{hoaglin83_MAD}, defined following \citet{Brammer2008},
\begin{equation}
\sigma_{\rm NMAD}=1.48\times\mathrm{median}\left(\frac{\vert \Delta z-\mathrm{median}(\Delta z)\vert}{1+z_\mathrm{spec}}\right),
\label{eq:MAD}
\end{equation}
\noindent as it is less sensitive to outliers compared to other definitions \citep[e.g.][]{Ilbert2006}. We additionally quantify an outlier fraction as the fraction of objects with $|z_{\rm phot} - z_{\rm spec}| \geq 0.15\,(1+z_{\rm spec})$.

In general, the \zphot{} performance in comparison to \zspec{} is equally good and essentially agnostic to the photometric color aperture size. This, however, is expected as objects in our spectroscopic sample are notably brighter than many of the sources found in this new, deep imaging. This has two consequences: firstly, the assessable \zphot{} performance does not depend strongly on source magnitude and secondly, the \zphot{} performance reflects that of bright and easy-to-measure sources.

Unsurprisingly, we find significantly better \zphot{} performance using the \texttt{SFHz\_CORR} SED templates  for higher redshift sources $z\gtrsim1$ compared to compared to the default \texttt{FSPS\_FULL} that produces about twice as many catastrophic \zphot{} underestimates. Therefore the addition of $z$-dependent star-formation histories and realistic line emission serves to enhance the ability of \eazy{} to correctly recover the \zphot{} of known, distant spectroscopic sources. However, we find an unexpected \zphot{} bias for $z\lesssim1$ sources when using \texttt{SFHz\_CORR} that is not present when using \texttt{FSPS\_FULL}. The sources of this bias is not yet understood, but given the success with \texttt{FSPS\_FULL} seems unlikely to be driven by issues in photometry. Further exploration of this effect will be left to future work when the field will be complemented with a slew of grism redshifts and medium band data as part of several Cycle 2 programs, and so here we limit our use of EA$z$Y to basic photometric validation in the following section. For simplicity in this early catalog aimed at studying the high-$z$ Universe, we report only \zphot{} computed with \texttt{SFHz\_CORR} templates and caution that the redshifts of $z\lesssim1$ objects may be underestimated.

As shown in Figure~\ref{fig:zphot_zspec}, we achieve a $\sigma_{\rm NMAD}=0.0301$ after removing the 10.8\% outlying sources, some of which are likely due to wrongly identified emission lines in the spectra. Figure~\ref{fig:z_dist} shows a comparison of the distribution of the total \zphot{} sample to the redshifts of the generally lower-$z$, bright spectroscopic sample. It is also worth noting that applying iterative zeropoint corrections does not significantly improve the overall performance as assessed by our limited spectroscopic sample. This also simplifies interpreting differences between our photometry and the predicted photometry from \eazy{} as discussed below in Section~\ref{sec:validation}.

For transparency, we provide the \zphot{} and several common rest-frame fluxes (e.g. $UVJ$ of \citealt{Williams2009} and $ugi_s$ of \citealt{Antwi-Danso2022}) for all objects in the catalog. However, more sophisticated methods utilizing extensive physically-based priors and advanced sampling techniques will enable even more robust \zphot{} and physical parameters (e.g. stellar mass). A forthcoming paper by B. Wang et al. (submitted) will detail how we have applied Prospector-$\beta$ \citep{Johnson2021, Wang2023} to this end.

\subsection{Validation of $JWST$ photometry}
\label{sec:validation}

\begin{figure*}
    \centering
    \includegraphics[width=\textwidth]{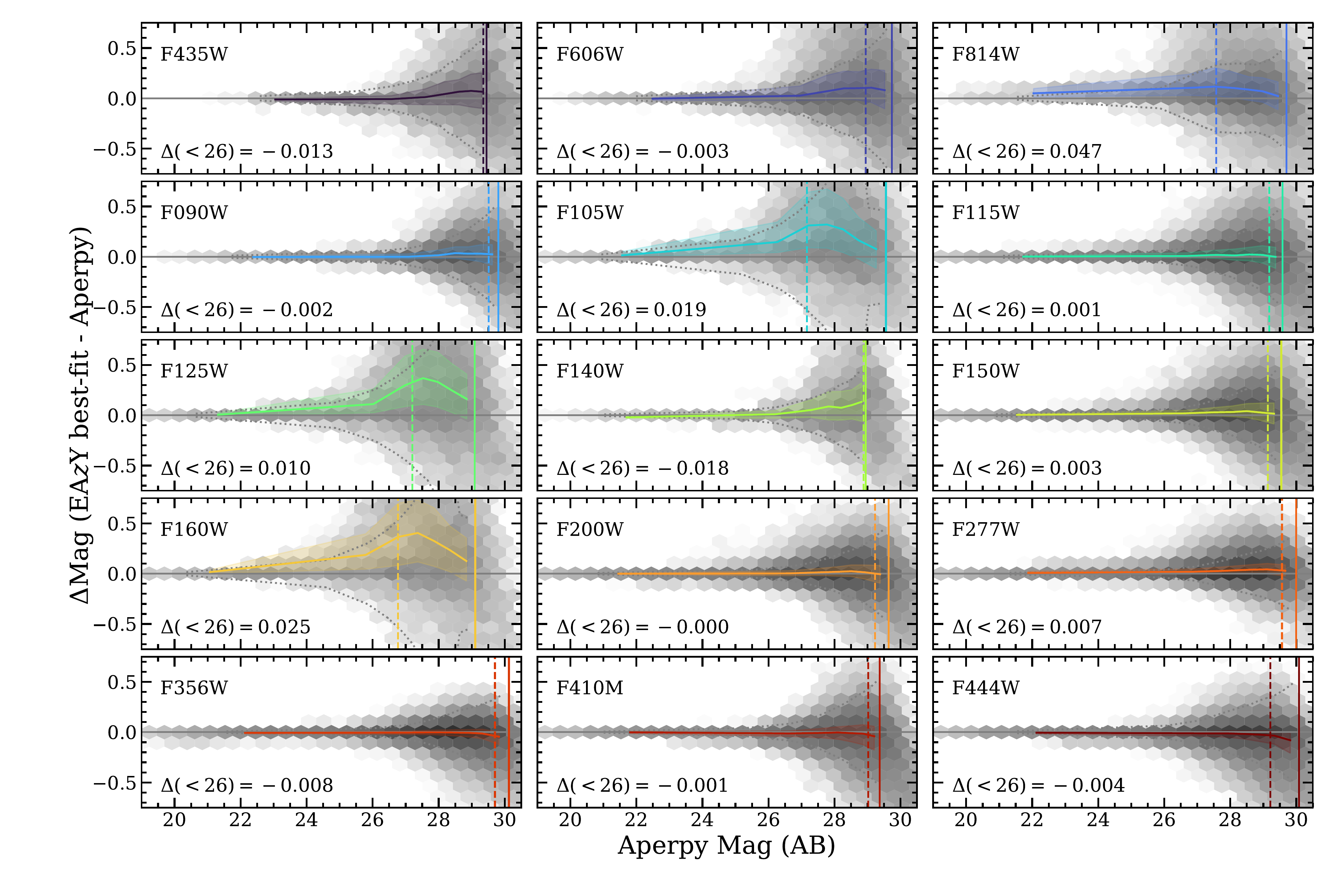}
    \caption{Photometry measured in the super catalog of this work compared against predicted fluxes from \eazy{} in all 15 available bands, based on \texttt{SFHz\_CORR} templates. In each panel, the difference in AB magnitude  ($\Delta$\,Mag) as a function of observed magnitude is summarized by the log$_{10}$-scaled 2D grey density histogram. Colored curves indicate binned medians with two-sided envelopes enclosing 68\% of sources per bin. The overall median offset is labeled on each panel computed on all magnitudes up to the magnitude limit of the deepest 10\% of the corresponding image indicated by the colored solid vertical line; that of the median depth is indicated by the colored dashed vertical line. The typical photometric error is indicated by the dotted grey curves.
    }
    \label{fig:dmag_model}
\end{figure*}

Given that \JWST is a relatively new facility, much is still to be learned about its performance. Consequently, photometric extractions of \JWST imaging in the literature are only just becoming available and so the traditional demonstration of comparing our on-cluster \JWST photometry to the published literature is not possible (see \citet{Merlin2022} and Appendix~\ref{app:catalog_columns}).

Thankfully, computing \zphot{} for every object provides access to the observed-frame fluxes predicted by the corresponding best-fit SED template.\footnote{\noindent Note that since \eazy{} uses linear combination of basis templates, the resulting best-fit model is not limited to physically allowed solutions.} While this is arguably a circular exercise, the SED template combinations allowed by \eazy{} still only span a finite volume in color-color space. By attempting to minimize the $\chi^{2}$ of the fit, \eazy{} is effectively maximizing its agreement with the error-weighted colors that we provide it; yet large-scale biases in individual filters should stand out given our well-sampled 15 filter SEDs.

Figure~\ref{fig:dmag_model} shows the photometric agreement $\Delta\,{\rm Mag}$ between our measured fluxes and those predicted by the best-fit model from \eazy{} using the best fit template set. While there are no significant offsets, we note that in a handful of filters the difference trends upwards at the 50$^{th}$ percentile depth but flattens towards the 10$^{th}$ percentile depth. This behavior correlates with the variation in depth and so can be readily explained by differences in the photometric performance of similarly bright objects with a range (or bimodality) of signal-to-noise. We also investigate $\Delta\,{\rm Mag}$ as a function of $z$, finding similar levels of agreement even at $z\gtrsim6$. 

Although this is a relatively basic comparison, it nonetheless demonstrates that our photometry appears reasonable and that the SEDs of most objects are describable by \eazy{}. Comparisons to literature \HST and recent \JWST photometric catalogs are included in Appendix~\ref{app:comparison}.

\section{Summary}

We present a first generation of high angular resolution space-based photometric catalogs of the strong lens cluster Abell~2744, including a combination of archival \HST imaging in 7 bands with public \JWST imaging in 8 bands from 3 programs (UNCOVER, GLASS/ERS, and a DDT program). With an ultra-deep noise-equalized F277W+F356W+F444W detection image at the native \JWST resolution (Figure~\ref{fig:detection_depths}), we present 0.4-4.4$\mu$m panchromatic coverage of 61\,648 sources within the extended cluster region, including two newly detailed structures to the north and northwest of the cluster heart. After removing stars, artifacts, and other spurious sources, we present reliable photometry for 55\,613 galaxy candidates. 

With this paper, we release the UNCOVER photometric catalogs derived from small (0.32\arcsec{} diameter) to large (1.40\arcsec{} diameter) circular apertures, including a straight-forward aperture-combined `super' catalog for rapid, go-to use. Aperture photometry is measured on images PSF-matched to \JWST/F444W resolution that have been cleaned of contaminating light from bCGs and ICL. The photometry is corrected to total based on the ratio of flux within a Kron-like aperture relative to a circular aperture, plus an additional correction of order 5-20\% for missing light beyond the Kron radius as determined from the PSF curve of growth in the F444W filter. Details including how to access the catalogs, column descriptions, and general use recommendations can be found in Appendix~\ref{app:catalog_columns}. Catalogs are available online.\footref{foot:catalog} The software used to produce these catalogs, \texttt{aperpy}, is generally applicable to any \JWST/NIRCam data and is freely available.\footref{foot:aperpy}

The UNCOVER photometric catalogs are among the deepest catalogs publicly available, reaching effective 5$\sigma$ depths greater than 29\,AB in all 15 bands for the 0.32\arcsec{} diameter apertures. These depths do not account for extra magnification factors from strong gravitational lensing for those background galaxies in optimal configurations. When combining the survey depths with strong lensing, UNCOVER is the deepest view into our Universe to date.

%% IMPORTANT! The old "\acknowledgment" command has be depreciated. It was
%% not robust enough to handle our new dual anonymous review requirements and
%% thus been replaced with the acknowledgment environment. If you try to 
%% compile with \acknowledgment you will get an error print to the screen
%% and in the compiled pdf.
%% 
%% Also note that the akcnowlodgment environment does not support long amounts of text. If you have a lot of people and institutions to acknowledge, do not use this command. Instead, create a new \section{Acknowledgments}.
% \begin{acknowledgments}
\section*{Acknowledgments}
The authors would like to acknowledge the generosity of
Vasily Kokorev, as well as Diego Paris, Adriano Fontana, Emiliano Merlin, and Tommaso Treu for making their photometry available for comparison and for many useful discussions that improved our catalogs. We would also like to thank Olivier Ilbert for comparisons between \Spitzer/IRAC and \JWST/NIRCam bands. We are also grateful for the many helpful and constructive comments from the anonymous referee.

This work is based in part on observations made with the NASA/ESA/CSA \emph{James Webb Space Telescope}. The data were obtained from the Mikulski Archive for Space Telescopes at the Space Telescope Science Institute, which is operated by the Association of Universities for Research in Astronomy, Inc., under NASA contract NAS 5-03127 for \JWST. These observations are associated with JWST-GO-2561, JWST-ERS-1324, and JWST-DD-2756. Support for program JWST-GO-2561 was provided by NASA through a grant from the Space Telescope Science Institute, which is operated by the Associations of Universities for Research in Astronomy, Incorporated, under NASA contract NAS 5-03127. Financial support for this program is gratefully acknowledged. This research is also based on observations made with the NASA/ESA \emph{Hubble Space Telescope} obtained from the Space Telescope Science Institute, which is operated by the Association of Universities for Research in Astronomy, Inc., under NASA contract NAS 5–26555. These observations are associated with programs HST-GO-11689, HST-GO-13386, HST-GO/DD-13495, HST-GO-13389, HST-GO-15117, and HST-GO/DD-17231. All of the data presented in this paper were obtained from the Mikulski Archive for Space Telescopes (MAST) at the Space Telescope Science Institute. The specific observations used to produce these catalogs can be accessed via \dataset[10.17909/nftp-e621]{http://dx.doi.org/10.17909/nftp-e621}.

This work has made use of data from the European Space Agency (ESA) mission
{\it Gaia} (\url{https://www.cosmos.esa.int/gaia}), processed by the {\it Gaia}
Data Processing and Analysis Consortium (DPAC,
\url{https://www.cosmos.esa.int/web/gaia/dpac/consortium}). Funding for the DPAC
has been provided by national institutions, in particular the institutions
participating in the {\it Gaia} Multilateral Agreement.

Cloud-based data processing and file storage for this work is provided by the AWS Cloud Credits for Research program. This research has made use of the NASA/IPAC Extragalactic Database (NED), which is funded by the National Aeronautics and Space Administration and operated by the California Institute of Technology. 

The Cosmic Dawn Center is funded by the Danish National Research Foundation (DNRF) under grant \#140. KEW gratefully acknowledges funding from HST-GO-16259, HST-GO-15663, and the Alfred P. Sloan Foundation Grant FG-2019-12514 to support the development of \texttt{aperpy}. RB acknowledges support from the Research Corporation for Scientific Advancement (RCSA) Cottrell Scholar Award ID No: 27587. LF and AZ acknowledge support by Grant No. 2020750 from the United States-Israel Binational Science Foundation (BSF) and Grant No.\ 2109066 from the United States National Science Foundation (NSF), and by the Ministry of Science \& Technology, Israel. PD acknowledges support from the NWO grant 016.VIDI.189.162 (``ODIN") and from the European Commission's and University of Groningen's CO-FUND Rosalind Franklin program. HA acknowledges support from CNES (Centre National d'Etudes Spatiales). RS acknowledges an STFC Ernest Rutherford Fellowship (ST/S004831/1). MS acknowledges support from the CIDEGENT/2021/059 grant, from project PID2019-109592GB-I00/AEI/10.13039/501100011033 from the Spanish Ministerio de Ciencia e Innovaci\'on - Agencia Estatal de Investigaci\'on, and from Proyecto ASFAE/2022/025 del Ministerio de Ciencia y Innovación en el marco del Plan de Recuperación, Transformación y Resiliencia del Gobierno de Espa\~na.

% \end{acknowledgments}

%% To help institutions obtain information on the effectiveness of their 
%% telescopes the AAS Journals has created a group of keywords for telescope 
%% facilities.
%
%% Following the acknowledgments section, use the following syntax and the
%% \facility{} or \facilities{} macros to list the keywords of facilities used 
%% in the research for the paper.  Each keyword is check against the master 
%% list during copy editing.  Individual instruments can be provided in 
%% parentheses, after the keyword, but they are not verified.

% \vspace{5mm}
\facilities{\JWST (NIRCam, NIRSpec, and NIRISS), \HST (ACS and WFC3), GAIA}

%% Similar to \facility{}, there is the optional \software command to allow 
%% authors a place to specify which programs were used during the creation of 
%% the manuscript. Authors should list each code and include either a
%% citation or url to the code inside ()s when available.

\software{\textsc{astropy} \citep{astropy:2013,astropy:2018,astropy:2022}, 
        \textsc{Source Extractor} \citep{SE},
        \textsc{SEP} \citep{Barbary2016},
        \textsc{extinction} \citep{extinction},
        \textsc{SFDMap} \citep[\url{github.com/kbarbary/sfdmap}]{Schlegel1998,Schlafly2011},
        \textsc{WebbPSF} \citep{Perrin2012,Perrin2014},
        \eazy{} \citep{Brammer2008},
        \textsc{Pypher} \citep{Boucaud2016},
        \textsc{Photutils} \citep{photutils},
        \textsc{astrodrizzle} \citep{Gonzaga12},
        \grizli{} (\url{github.com/gbrammer/grizli}),
        \textsc{numpy} \citep{numpy2011},
        \textsc{matplotlib} \citep{matplotlib2007}
          }

%% Appendix material should be preceded with a single \appendix command.
%% There should be a \section command for each appendix. Mark appendix
%% subsections with the same markup you use in the main body of the paper.

%% Each Appendix (indicated with \section) will be lettered A, B, C, etc.
%% The equation counter will reset when it encounters the \appendix
%% command and will number appendix equations (A1), (A2), etc. The
%% Figure and Table counter will not reset.

\appendix

\section{Photometric Comparisons}
\label{app:comparison}

\begin{figure*}
    \centering
    \includegraphics[width=0.95\textwidth]{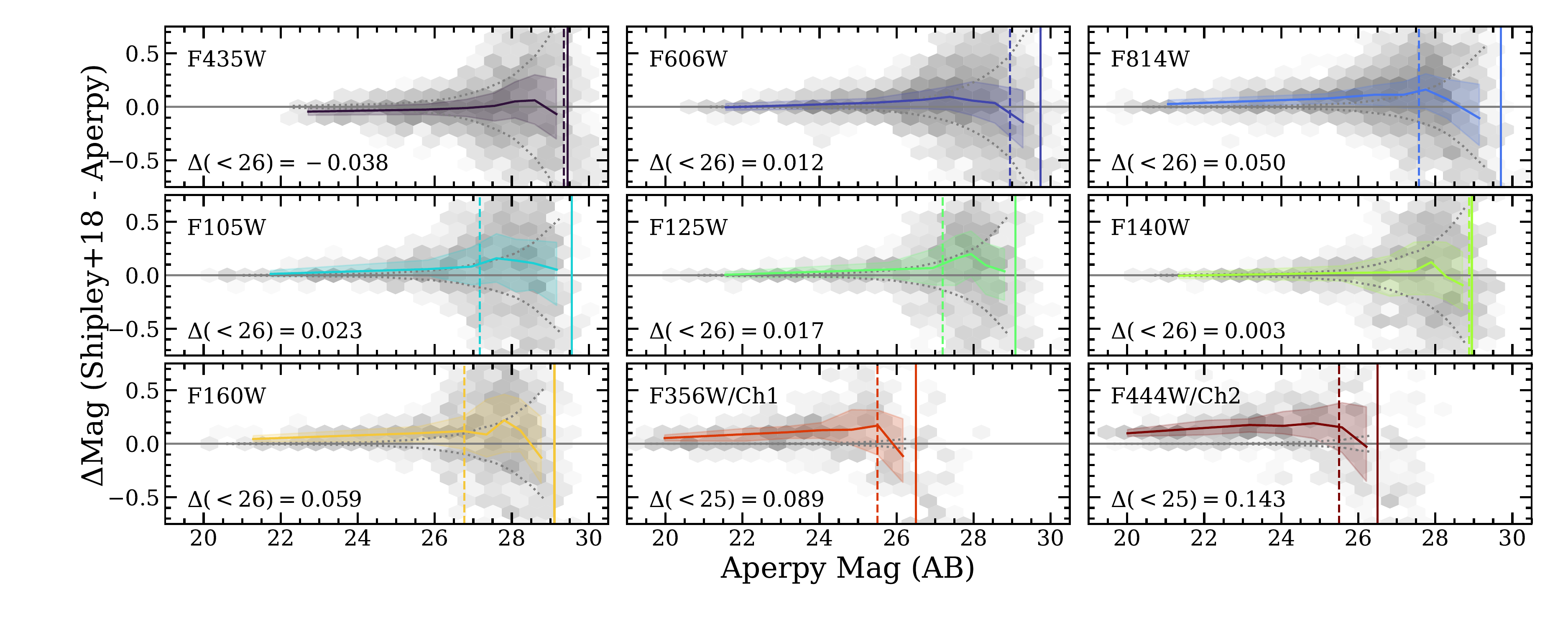}
    \caption{Photometry measured in this work based on corrected 0.7\arcsec{} apertures compared to that of \citet{shipley:18} from the apertures of the same diameter} in the 7 common \HST bands and 2 IRAC bands similar to those in NIRCam. Format follows that of Figure~\ref{fig:dmag_model}.
    \label{fig:dmag_shipley}
\end{figure*}

\begin{figure*}
    \centering
    \includegraphics[width=0.95\textwidth]{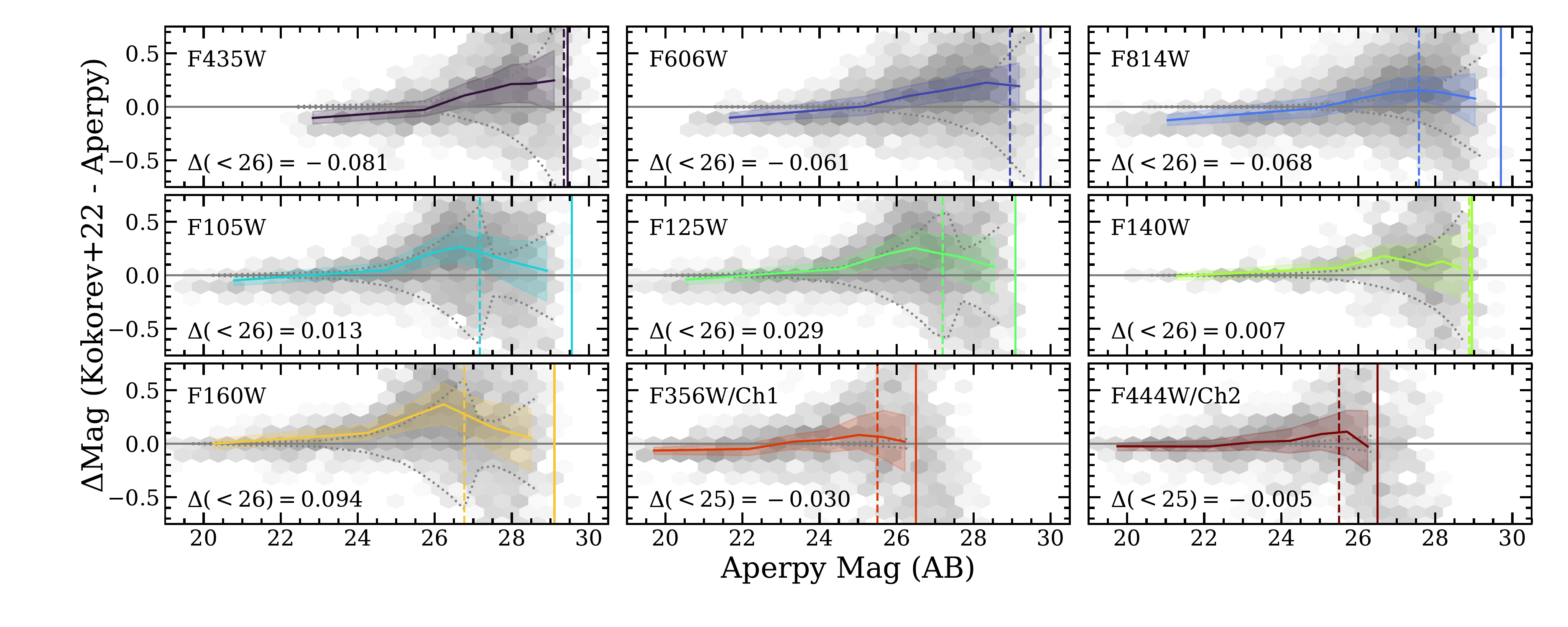}
    \caption{Photometry measured in this work based on corrected 0.7\arcsec{} apertures compared to that of \citet{Kokorev2022} from the apertures of the same diameter} in the 7 common \HST bands and 2 IRAC bands similar to those in NIRCam. Format follows that of Figure~\ref{fig:dmag_model}.
    \label{fig:dmag_kokorev}
\end{figure*}

\begin{figure*}
    \centering
    \includegraphics[width=0.95\textwidth]{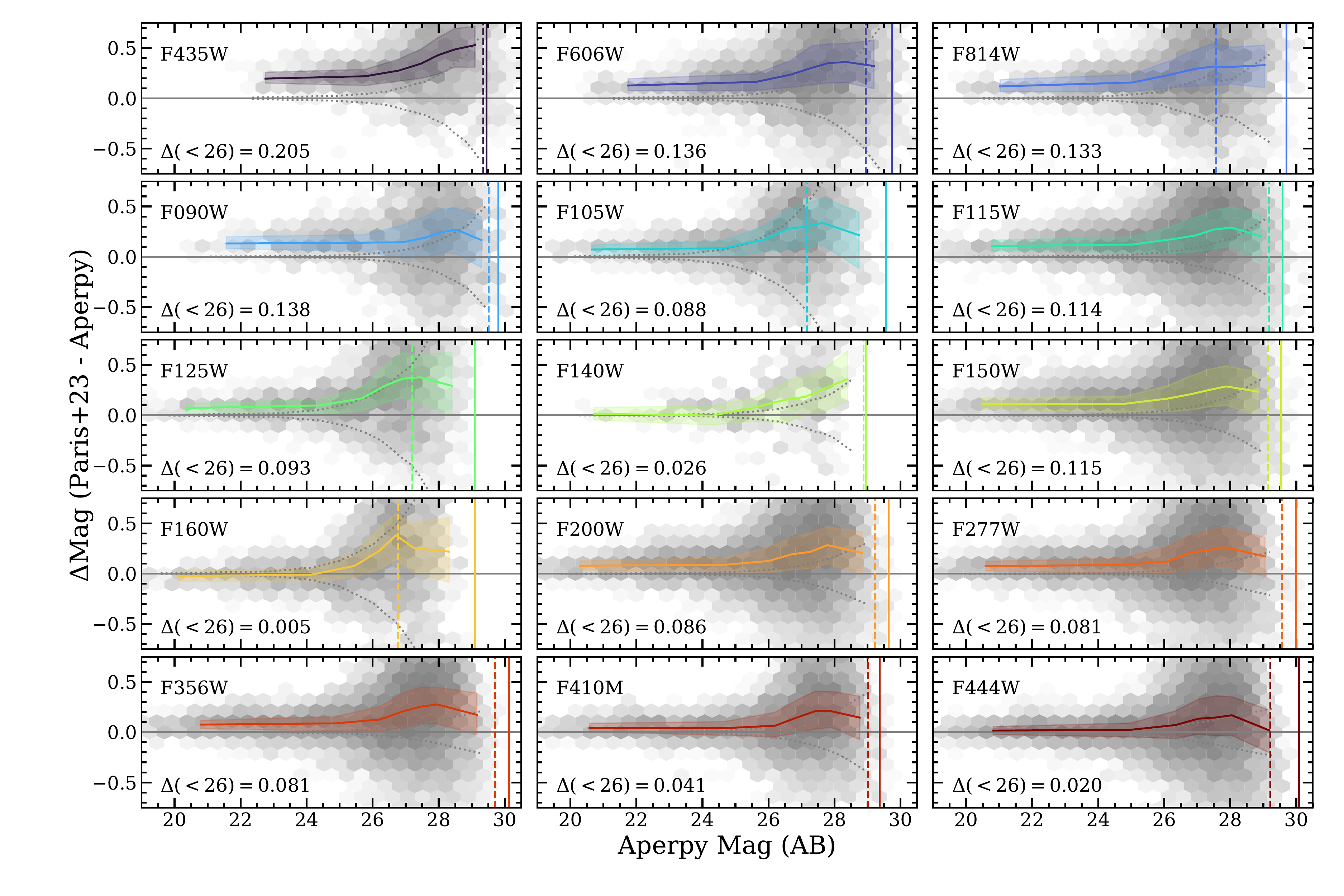}
    \caption{Photometry measured in this work based on corrected 0.48\arcsec{} apertures compared to that of \citet{Paris2023} from 0.42\arcsec{} apertures} in the 7 common \HST bands and 8 common \JWST bands. Format follows that of Figure~\ref{fig:dmag_model}.
    \label{fig:dmag_glass}
\end{figure*}

Although comparisons with predicted model fluxes from \eazy{} are useful (see Figure~\ref{fig:dmag_model}), we can also leverage the existing \HST catalogs to make comparisons for ACS and WFC3 derived photometry. Comparisons for all 7 common bands for \citet{shipley:18} and \citet{Kokorev2022} are shown in Figures~\ref{fig:dmag_shipley} and \ref{fig:dmag_kokorev}, respectively. Objects that are known blends, stars, or artifacts are not shown. We choose to show the total fluxes derived from 0.7\arcsec{} diameter apertures to be consistent with their choice of aperture size. We also compare to \Spitzer/IRAC as the filter profiles of NIRCam F356W and F444W are sufficiently similar to those of IRAC Channel~1 and 2, noting that there likely exist systematics on the 5\% level owing to the exact filter profiles and the method of photometric extraction from the literature IRAC measurements. For the comparison of IRAC to NIRCAm we adopt the shallower $25.5-26.5\,AB$ depths of the IRAC data that the photometry from \citeauthor{shipley:18} and \citeauthor{Kokorev2022} are based on.

The agreement with literature photometry is generally excellent. Compared to \citeauthor{shipley:18}, we generally find good agreement below 6\% for all \HST bands with somewhat larger offsets compared to \Spitzer/IRAC. Compared to \citeauthor{Kokorev2022}, we again generally find good agreement except for \HST/ACS bands where in our photometry bright objects are $\sim10$\% fainter and faint sources 0.2-0.3\,mag brighter. However, as detailed in \citet{Kokorev2022}, the authors elected to forgo PSF homogenization which makes this comparison particularly hazardous for faint, point-like sources. We achieve a surprisingly good agreement with the IRAC photometry from \citeauthor{Kokorev2022} despite the order of magnitude difference in resolution. As seen before, the bimodailty in depth for some of the \HST bands drives up $\Delta\,{\rm Mag}$ for sources in the shallow areas.

At the time of this writing there are no fully peer-reviewed catalogs with \JWST photometry over Abell~2744 produced with in-flight calibrations (see \citealt{Merlin2022}). Thankfully, we are fortunate in that we were able to compare with an updated photometric catalog developed by the GLASS team \citep{Paris2023}. Both catalogs are based on roughly the same public datasets from \HST and \JWST, though \citeauthor{Paris2023} performs an independent data reduction and analysis to ours. Several notable differences between the two catalogs include (1) independent image reduction pipelines, (2) source detection (the GLASS catalog is F444W detected, whereas we use a noise-equalized LW detection), (3) the treatment of bCG/ICL (modeled out herein, and not in GLASS), and (4) different software and approaches to PSF modelling, PSF homogenization, aperture photometry, and total flux corrections. Figure~\ref{fig:dmag_glass} shows the photometric comparison in all 15 common filters for \HST and \JWST, adopting the 0.42\arcsec{} corrected photometry from GLASS compared to our 0.48\arcsec{} corrected photometry. The GLASS photometry is not already corrected for Galactic attenuation, and so we apply a correction consistent with our catalog (see Section~\ref{sec:photometry}). We note that while the depths indicated correspond to those in Table~\ref{tab:depths} measured in this work, many of the faintest sources in our catalog do not have matches to those in GLASS due to their shallower F444W only detection (see Figure~\ref{fig:depthcurves}).

Given the considerable differences in the construction of the two catalogs, the photometric agreement is reasonable. The agreement is worst in the \HST/ACS where photometry from \citeauthor{Paris2023} is fainter by 10-20\% relative to ours and that of \citeauthor{shipley:18} or \citeauthor{Kokorev2022}. However, while photometry from GLASS is $\sim$10\% fainter for bluer \HST/WFC3 bands, \HST/WFC3 F140W and F160W photometry is in excellent agreement. We suspect that the observed differences in \HST photometry are driven by both the choice of PSF and the PSF-matching techniques. The \JWST/NIRCam photometry from GLASS is similarly fainter by $\sim10\%$ in SW bands, and $<8\%$ in LW bands. The result of these differences is that galaxies in the GLASS catalog are systematically redder than measured in this work.

\section{JWST/NIRCam PSF Profiles}
\label{app:nircam_psfs}

\begin{figure*}
    \centering
    \includegraphics[width=0.95\textwidth]{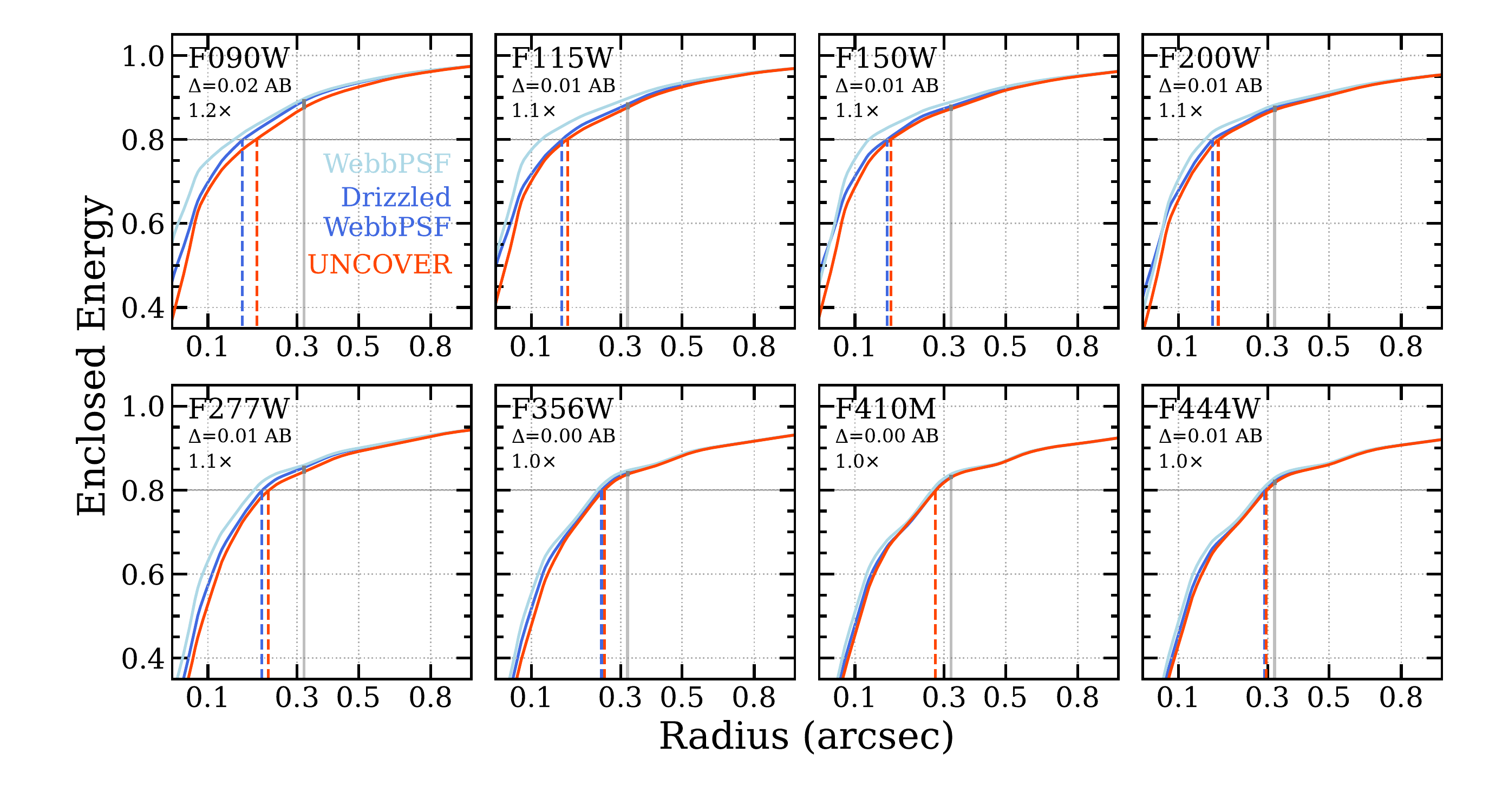}
    \caption{Enclosed energy (EE) of the pointspread functions (PSFs) reported by WebbPSF (blue),  with drizzling (light blue), and those determined in this work by stacking stars (orange). Yet even after accounting for drizzling, our observed PSFs are still wider at fixed EE, e.g. 80\% (colored dashed lines) and, equivalently, they have less EE at fixed aperture diameter, e.g. 0.32" (vertical grey lines). The corresponding photometric difference, however, is relatively small: $0.01-0.02$\,AB for SW and $\lesssim0.01$ for LW.}
    \label{fig:psfs_cogs}
\end{figure*}

\JWST is a new facility and so it is worthwhile investigating the point-spread functions (PSFs) predicted by WebbPSF to the profiles of bone fide point-sources within the UNCOVER field. The generation of observed PSFs from point-sources within the images follow the description in Section~\ref{sec:psfmatching}, noting again that they are normalized at 4\arcsec{} diameter to match the tabulated enclosed energy (EE) provided by STScI. We adopt single \JWST PSFs generated by \textsc{WebbPSF} corresponding to 5 November, 2022 near the expected mid-point of the planned UNCOVER visits\footnote{\noindent Due to a guide star acquisition failure on 31 October, visit 1.1 was repeated successfully on 15 November.} at the nominal PA of UNCOVER (31.4\textdegree). Importantly, we set \texttt{normalization}\,$=$\,\texttt{exit\_pupil} so that the 4\arcsec{} FOV stamps are normalized such that the energy at large radii is accounted for correctly. Comparisons of our WebbPSF PSFs to simulated PSFs and tabulated EE measurements provided by STScI show sub-percent agreement at all radii. We also measured the EE of observed bright stars from single exposures taken by the Absolute Flux Calibration Program of \citet{Gordon2022} and hosted on MAST, also finding sub-percent agreement with WebbPSF.

Figure~\ref{fig:psfs_cogs} shows the curves of growth predicted by WebbPSF as the energy enclosed as a function of diameter from the center (blue). These are single exposure predictions, and so to account for PSF broadening introduced by imaging reduction we consistently drizzle each PSF and extract a representative stamp that is spatially averaged across the FOV (light blue). While the drizzled PSFs are only slightly broader in most cases, the EEs of our observed PSFs from real stars are still broader. These shape differences translate into photometric offsets: $0.01-0.02$\,AB for SW and $\lesssim0.01$ for LW. These differences may be driven by missing elements in the computation of the WebbPSF, or minor inconsistencies in the drizzling procedure. As such, we recommend that, where possible, observed PSFs should be used for measurements on NIRCam imaging.

\section{JWST/NIRCam PSF Stability}
\label{app:psf_stability}

\begin{figure*}
    \centering
    \includegraphics[width=0.95\textwidth]{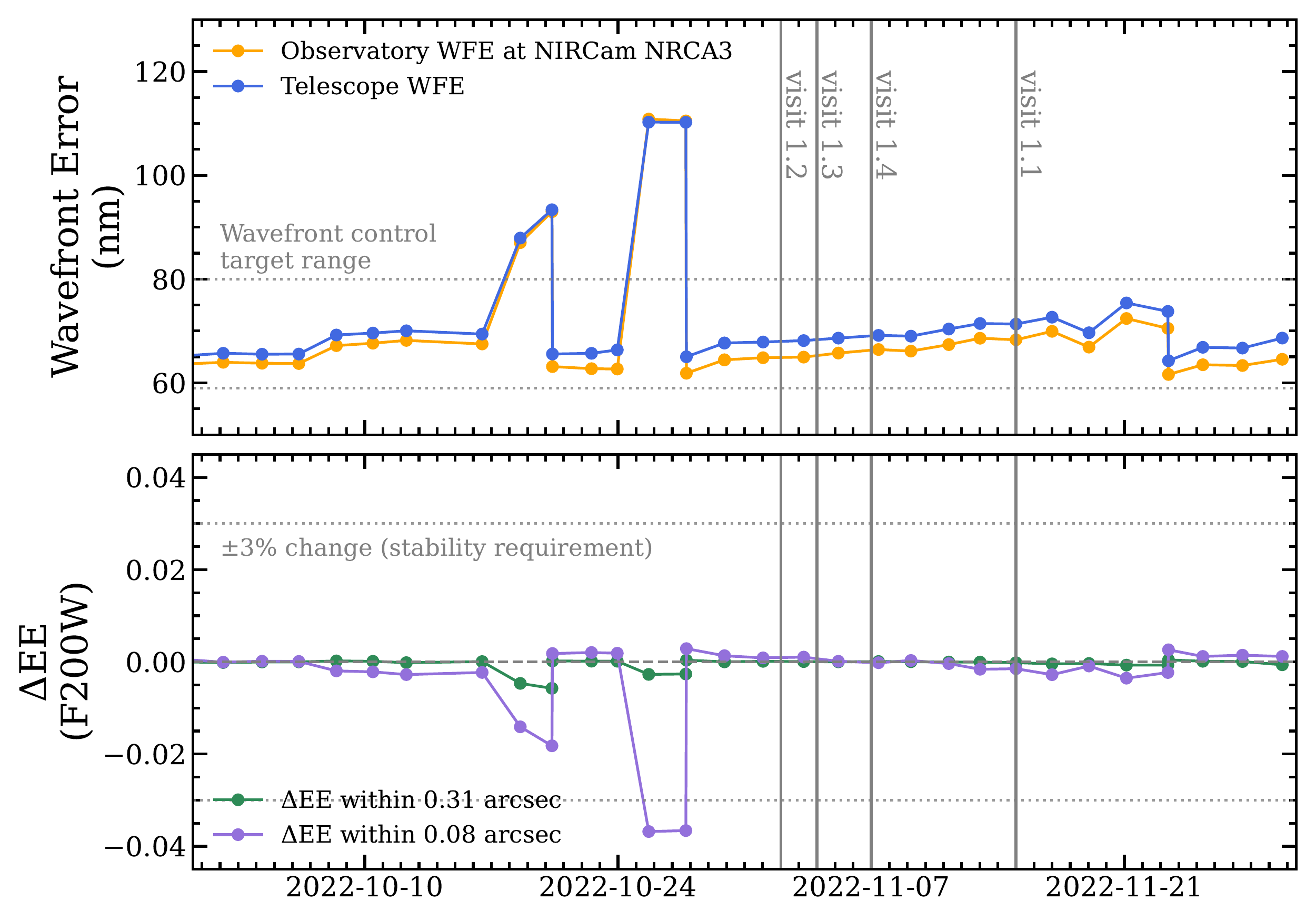}
    \caption{Evolution of the wavefront error (WFE) for \JWST generated from \textsc{WebbPSF} from 1 October to 30 November, 2022. \textit{Top}: the measured WFE for NIRCam Module A3 (NRCA3, yellow) is shown along with that of the general telescope facility (blue). \textit{Bottom}: the enclosed energy (EE) of the F200W PSF in 0.31 and 0.08\arcsec{} apertures (green and purple, respectively) relative to the median. Relevant control and stability criteria are indicated. behavior during the four UNCOVER visits is effectively unchanged, with relative $\mathrm{EE}\,\lesssim\,0.1$\%.}
    \label{fig:wsstrending}
\end{figure*}

\begin{figure*}
    \centering
    \includegraphics[width=0.95\textwidth]{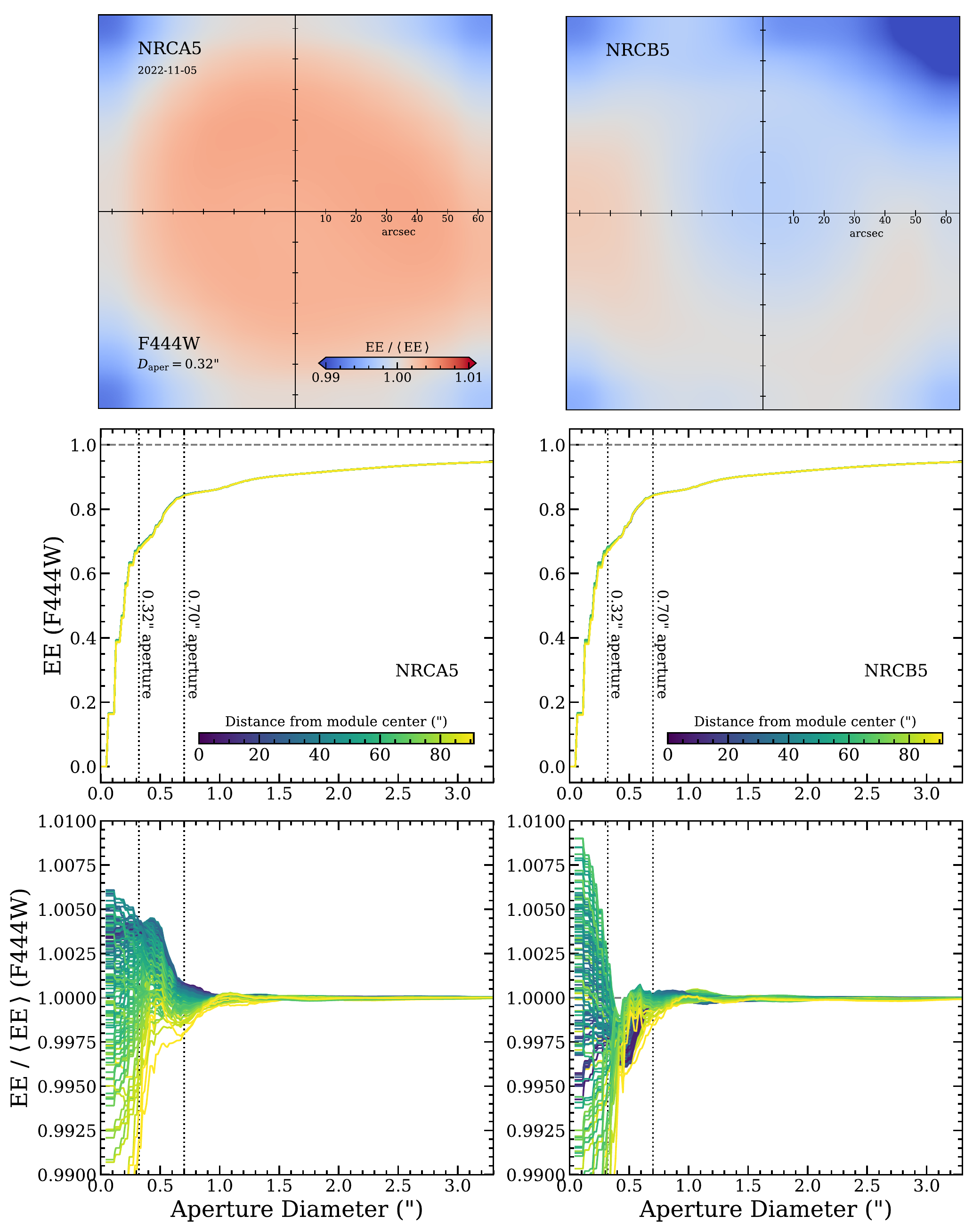}
    \caption{Behavior of \JWST/NIRCam as measured by the wavefront sensor on 5 November, 2022, from which our \JWST PSFs are derived, based on a grid of 144 PSF realizations generated with \textsc{WebbPSF}. \textit{Top}: the spatial dependence of the energy enclosed in 0.32\arcsec{} apertures for NIRCam Module A5 (NRCA5) and B5 (NRCB5) in F444W relative to that of the average PSF. \textit{Middle}: enclosed energies (EE) as a function of aperture size colored by distance from the respective module center. \textit{Bottom}: enclosed energies as a function of aperture size relative to that of the average PSF for each module, colored as before. PSF variability across NIRCam for F444W is minimal with a relative $\mathrm{EE}\,\lesssim\,0.1$\% in even small 0.32\arcsec{} apertures.}
    \label{fig:wssf444w}
\end{figure*}

While \JWST is already proving to be a revolutionary facility to advance nearly all areas of astrophysics, the typical behavior of the telescope resolution, both in time and across the detectors, remains a concern. Unlike the generally stable PSF enjoyed by \HST, the moving hexagonal mirrors of \JWST means that the PSF can be highly variable. Thankfully, the Wavefront Sensor (WSS) samples the observatory mirrors on an approximate two-day cadence with sufficient detail to reconstruct the PSF at any detector location in any band. This is most easily accessed by \textsc{WebbPSF}, which includes near-live reports from the WSS to enable PSF reconstruction on the fly, as well as tools to visualize the typical PSF behavior with time. 

\citet{Nardiello2022} recently presented a detailed analysis of the PSF variability of NIRCam finding $\sim$9\% variation across their FOV and 3-4\% over multi-epoch exposures based on peak-to-peak variations measured from PSF residuals. However, their results are concerned with the shape of the PSF (especially its core) which is directly relevant to their PSF-fitting photometry of dense star clusters (see also \citealt{Zhuang2023}). The power of aperture photometry, crucially, is that it is sensitive only to the energy enclosed by a given aperture and not about precisely where that energy is located \textit{within} the aperture. By now studying the variation of the enclosed energy we can not only better understand the aperture \textit{photometric} impact of these variations, but do so in the context of our particular observations from UNCOVER. 

Our photometry is derived from images PSF-matched to F444W and so their accuracy strongly depends on the F444W PSF characteristics. We adopt PSFs from WebbPSF in the same fasion as described above in Appendix~\ref{app:nircam_psfs}. In the following analysis we generate two regular grids of 144 PSF samples for each NIRCam LW detector: NRCA5 and NRCB5.  Here we focus on the UNCOVER dataset only leaving the characterization of the PSF behavior during the GLASS and DDT visits to future work. Although shown below only for F444W, we have repeated the analysis for all NIRCam bands and find similar results. 

Figure~\ref{fig:wsstrending} quantifies the wavefront error (WFE) from 1 October to 30 November, 2022, during which the UNCOVER program was executed. While there were significant anomalies in the WFE at the end of October, all four UNCOVER visits were executed with nominal PSF behavior with a WFE\,$\approx 65-75$\,nm, well within the target control range.\footnote{\noindent \url{https://jwst-docs.stsci.edu/jwst-observatory-hardware/jwst-wavefront-sensing-and-control}} Of even greater significance to this work, the relative enclosed energy $\Delta$EE during that same window was effectively static meaning that any uncertainty due to the time evolution of the F444W PSF when measuring photometry can be safely ignored.  

Figure~\ref{fig:wssf444w} provides insight into the spatial variability of the PSF across NRCA5 and NRCB5 on 5 November. Instead of examining FWHM as provided by \textsc{WebbPSF}, we measure enclosed energy (EE) as it is more pertinent to assessing uncertanties in aperture photometry. While the EE relative to that of the average PSF $\langle\,\mathrm{EE}\,\rangle$ differs between NCRA5 and NCRB5, we find that the spatial variation is generally smooth and increases with decreasing aperture size such that the variability in EE is $\ll\,0.1$\% in 0.70\arcsec{} apertures and $\sim1$\% at 0.32\arcsec{}. 

We caution, however, that these results do not necessarily apply to model-based photometric techniques (e.g. GALSIM, \citealt{Rowe2015}) because although the summed energy at fixed radius is relatively constant, the exact shape of the PSF itself may vary enough that the spatial dependence may need to be taking into account, as stressed by  \citeauthor{Nardiello2022}.

Together, these two pieces of evidence point to a favourably small contribution to the error budget from the \JWST PSF behavior for UNCOVER -- both in time evolution and spatial variation -- such that we can confidently neglect these uncertainties in our photometric catalogs.

\section{Catalog Column Descriptions}
\label{app:catalog_columns}

This paper is accompanied by a `super' photometric catalog and five others computed from individual aperture sizes from 0.32\arcsec{} to 1.40\arcsec{}. Each is derived from the same bCG-subtracted images which have been PSF-matched to that of F444W, and share the same LW-selected objects. Fluxes are reported in total as described in Section~\ref{sec:photometry}, accounting for any additional light outside the Kron aperture. While the columns included in the catalogs are described in Table~\ref{table:cat_cols}, we also include a dedicated \texttt{README} file with the catalogs in the case of future changes.

We strongly recommend using photometry from the super catalog which adopts 0.32\arcsec{} diameter apertures for small objects (e.g., high-$z$ galaxies) that make up the majority of the catalog and larger 0.48-1.40\arcsec{} apertures for the relatively fewer bright, extended objects at lower-$z$. The photometry is reported in $F_\nu$ in units of 10\,nJy corresponding to a zeropoint of 28.9 AB. \zphot{}, rest-frame fluxes, and magnification estimates are reported for all objects. No quantities are corrected for magnification. Clean samples of galaxies can be readily identified with \texttt{USE\_PHOT}=1 (see Section~\ref{sec:flags}).

\setlength{\tabcolsep}{3pt}
\renewcommand{\arraystretch}{0.85}

\begin{table*}[!th]
\centering
\caption{Catalog columns}\label{table:cat_cols}
\begin{tabular}{ll}
\hline \hline
\noalign{\smallskip}
Column name & Description \\
\hline
\noalign{\smallskip}
\texttt{id} & Unique identifier \\
\texttt{x/y}    &  X/Y centroid in image coordinates\\
\texttt{ra}  & RA J2000 (degrees)\\
\texttt{dec} & Dec J2000 (degrees)\\
\texttt{ebv\_mw} & Line-of-sight E(B-V), already applied, computed from \cite{Schlafly2011}\\
\texttt{faper\_f277w+f356w+f444w} & Flux computed on the F444W-matched LW stack in circular apertures ($F_\nu$, zeropoint = 28.9)\\
\texttt{eaper\_f277w+f356w+f444w} & $1\,\sigma$ flux error computed from empty apertures on F444W-matched LW stack ($F_\nu$, zeropoint = 28.9)\\
\texttt{fauto\_f277w+f356w+f444w} & Flux computed in Kron ellipse on F444W-matched LW stack ($F_\nu$, zeropoint = 28.9)\\
\texttt{f\_X}   & Total flux for each filter X ($F_\nu$, zeropoint = 28.9)\\
\texttt{e\_X}  & $1\,\sigma$ flux error for each filter X ($F_\nu$, zeropoint = 28.9)\\
\texttt{w\_X}  & Weight relative to the maximum within image X (see text)\\
\texttt{tot\_cor}  & Aperturre-to-total correction (including both aperture-to-Kron and Kron-to-total) \\
\texttt{kron\_radius}  & Kron radius factor (\se{}-like, unitless)\\
\texttt{kron\_radius\_circ}  & Circularized Kron radius (arcsec)\\
\texttt{use\_circle}    & 1 for objects without Kron correction (faint and/or blended) \\
\texttt{flag\_kron}    & 1 for blended objects without Kron correction \\
\texttt{iso\_area}    & isophotal area based on the source segment, in square arcseconds \\
\texttt{a\_image} & Semi-major axis (pixels)\\
\texttt{b\_image}  & Semi-minor axis (pixels)\\
\texttt{theta\_J2000} & Position angle of the major axis (counter-clockwise, measured from East)\\ 
\texttt{flux\_radius} & Radius containing 50\% of the flux with neighbors masked (arcsec)\\ 
\texttt{use\_phot}    & 1 for reliable sources, 0 if any of the following \texttt{flag\_X}=1 \\
\texttt{flag\_nophot} & 1 if the object has no viable photometry in any band at this aperture size \\
\texttt{flag\_lowsnr} & 1 if source has aperture SNR$ < 3$ in LW stack \\
\texttt{flag\_star} & 1 if source is identified as a star (see Section~\ref{sec:stars}) \\
\texttt{flag\_artifact} & 1 if source is identified as an artifact (see Section~\ref{sec:flags}) \\
\texttt{flag\_nearbcg} & 1 if within 3\arcsec{} of a known bCG (see Section~\ref{sec:flags}) \\
\texttt{z\_spec}  & Spectroscopic redshift, where available (pre-\JWST only) \\
\texttt{id\_DR1} & ID of the source in the DR1 catalog release, if available \\
\texttt{match\_radius\_DR1} & Distance to DR1 catalog release match, if available (any within 0.08\arcsec{}) \\
\texttt{id\_msa} & Corresponding object ID from UNCOVER MSA Spectral Catalog of Price et al. (in prep.), if available  \\
\texttt{match\_radius\_msa} & Distance to object in the UNCOVER MSA catalog, if available (any within 0.24\arcsec{}) \\
\texttt{id\_alma} & ID of the source in the ALMA Catalog of Fujimoto et al. (in prep.), if available. \\
\texttt{f\_alma} & ALMA flux (\texttt{NaN} for non-detections; $F_\nu$, zeropoint = 28.9) \\ 
\texttt{e\_alma} & ALMA flux uncertainty (1$\sigma$ upper limit for non-detections; $F_\nu$, zeropoint = 28.9) \\
\texttt{match\_radius\_alma} & Distance to object in the ALMA catalog (nearest within 1\arcsec{}) \\
\texttt{use\_aper}  & Aperture diameter used to measure fluxes, for super catalog only (arcsec) \\
\texttt{nusefilt} & Number of filters used in the fit \\
\texttt{z\_phot} & Best-fit photometric redshift via minimum $\chi^2$ using \texttt{SFHz\_CORR} templates\\
\texttt{z\_phot\_chi2} & Total $\chi^2$ at \texttt{z\_phot} \\
\texttt{z025/160/500/840/975} & Redshift posterior percentiles, e.g. z025 $\to$ 2.5\% \\
\texttt{restU/V/J} & Rest-frame $U$/$V$/$J$-band flux ($F_\nu$, zeropoint = 28.9) \\
\texttt{restU/V/J\_err} & Rest-frame $U$/$V$/$J$-band flux uncertainty ($F_\nu$, zeropoint = 28.9) \\
\texttt{restus/gs/is} & Rest-frame synthetic $u$/$g$/$i$-band flux ($F_\nu$, zeropoint = 28.9) \\
\texttt{restus/gs/is\_err} & Rest-frame synthetic $u$/$g$/$i$-band flux uncertainty ($F_\nu$, zeropoint = 28.9) \\
\texttt{star\_min\_chi2} & $\chi^2$ of best-fit stellar template \\
\texttt{mu} & Gravitational magnification $\mu$ (best-fit; = 1 for foreground objects) \\
\texttt{mu025/160/840/975} & Total magnification $\mu=\mu_{\mathrm{r}}\mu_{\mathrm{t}}$ posterior percentiles  \\
\texttt{mu\_r} & Radial magnification (best-fit; = 1 for foreground objects) \\
\texttt{mu\_r025/160/840/975} & Radial magnification $\mu_{\mathrm{r}}$ posterior percentiles  \\
\texttt{mu\_t} & Tangential magnification (best-fit; = 1 for foreground objects) \\
\texttt{mu\_t025/160/840/975} & Tangential magnification $\mu_{\mathrm{t}}$ posterior percentiles  \\
\texttt{shearone} & Shear $\gamma_1$ (best-fit; = 1 for foreground objects) \\
\texttt{shearone025/160/840/975} & Shear $\gamma_1$ posterior percentiles \\
\texttt{sheartwo} & Shear $\gamma_2$ (best-fit; = 1 for foreground objects) \\
\texttt{sheartwo025/160/840/975} & Shear $\gamma_2$ posterior percentiles  \\

\noalign{\smallskip}
\hline
\noalign{\smallskip}
%\multicolumn{2}{l}{Y = reference band (F140W for AEGIS and GOODS-N, F160W for COSMOS, GOODS-S and UDS)}\\
\multicolumn{2}{l}{X = filter name, as defined in Section~\ref{sec:data}.
Synthetic rest-frame $ugi_{\rm s}$ filters are detailed in \citet{Antwi-Danso2022}.}
\end{tabular}
\label{tab:catalog}
\end{table*}

\bibliography{uncover_catalog}{}

\begin{thebibliography}{}
\expandafter\ifx\csname natexlab\endcsname\relax\def\natexlab#1{#1}\fi
\providecommand{\url}[1]{\href{#1}{#1}}
\providecommand{\dodoi}[1]{doi:~\href{http://doi.org/#1}{\nolinkurl{#1}}}
\providecommand{\doeprint}[1]{\href{http://ascl.net/#1}{\nolinkurl{http://ascl.net/#1}}}
\providecommand{\doarXiv}[1]{\href{https://arxiv.org/abs/#1}{\nolinkurl{https://arxiv.org/abs/#1}}}

\bibitem[{{Abbott} {et~al.}(2018){Abbott}, {Abdalla}, {Allam}, {Amara},
  {Annis}, {Asorey}, {Avila}, {Ballester}, {Banerji}, {Barkhouse}, \&
  et~al.}]{Abbott2018}
{Abbott}, T.~M.~C., {Abdalla}, F.~B., {Allam}, S., {et~al.} 2018, \apjs, 239,
  18, \dodoi{10.3847/1538-4365/aae9f0}

\bibitem[{{Adams} {et~al.}(2023){Adams}, {Conselice}, {Ferreira}, {Austin},
  {Trussler}, {Juod{\v{z}}balis}, {Wilkins}, {Caruana}, {Dayal}, {Verma}, \&
  {Vijayan}}]{Adams2023}
{Adams}, N.~J., {Conselice}, C.~J., {Ferreira}, L., {et~al.} 2023, \mnras, 518,
  4755, \dodoi{10.1093/mnras/stac3347}

\bibitem[{{Aihara} {et~al.}(2018){Aihara}, {Arimoto}, {Armstrong}, {Arnouts},
  {Bahcall}, {Bickerton}, {Bosch}, {Bundy}, {Capak}, {Chan}, {Chiba}, {Coupon},
  {Egami}, {Enoki}, {Finet}, {Fujimori}, {Fujimoto}, {Furusawa}, {Furusawa},
  {Goto}, {Goulding}, {Greco}, {Greene}, {Gunn}, {Hamana}, {Harikane},
  {Hashimoto}, {Hattori}, {Hayashi}, {Hayashi}, {He{\l}miniak}, {Higuchi},
  {Hikage}, {Ho}, {Hsieh}, {Huang}, {Huang}, {Ikeda}, {Imanishi}, {Inoue},
  {Iwasawa}, {Iwata}, {Jaelani}, {Jian}, {Kamata}, {Karoji}, {Kashikawa},
  {Katayama}, {Kawanomoto}, {Kayo}, {Koda}, {Koike}, {Kojima}, {Komiyama},
  {Konno}, {Koshida}, {Koyama}, {Kusakabe}, {Leauthaud}, {Lee}, {Lin}, {Lin},
  {Lupton}, {Mandelbaum}, {Matsuoka}, {Medezinski}, {Mineo}, {Miyama},
  {Miyatake}, {Miyazaki}, {Momose}, {More}, {More}, {Moritani}, {Moriya},
  {Morokuma}, {Mukae}, {Murata}, {Murayama}, {Nagao}, {Nakata}, {Niida},
  {Niikura}, {Nishizawa}, {Obuchi}, {Oguri}, {Oishi}, {Okabe}, {Okamoto},
  {Okura}, {Ono}, {Onodera}, {Onoue}, {Osato}, {Ouchi}, {Price}, {Pyo}, {Sako},
  {Sawicki}, {Shibuya}, {Shimasaku}, {Shimono}, {Shirasaki}, {Silverman},
  {Simet}, {Speagle}, {Spergel}, {Strauss}, {Sugahara}, {Sugiyama}, {Suto},
  {Suyu}, {Suzuki}, {Tait}, {Takada}, {Takata}, {Tamura}, {Tanaka}, {Tanaka},
  {Tanaka}, {Tanaka}, {Terai}, {Terashima}, {Toba}, {Tominaga}, {Toshikawa},
  {Turner}, {Uchida}, {Uchiyama}, {Umetsu}, {Uraguchi}, {Urata}, {Usuda},
  {Utsumi}, {Wang}, {Wang}, {Wong}, {Yabe}, {Yamada}, {Yamanoi}, {Yasuda},
  {Yeh}, {Yonehara}, \& {Yuma}}]{Aihara2018}
{Aihara}, H., {Arimoto}, N., {Armstrong}, R., {et~al.} 2018, \pasj, 70, S4,
  \dodoi{10.1093/pasj/psx066}

\bibitem[{{Allard} {et~al.}(2012){Allard}, {Homeier}, \&
  {Freytag}}]{Allard_2012}
{Allard}, F., {Homeier}, D., \& {Freytag}, B. 2012, Philosophical Transactions
  of the Royal Society of London Series A, 370, 2765,
  \dodoi{10.1098/rsta.2011.0269}

\bibitem[{{Antwi-Danso} {et~al.}(2022){Antwi-Danso}, {Papovich}, {Leja},
  {Marchesini}, {Marsan}, {Martis}, {Labb{\'e}}, {Muzzin}, {Glazebrook},
  {Straatman}, \& {Tran}}]{Antwi-Danso2022}
{Antwi-Danso}, J., {Papovich}, C., {Leja}, J., {et~al.} 2022, arXiv e-prints,
  arXiv:2207.07170.
\newblock \doarXiv{2207.07170}

\bibitem[{{Astropy Collaboration} {et~al.}(2013){Astropy Collaboration},
  {Robitaille}, {Tollerud}, {Greenfield}, {Droettboom}, {Bray}, {Aldcroft},
  {Davis}, {Ginsburg}, {Price-Whelan}, {Kerzendorf}, {Conley}, {Crighton},
  {Barbary}, {Muna}, {Ferguson}, {Grollier}, {Parikh}, {Nair}, {Unther},
  {Deil}, {Woillez}, {Conseil}, {Kramer}, {Turner}, {Singer}, {Fox}, {Weaver},
  {Zabalza}, {Edwards}, {Azalee Bostroem}, {Burke}, {Casey}, {Crawford},
  {Dencheva}, {Ely}, {Jenness}, {Labrie}, {Lim}, {Pierfederici}, {Pontzen},
  {Ptak}, {Refsdal}, {Servillat}, \& {Streicher}}]{astropy:2013}
{Astropy Collaboration}, {Robitaille}, T.~P., {Tollerud}, E.~J., {et~al.} 2013,
  \aap, 558, A33, \dodoi{10.1051/0004-6361/201322068}

\bibitem[{{Astropy Collaboration} {et~al.}(2018){Astropy Collaboration},
  {Price-Whelan}, {Sip{\H{o}}cz}, {G{\"u}nther}, {Lim}, {Crawford}, {Conseil},
  {Shupe}, {Craig}, {Dencheva}, {Ginsburg}, {Vand erPlas}, {Bradley},
  {P{\'e}rez-Su{\'a}rez}, {de Val-Borro}, {Aldcroft}, {Cruz}, {Robitaille},
  {Tollerud}, {Ardelean}, {Babej}, {Bach}, {Bachetti}, {Bakanov}, {Bamford},
  {Barentsen}, {Barmby}, {Baumbach}, {Berry}, {Biscani}, {Boquien}, {Bostroem},
  {Bouma}, {Brammer}, {Bray}, {Breytenbach}, {Buddelmeijer}, {Burke},
  {Calderone}, {Cano Rodr{\'\i}guez}, {Cara}, {Cardoso}, {Cheedella}, {Copin},
  {Corrales}, {Crichton}, {D'Avella}, {Deil}, {Depagne}, {Dietrich}, {Donath},
  {Droettboom}, {Earl}, {Erben}, {Fabbro}, {Ferreira}, {Finethy}, {Fox},
  {Garrison}, {Gibbons}, {Goldstein}, {Gommers}, {Greco}, {Greenfield},
  {Groener}, {Grollier}, {Hagen}, {Hirst}, {Homeier}, {Horton}, {Hosseinzadeh},
  {Hu}, {Hunkeler}, {Ivezi{\'c}}, {Jain}, {Jenness}, {Kanarek}, {Kendrew},
  {Kern}, {Kerzendorf}, {Khvalko}, {King}, {Kirkby}, {Kulkarni}, {Kumar},
  {Lee}, {Lenz}, {Littlefair}, {Ma}, {Macleod}, {Mastropietro}, {McCully},
  {Montagnac}, {Morris}, {Mueller}, {Mumford}, {Muna}, {Murphy}, {Nelson},
  {Nguyen}, {Ninan}, {N{\"o}the}, {Ogaz}, {Oh}, {Parejko}, {Parley}, {Pascual},
  {Patil}, {Patil}, {Plunkett}, {Prochaska}, {Rastogi}, {Reddy Janga},
  {Sabater}, {Sakurikar}, {Seifert}, {Sherbert}, {Sherwood-Taylor}, {Shih},
  {Sick}, {Silbiger}, {Singanamalla}, {Singer}, {Sladen}, {Sooley},
  {Sornarajah}, {Streicher}, {Teuben}, {Thomas}, {Tremblay}, {Turner},
  {Terr{\'o}n}, {van Kerkwijk}, {de la Vega}, {Watkins}, {Weaver}, {Whitmore},
  {Woillez}, {Zabalza}, \& {Astropy Contributors}}]{astropy:2018}
{Astropy Collaboration}, {Price-Whelan}, A.~M., {Sip{\H{o}}cz}, B.~M., {et~al.}
  2018, \aj, 156, 123, \dodoi{10.3847/1538-3881/aabc4f}

\bibitem[{{Astropy Collaboration} {et~al.}(2022){Astropy Collaboration},
  {Price-Whelan}, {Lim}, {Earl}, {Starkman}, {Bradley}, {Shupe}, {Patil},
  {Corrales}, {Brasseur}, {N{"o}the}, {Donath}, {Tollerud}, {Morris},
  {Ginsburg}, {Vaher}, {Weaver}, {Tocknell}, {Jamieson}, {van Kerkwijk},
  {Robitaille}, {Merry}, {Bachetti}, {G{"u}nther}, {Aldcroft},
  {Alvarado-Montes}, {Archibald}, {B{'o}di}, {Bapat}, {Barentsen}, {Baz{'a}n},
  {Biswas}, {Boquien}, {Burke}, {Cara}, {Cara}, {Conroy}, {Conseil}, {Craig},
  {Cross}, {Cruz}, {D'Eugenio}, {Dencheva}, {Devillepoix}, {Dietrich},
  {Eigenbrot}, {Erben}, {Ferreira}, {Foreman-Mackey}, {Fox}, {Freij}, {Garg},
  {Geda}, {Glattly}, {Gondhalekar}, {Gordon}, {Grant}, {Greenfield}, {Groener},
  {Guest}, {Gurovich}, {Handberg}, {Hart}, {Hatfield-Dodds}, {Homeier},
  {Hosseinzadeh}, {Jenness}, {Jones}, {Joseph}, {Kalmbach}, {Karamehmetoglu},
  {Ka{l}uszy{'n}ski}, {Kelley}, {Kern}, {Kerzendorf}, {Koch}, {Kulumani},
  {Lee}, {Ly}, {Ma}, {MacBride}, {Maljaars}, {Muna}, {Murphy}, {Norman},
  {O'Steen}, {Oman}, {Pacifici}, {Pascual}, {Pascual-Granado}, {Patil},
  {Perren}, {Pickering}, {Rastogi}, {Roulston}, {Ryan}, {Rykoff}, {Sabater},
  {Sakurikar}, {Salgado}, {Sanghi}, {Saunders}, {Savchenko}, {Schwardt},
  {Seifert-Eckert}, {Shih}, {Jain}, {Shukla}, {Sick}, {Simpson},
  {Singanamalla}, {Singer}, {Singhal}, {Sinha}, {Sip{H{o}}cz}, {Spitler},
  {Stansby}, {Streicher}, {{{S}}umak}, {Swinbank}, {Taranu}, {Tewary},
  {Tremblay}, {Val-Borro}, {Van Kooten}, {Vasovi{'c}}, {Verma}, {de Miranda
  Cardoso}, {Williams}, {Wilson}, {Winkel}, {Wood-Vasey}, {Xue}, {Yoachim},
  {Zhang}, {Zonca}, \& {Astropy Project Contributors}}]{astropy:2022}
{Astropy Collaboration}, {Price-Whelan}, A.~M., {Lim}, P.~L., {et~al.} 2022,
  apj, 935, 167, \dodoi{10.3847/1538-4357/ac7c74}

\bibitem[{{Atek} {et~al.}(2018){Atek}, {Richard}, {Kneib}, \&
  {Schaerer}}]{Atek2018}
{Atek}, H., {Richard}, J., {Kneib}, J.-P., \& {Schaerer}, D. 2018, \mnras, 479,
  5184, \dodoi{10.1093/mnras/sty1820}

\bibitem[{{Atek} {et~al.}(2023){Atek}, {Shuntov}, {Furtak}, {Richard}, {Kneib},
  {Mahler}, {Zitrin}, {McCracken}, {Charlot}, {Chevallard}, \&
  {Chemerynska}}]{Atek2023}
{Atek}, H., {Shuntov}, M., {Furtak}, L.~J., {et~al.} 2023, \mnras, 519, 1201,
  \dodoi{10.1093/mnras/stac3144}

\bibitem[{{Barbary}(2016)}]{Barbary2016}
{Barbary}, K. 2016, The Journal of Open Source Software, 1, 58,
  \dodoi{10.21105/joss.00058}

\bibitem[{Barbary(2016)}]{extinction}
Barbary, K. 2016, extinction v0.3.0,  Zenodo, \dodoi{10.5281/zenodo.804967}

\bibitem[{{Beckwith} {et~al.}(2006){Beckwith}, {Stiavelli}, {Koekemoer},
  {Caldwell}, {Ferguson}, {Hook}, {Lucas}, {Bergeron}, {Corbin}, {Jogee},
  {Panagia}, {Robberto}, {Royle}, {Somerville}, \& {Sosey}}]{Beckwith2006}
{Beckwith}, S. V.~W., {Stiavelli}, M., {Koekemoer}, A.~M., {et~al.} 2006, \aj,
  132, 1729, \dodoi{10.1086/507302}

\bibitem[{{Bergamini} {et~al.}(2023){Bergamini}, {Acebron}, {Grillo}, {Rosati},
  {Caminha}, {Mercurio}, {Vanzella}, {Mason}, {Treu}, {Angora}, {Brammer},
  {Meneghetti}, {Nonino}, {Boyett}, {Brada{\v{c}}}, {Castellano}, {Fontana},
  {Morishita}, {Paris}, {Prieto-Lyon}, {Roberts-Borsani}, {Roy}, {Santini},
  {Vulcani}, {Wang}, \& {Yang}}]{Bergamini2023}
{Bergamini}, P., {Acebron}, A., {Grillo}, C., {et~al.} 2023, \apj, 952, 84,
  \dodoi{10.3847/1538-4357/acd643}

\bibitem[{{Bertin} \& {Arnouts}(1996)}]{SE}
{Bertin}, E., \& {Arnouts}, S. 1996, \aaps, 117, 393,
  \dodoi{10.1051/aas:1996164}

\bibitem[{{Bezanson} {et~al.}(2022){Bezanson}, {Labbe}, {Whitaker}, {Leja},
  {Price}, {Franx}, {Brammer}, {Marchesini}, {Zitrin}, {Wang}, {Weaver},
  {Furtak}, {Atek}, {Coe}, {Cutler}, {Dayal}, {van Dokkum}, {Feldmann},
  {Forster Schreiber}, {Fujimoto}, {Geha}, {Glazebrook}, {de Graaff}, {Juneau},
  {Kassin}, {Kriek}, {Khullar}, {Greene}, {Maseda}, {Oesch}, {Smit},
  {Stefanon}, {Taylor}, \& {Williams}}]{bezanson:22}
{Bezanson}, R., {Labbe}, I., {Whitaker}, K.~E., {et~al.} 2022, arXiv e-prints,
  arXiv:2212.04026.
\newblock \doarXiv{2212.04026}

\bibitem[{{Bhatawdekar} {et~al.}(2019){Bhatawdekar}, {Conselice},
  {Margalef-Bentabol}, \& {Duncan}}]{Bhatawdekar2019}
{Bhatawdekar}, R., {Conselice}, C.~J., {Margalef-Bentabol}, B., \& {Duncan}, K.
  2019, \mnras, 486, 3805, \dodoi{10.1093/mnras/stz866}

\bibitem[{{Boucaud} {et~al.}(2016){Boucaud}, {Bocchio}, {Abergel}, {Orieux},
  {Dole}, \& {Hadj-Youcef}}]{Boucaud2016}
{Boucaud}, A., {Bocchio}, M., {Abergel}, A., {et~al.} 2016, \aap, 596, A63,
  \dodoi{10.1051/0004-6361/201629080}

\bibitem[{{Bouwens} {et~al.}(2022){Bouwens}, {Illingworth}, {Ellis}, {Oesch},
  \& {Stefanon}}]{Bouwens2022}
{Bouwens}, R.~J., {Illingworth}, G., {Ellis}, R.~S., {Oesch}, P., \&
  {Stefanon}, M. 2022, \apj, 940, 55, \dodoi{10.3847/1538-4357/ac86d1}

\bibitem[{{Bouwens} {et~al.}(2017){Bouwens}, {Oesch}, {Illingworth}, {Ellis},
  \& {Stefanon}}]{Bouwens2017}
{Bouwens}, R.~J., {Oesch}, P.~A., {Illingworth}, G.~D., {Ellis}, R.~S., \&
  {Stefanon}, M. 2017, \apj, 843, 129, \dodoi{10.3847/1538-4357/aa70a4}

\bibitem[{{Bouwens} {et~al.}(2011){Bouwens}, {Illingworth}, {Oesch},
  {Labb{\'e}}, {Trenti}, {van Dokkum}, {Franx}, {Stiavelli}, {Carollo},
  {Magee}, \& {Gonzalez}}]{Bouwens2011}
{Bouwens}, R.~J., {Illingworth}, G.~D., {Oesch}, P.~A., {et~al.} 2011, \apj,
  737, 90, \dodoi{10.1088/0004-637X/737/2/90}

\bibitem[{Bradley {et~al.}(2022)Bradley, Sipőcz, Robitaille, Tollerud,
  Vinícius, Deil, Barbary, Wilson, Busko, Donath, Günther, Cara, Lim,
  Meßlinger, Conseil, Bostroem, Droettboom, Bray, Bratholm, Barentsen, Craig,
  Rathi, Pascual, Perren, Georgiev, de~Val-Borro, Kerzendorf, Bach, Quint, \&
  Souchereau}]{photutils}
Bradley, L., Sipőcz, B., Robitaille, T., {et~al.} 2022, astropy/photutils:
  1.5.0, 1.5.0,  Zenodo, \dodoi{10.5281/zenodo.6825092}

\bibitem[{{Bradley} {et~al.}(2022){Bradley}, {Coe}, {Brammer}, {Furtak},
  {Larson}, {Andrade-Santos}, {Bhatawdekar}, {Bradac}, {Broadhurst}, {Carnall},
  {Conselice}, {Diego}, {Frye}, {Fujimoto}, {Y. -Y Hsiao}, {Hutchison}, {Jung},
  {Mahler}, {McCandliss}, {Oguri}, {Postman}, {Sharon}, {Trenti}, {Vanzella},
  {Welch}, {Windhorst}, \& {Zitrin}}]{Bradley2022}
{Bradley}, L.~D., {Coe}, D., {Brammer}, G., {et~al.} 2022, arXiv e-prints,
  arXiv:2210.01777.
\newblock \doarXiv{2210.01777}

\bibitem[{{Brammer}(2019)}]{Brammer2019}
{Brammer}, G. 2019, {Grizli: Grism redshift and line analysis software},
  Astrophysics Source Code Library, record ascl:1905.001.
\newblock \doeprint{1905.001}

\bibitem[{{Brammer} {et~al.}(2008){Brammer}, {van Dokkum}, \&
  {Coppi}}]{Brammer2008}
{Brammer}, G.~B., {van Dokkum}, P.~G., \& {Coppi}, P. 2008, \apj, 686, 1503,
  \dodoi{10.1086/591786}

\bibitem[{{Broadhurst} {et~al.}(1995){Broadhurst}, {Taylor}, \&
  {Peacock}}]{Broadhurst1995}
{Broadhurst}, T.~J., {Taylor}, A.~N., \& {Peacock}, J.~A. 1995, \apj, 438, 49,
  \dodoi{10.1086/175053}

\bibitem[{{Carnall} {et~al.}(2023){Carnall}, {Begley}, {McLeod}, {Hamadouche},
  {Donnan}, {McLure}, {Dunlop}, {Milvang-Jensen}, {Bondestam}, {Cullen},
  {Jewell}, \& {Pollock}}]{Carnall2023}
{Carnall}, A.~C., {Begley}, R., {McLeod}, D.~J., {et~al.} 2023, \mnras, 518,
  L45, \dodoi{10.1093/mnrasl/slac136}

\bibitem[{{Coe} {et~al.}(2013){Coe}, {Zitrin}, {Carrasco}, {Shu}, {Zheng},
  {Postman}, {Bradley}, {Koekemoer}, {Bouwens}, {Broadhurst}, {Monna}, {Host},
  {Moustakas}, {Ford}, {Moustakas}, {van der Wel}, {Donahue}, {Rodney},
  {Ben{\'\i}tez}, {Jouvel}, {Seitz}, {Kelson}, \& {Rosati}}]{Coe2013}
{Coe}, D., {Zitrin}, A., {Carrasco}, M., {et~al.} 2013, \apj, 762, 32,
  \dodoi{10.1088/0004-637X/762/1/32}

\bibitem[{{Coe} {et~al.}(2019){Coe}, {Salmon}, {Brada{\v{c}}}, {Bradley},
  {Sharon}, {Zitrin}, {Acebron}, {Cerny}, {Cibirka}, {Strait},
  {Paterno-Mahler}, {Mahler}, {Avila}, {Ogaz}, {Huang}, {Pelliccia}, {Stark},
  {Mainali}, {Oesch}, {Trenti}, {Carrasco}, {Dawson}, {Rodney}, {Strolger},
  {Riess}, {Jones}, {Frye}, {Czakon}, {Umetsu}, {Vulcani}, {Graur}, {Jha},
  {Graham}, {Molino}, {Nonino}, {Hjorth}, {Selsing}, {Christensen},
  {Kikuchihara}, {Ouchi}, {Oguri}, {Welch}, {Lemaux}, {Andrade-Santos}, {Hoag},
  {Johnson}, {Peterson}, {Past}, {Fox}, {Agulli}, {Livermore}, {Ryan}, {Lam},
  {Sendra-Server}, {Toft}, {Lovisari}, \& {Su}}]{Coe2019}
{Coe}, D., {Salmon}, B., {Brada{\v{c}}}, M., {et~al.} 2019, \apj, 884, 85,
  \dodoi{10.3847/1538-4357/ab412b}

\bibitem[{{Ferrarese} {et~al.}(2006){Ferrarese}, {C{\^o}t{\'e}}, {Jord{\'a}n},
  {Peng}, {Blakeslee}, {Piatek}, {Mei}, {Merritt}, {Milosavljevi{\'c}},
  {Tonry}, \& {West}}]{ferrarese:06}
{Ferrarese}, L., {C{\^o}t{\'e}}, P., {Jord{\'a}n}, A., {et~al.} 2006, \apjs,
  164, 334, \dodoi{10.1086/501350}

\bibitem[{{Finkelstein} {et~al.}(2015){Finkelstein}, {Ryan}, {Papovich},
  {Dickinson}, {Song}, {Somerville}, {Ferguson}, {Salmon}, {Giavalisco},
  {Koekemoer}, {Ashby}, {Behroozi}, {Castellano}, {Dunlop}, {Faber}, {Fazio},
  {Fontana}, {Grogin}, {Hathi}, {Jaacks}, {Kocevski}, {Livermore}, {McLure},
  {Merlin}, {Mobasher}, {Newman}, {Rafelski}, {Tilvi}, \&
  {Willner}}]{Finkelstein2015}
{Finkelstein}, S.~L., {Ryan}, Russell~E., J., {Papovich}, C., {et~al.} 2015,
  \apj, 810, 71, \dodoi{10.1088/0004-637X/810/1/71}

\bibitem[{{Fitzpatrick} \& {Massa}(2007)}]{Fitzpatrick2007}
{Fitzpatrick}, E.~L., \& {Massa}, D. 2007, \apj, 663, 320,
  \dodoi{10.1086/518158}

\bibitem[{{Fox} {et~al.}(2022){Fox}, {Mahler}, {Sharon}, \& {Remolina
  Gonz{\'a}lez}}]{Fox2022}
{Fox}, C., {Mahler}, G., {Sharon}, K., \& {Remolina Gonz{\'a}lez}, J.~D. 2022,
  \apj, 928, 87, \dodoi{10.3847/1538-4357/ac5024}

\bibitem[{{Furtak} {et~al.}(2021){Furtak}, {Atek}, {Lehnert}, {Chevallard}, \&
  {Charlot}}]{Furtak2021}
{Furtak}, L.~J., {Atek}, H., {Lehnert}, M.~D., {Chevallard}, J., \& {Charlot},
  S. 2021, \mnras, 501, 1568, \dodoi{10.1093/mnras/staa3760}

\bibitem[{{Furtak} {et~al.}(2022){Furtak}, {Shuntov}, {Atek}, {Zitrin},
  {Richard}, {Lehnert}, \& {Chevallard}}]{Furtak2022a}
{Furtak}, L.~J., {Shuntov}, M., {Atek}, H., {et~al.} 2022, \mnras,
  \dodoi{10.1093/mnras/stac3717}

\bibitem[{{Furtak} {et~al.}(2023){Furtak}, {Zitrin}, {Weaver}, {Atek},
  {Bezanson}, {Labb{\'e}}, {Whitaker}, {Leja}, {Price}, {Brammer}, {Wang},
  {Marchesini}, {Pan}, {Dayal}, {van Dokkum}, {Feldmann}, {Fujimoto}, {Franx},
  {Khullar}, {Nelson}, \& {Mowla}}]{Furtak2022}
{Furtak}, L.~J., {Zitrin}, A., {Weaver}, J.~R., {et~al.} 2023, \mnras, 523,
  4568, \dodoi{10.1093/mnras/stad1627}

\bibitem[{{Gaia Collaboration} {et~al.}(2016){Gaia Collaboration}, {Prusti},
  {de Bruijne}, {Brown}, {Vallenari}, {Babusiaux}, {Bailer-Jones}, {Bastian},
  {Biermann}, {Evans}, {Eyer}, {Jansen}, {Jordi}, {Klioner}, {Lammers},
  {Lindegren}, {Luri}, {Mignard}, {Milligan}, {Panem}, {Poinsignon},
  {Pourbaix}, {Randich}, {Sarri}, {Sartoretti}, {Siddiqui}, {Soubiran},
  {Valette}, {van Leeuwen}, {Walton}, {Aerts}, {Arenou}, {Cropper}, {Drimmel},
  {H{\o}g}, {Katz}, {Lattanzi}, {O'Mullane}, {Grebel}, {Holland}, {Huc},
  {Passot}, {Bramante}, {Cacciari}, {Casta{\~n}eda}, {Chaoul}, {Cheek}, {De
  Angeli}, {Fabricius}, {Guerra}, {Hern{\'a}ndez}, {Jean-Antoine-Piccolo},
  {Masana}, {Messineo}, {Mowlavi}, {Nienartowicz}, {Ord{\'o}{\~n}ez-Blanco},
  {Panuzzo}, {Portell}, {Richards}, {Riello}, {Seabroke}, {Tanga},
  {Th{\'e}venin}, {Torra}, {Els}, {Gracia-Abril}, {Comoretto},
  {Garcia-Reinaldos}, {Lock}, {Mercier}, {Altmann}, {Andrae}, {Astraatmadja},
  {Bellas-Velidis}, {Benson}, {Berthier}, {Blomme}, {Busso}, {Carry},
  {Cellino}, {Clementini}, {Cowell}, {Creevey}, {Cuypers}, {Davidson}, {De
  Ridder}, {de Torres}, {Delchambre}, {Dell'Oro}, {Ducourant}, {Fr{\'e}mat},
  {Garc{\'\i}a-Torres}, {Gosset}, {Halbwachs}, {Hambly}, {Harrison}, {Hauser},
  {Hestroffer}, {Hodgkin}, {Huckle}, {Hutton}, {Jasniewicz}, {Jordan},
  {Kontizas}, {Korn}, {Lanzafame}, {Manteiga}, {Moitinho}, {Muinonen},
  {Osinde}, {Pancino}, {Pauwels}, {Petit}, {Recio-Blanco}, {Robin}, {Sarro},
  {Siopis}, {Smith}, {Smith}, {Sozzetti}, {Thuillot}, {van Reeven}, {Viala},
  {Abbas}, {Abreu Aramburu}, {Accart}, {Aguado}, {Allan}, {Allasia},
  {Altavilla}, {{\'A}lvarez}, {Alves}, {Anderson}, {Andrei}, {Anglada Varela},
  {Antiche}, {Antoja}, {Ant{\'o}n}, {Arcay}, {Atzei}, {Ayache}, {Bach},
  {Baker}, {Balaguer-N{\'u}{\~n}ez}, {Barache}, {Barata}, {Barbier}, {Barblan},
  {Baroni}, {Barrado y Navascu{\'e}s}, {Barros}, {Barstow}, {Becciani},
  {Bellazzini}, {Bellei}, {Bello Garc{\'\i}a}, {Belokurov}, {Bendjoya},
  {Berihuete}, {Bianchi}, {Bienaym{\'e}}, {Billebaud}, {Blagorodnova},
  {Blanco-Cuaresma}, {Boch}, {Bombrun}, {Borrachero}, {Bouquillon}, {Bourda},
  {Bouy}, {Bragaglia}, {Breddels}, {Brouillet}, {Br{\"u}semeister},
  {Bucciarelli}, {Budnik}, {Burgess}, {Burgon}, {Burlacu}, {Busonero}, {Buzzi},
  {Caffau}, {Cambras}, {Campbell}, {Cancelliere}, {Cantat-Gaudin}, {Carlucci},
  {Carrasco}, {Castellani}, {Charlot}, {Charnas}, {Charvet}, {Chassat},
  {Chiavassa}, {Clotet}, {Cocozza}, {Collins}, {Collins}, {Costigan}, {Crifo},
  {Cross}, {Crosta}, {Crowley}, {Dafonte}, {Damerdji}, {Dapergolas}, {David},
  {David}, {De Cat}, {de Felice}, {de Laverny}, {De Luise}, {De March}, {de
  Martino}, {de Souza}, {Debosscher}, {del Pozo}, {Delbo}, {Delgado},
  {Delgado}, {di Marco}, {Di Matteo}, {Diakite}, {Distefano}, {Dolding}, {Dos
  Anjos}, {Drazinos}, {Dur{\'a}n}, {Dzigan}, {Ecale}, {Edvardsson}, {Enke},
  {Erdmann}, {Escolar}, {Espina}, {Evans}, {Eynard Bontemps}, {Fabre},
  {Fabrizio}, {Faigler}, {Falc{\~a}o}, {Farr{\`a}s Casas}, {Faye}, {Federici},
  {Fedorets}, {Fern{\'a}ndez-Hern{\'a}ndez}, {Fernique}, {Fienga}, {Figueras},
  {Filippi}, {Findeisen}, {Fonti}, {Fouesneau}, {Fraile}, {Fraser}, {Fuchs},
  {Furnell}, {Gai}, {Galleti}, {Galluccio}, {Garabato}, {Garc{\'\i}a-Sedano},
  {Gar{\'e}}, {Garofalo}, {Garralda}, {Gavras}, {Gerssen}, {Geyer}, {Gilmore},
  {Girona}, {Giuffrida}, {Gomes}, {Gonz{\'a}lez-Marcos},
  {Gonz{\'a}lez-N{\'u}{\~n}ez}, {Gonz{\'a}lez-Vidal}, {Granvik}, {Guerrier},
  {Guillout}, {Guiraud}, {G{\'u}rpide}, {Guti{\'e}rrez-S{\'a}nchez}, {Guy},
  {Haigron}, {Hatzidimitriou}, {Haywood}, {Heiter}, {Helmi}, {Hobbs},
  {Hofmann}, {Holl}, {Holland}, {Hunt}, {Hypki}, {Icardi}, {Irwin}, {Jevardat
  de Fombelle}, {Jofr{\'e}}, {Jonker}, {Jorissen}, {Julbe}, {Karampelas},
  {Kochoska}, {Kohley}, {Kolenberg}, {Kontizas}, {Koposov}, {Kordopatis},
  {Koubsky}, {Kowalczyk}, {Krone-Martins}, {Kudryashova}, {Kull}, {Bachchan},
  {Lacoste-Seris}, {Lanza}, {Lavigne}, {Le Poncin-Lafitte}, {Lebreton},
  {Lebzelter}, {Leccia}, {Leclerc}, {Lecoeur-Taibi}, {Lemaitre}, {Lenhardt},
  {Leroux}, {Liao}, {Licata}, {Lindstr{\o}m}, {Lister}, {Livanou}, {Lobel},
  {L{\"o}ffler}, {L{\'o}pez}, {Lopez-Lozano}, {Lorenz}, {Loureiro},
  {MacDonald}, {Magalh{\~a}es Fernandes}, {Managau}, {Mann}, {Mantelet},
  {Marchal}, {Marchant}, {Marconi}, {Marie}, {Marinoni}, {Marrese},
  {Marschalk{\'o}}, {Marshall}, {Mart{\'\i}n-Fleitas}, {Martino}, {Mary},
  {Matijevi{\v{c}}}, {Mazeh}, {McMillan}, {Messina}, {Mestre}, {Michalik},
  {Millar}, {Miranda}, {Molina}, {Molinaro}, {Molinaro}, {Moln{\'a}r},
  {Moniez}, {Montegriffo}, {Monteiro}, {Mor}, {Mora}, {Morbidelli}, {Morel},
  {Morgenthaler}, {Morley}, {Morris}, {Mulone}, {Muraveva}, {Musella},
  {Narbonne}, {Nelemans}, {Nicastro}, {Noval}, {Ord{\'e}novic},
  {Ordieres-Mer{\'e}}, {Osborne}, {Pagani}, {Pagano}, {Pailler}, {Palacin},
  {Palaversa}, {Parsons}, {Paulsen}, {Pecoraro}, {Pedrosa}, {Pentik{\"a}inen},
  {Pereira}, {Pichon}, {Piersimoni}, {Pineau}, {Plachy}, {Plum}, {Poujoulet},
  {Pr{\v{s}}a}, {Pulone}, {Ragaini}, {Rago}, {Rambaux}, {Ramos-Lerate},
  {Ranalli}, {Rauw}, {Read}, {Regibo}, {Renk}, {Reyl{\'e}}, {Ribeiro},
  {Rimoldini}, {Ripepi}, {Riva}, {Rixon}, {Roelens}, {Romero-G{\'o}mez},
  {Rowell}, {Royer}, {Rudolph}, {Ruiz-Dern}, {Sadowski}, {Sagrist{\`a}
  Sell{\'e}s}, {Sahlmann}, {Salgado}, {Salguero}, {Sarasso}, {Savietto},
  {Schnorhk}, {Schultheis}, {Sciacca}, {Segol}, {Segovia}, {Segransan},
  {Serpell}, {Shih}, {Smareglia}, {Smart}, {Smith}, {Solano}, {Solitro},
  {Sordo}, {Soria Nieto}, {Souchay}, {Spagna}, {Spoto}, {Stampa}, {Steele},
  {Steidelm{\"u}ller}, {Stephenson}, {Stoev}, {Suess}, {S{\"u}veges}, {Surdej},
  {Szabados}, {Szegedi-Elek}, {Tapiador}, {Taris}, {Tauran}, {Taylor},
  {Teixeira}, {Terrett}, {Tingley}, {Trager}, {Turon}, {Ulla}, {Utrilla},
  {Valentini}, {van Elteren}, {Van Hemelryck}, {van Leeuwen}, {Varadi},
  {Vecchiato}, {Veljanoski}, {Via}, {Vicente}, {Vogt}, {Voss}, {Votruba},
  {Voutsinas}, {Walmsley}, {Weiler}, {Weingrill}, {Werner}, {Wevers},
  {Whitehead}, {Wyrzykowski}, {Yoldas}, {{\v{Z}}erjal}, {Zucker}, {Zurbach},
  {Zwitter}, {Alecu}, {Allen}, {Allende Prieto}, {Amorim},
  {Anglada-Escud{\'e}}, {Arsenijevic}, {Azaz}, {Balm}, {Beck}, {Bernstein},
  {Bigot}, {Bijaoui}, {Blasco}, {Bonfigli}, {Bono}, {Boudreault}, {Bressan},
  {Brown}, {Brunet}, {Bunclark}, {Buonanno}, {Butkevich}, {Carret}, {Carrion},
  {Chemin}, {Ch{\'e}reau}, {Corcione}, {Darmigny}, {de Boer}, {de Teodoro}, {de
  Zeeuw}, {Delle Luche}, {Domingues}, {Dubath}, {Fodor}, {Fr{\'e}zouls},
  {Fries}, {Fustes}, {Fyfe}, {Gallardo}, {Gallegos}, {Gardiol}, {Gebran},
  {Gomboc}, {G{\'o}mez}, {Grux}, {Gueguen}, {Heyrovsky}, {Hoar}, {Iannicola},
  {Isasi Parache}, {Janotto}, {Joliet}, {Jonckheere}, {Keil}, {Kim},
  {Klagyivik}, {Klar}, {Knude}, {Kochukhov}, {Kolka}, {Kos}, {Kutka}, {Lainey},
  {LeBouquin}, {Liu}, {Loreggia}, {Makarov}, {Marseille}, {Martayan},
  {Martinez-Rubi}, {Massart}, {Meynadier}, {Mignot}, {Munari}, {Nguyen},
  {Nordlander}, {Ocvirk}, {O'Flaherty}, {Olias Sanz}, {Ortiz}, {Osorio},
  {Oszkiewicz}, {Ouzounis}, {Palmer}, {Park}, {Pasquato}, {Peltzer}, {Peralta},
  {P{\'e}turaud}, {Pieniluoma}, {Pigozzi}, {Poels}, {Prat}, {Prod'homme},
  {Raison}, {Rebordao}, {Risquez}, {Rocca-Volmerange}, {Rosen}, {Ruiz-Fuertes},
  {Russo}, {Sembay}, {Serraller Vizcaino}, {Short}, {Siebert}, {Silva},
  {Sinachopoulos}, {Slezak}, {Soffel}, {Sosnowska}, {Strai{\v{z}}ys}, {ter
  Linden}, {Terrell}, {Theil}, {Tiede}, {Troisi}, {Tsalmantza}, {Tur},
  {Vaccari}, {Vachier}, {Valles}, {Van Hamme}, {Veltz}, {Virtanen}, {Wallut},
  {Wichmann}, {Wilkinson}, {Ziaeepour}, \& {Zschocke}}]{GaiaCollaboration2016}
{Gaia Collaboration}, {Prusti}, T., {de Bruijne}, J.~H.~J., {et~al.} 2016,
  \aap, 595, A1, \dodoi{10.1051/0004-6361/201629272}

\bibitem[{{Gaia Collaboration} {et~al.}(2022){Gaia Collaboration}, {Vallenari},
  {Brown}, {Prusti}, {de Bruijne}, {Arenou}, {Babusiaux}, {Biermann},
  {Creevey}, {Ducourant}, {Evans}, {Eyer}, {Guerra}, {Hutton}, {Jordi},
  {Klioner}, {Lammers}, {Lindegren}, {Luri}, {Mignard}, {Panem}, {Pourbaix},
  {Randich}, {Sartoretti}, {Soubiran}, {Tanga}, {Walton}, {Bailer-Jones},
  {Bastian}, {Drimmel}, {Jansen}, {Katz}, {Lattanzi}, {van Leeuwen}, {Bakker},
  {Cacciari}, {Casta{\~n}eda}, {De Angeli}, {Fabricius}, {Fouesneau},
  {Fr{\'e}mat}, {Galluccio}, {Guerrier}, {Heiter}, {Masana}, {Messineo},
  {Mowlavi}, {Nicolas}, {Nienartowicz}, {Pailler}, {Panuzzo}, {Riclet}, {Roux},
  {Seabroke}, {Sordo{\o}rcit}, {Th{\'e}venin}, {Gracia-Abril}, {Portell},
  {Teyssier}, {Altmann}, {Andrae}, {Audard}, {Bellas-Velidis}, {Benson},
  {Berthier}, {Blomme}, {Burgess}, {Busonero}, {Busso}, {C{\'a}novas}, {Carry},
  {Cellino}, {Cheek}, {Clementini}, {Damerdji}, {Davidson}, {de Teodoro},
  {Nu{\~n}ez Campos}, {Delchambre}, {Dell'Oro}, {Esquej},
  {Fern{\'a}ndez-Hern{\'a}ndez}, {Fraile}, {Garabato}, {Garc{\'\i}a-Lario},
  {Gosset}, {Haigron}, {Halbwachs}, {Hambly}, {Harrison}, {Hern{\'a}ndez},
  {Hestroffer}, {Hodgkin}, {Holl}, {Jan{\ss}en}, {Jevardat de Fombelle},
  {Jordan}, {Krone-Martins}, {Lanzafame}, {L{\"o}ffler}, {Marchal}, {Marrese},
  {Moitinho}, {Muinonen}, {Osborne}, {Pancino}, {Pauwels}, {Recio-Blanco},
  {Reyl{\'e}}, {Riello}, {Rimoldini}, {Roegiers}, {Rybizki}, {Sarro}, {Siopis},
  {Smith}, {Sozzetti}, {Utrilla}, {van Leeuwen}, {Abbas}, {{\'A}brah{\'a}m},
  {Abreu Aramburu}, {Aerts}, {Aguado}, {Ajaj}, {Aldea-Montero}, {Altavilla},
  {{\'A}lvarez}, {Alves}, {Anders}, {Anderson}, {Anglada Varela}, {Antoja},
  {Baines}, {Baker}, {Balaguer-N{\'u}{\~n}ez}, {Balbinot}, {Balog}, {Barache},
  {Barbato}, {Barros}, {Barstow}, {Bartolom{\'e}}, {Bassilana}, {Bauchet},
  {Becciani}, {Bellazzini}, {Berihuete}, {Bernet}, {Bertone}, {Bianchi},
  {Binnenfeld}, {Blanco-Cuaresma}, {Blazere}, {Boch}, {Bombrun}, {Bossini},
  {Bouquillon}, {Bragaglia}, {Bramante}, {Breedt}, {Bressan}, {Brouillet},
  {Brugaletta}, {Bucciarelli}, {Burlacu}, {Butkevich}, {Buzzi}, {Caffau},
  {Cancelliere}, {Cantat-Gaudin}, {Carballo}, {Carlucci}, {Carnerero},
  {Carrasco}, {Casamiquela}, {Castellani}, {Castro-Ginard}, {Chaoul},
  {Charlot}, {Chemin}, {Chiaramida}, {Chiavassa}, {Chornay}, {Comoretto},
  {Contursi}, {Cooper}, {Cornez}, {Cowell}, {Crifo}, {Cropper}, {Crosta},
  {Crowley}, {Dafonte}, {Dapergolas}, {David}, {David}, {de Laverny}, {De
  Luise}, {De March}, {De Ridder}, {de Souza}, {de Torres}, {del Peloso}, {del
  Pozo}, {Delbo}, {Delgado}, {Delisle}, {Demouchy}, {Dharmawardena}, {Di
  Matteo}, {Diakite}, {Diener}, {Distefano}, {Dolding}, {Edvardsson}, {Enke},
  {Fabre}, {Fabrizio}, {Faigler}, {Fedorets}, {Fernique}, {Fienga}, {Figueras},
  {Fournier}, {Fouron}, {Fragkoudi}, {Gai}, {Garcia-Gutierrez},
  {Garcia-Reinaldos}, {Garc{\'\i}a-Torres}, {Garofalo}, {Gavel}, {Gavras},
  {Gerlach}, {Geyer}, {Giacobbe}, {Gilmore}, {Girona}, {Giuffrida}, {Gomel},
  {Gomez}, {Gonz{\'a}lez-N{\'u}{\~n}ez}, {Gonz{\'a}lez-Santamar{\'\i}a},
  {Gonz{\'a}lez-Vidal}, {Granvik}, {Guillout}, {Guiraud},
  {Guti{\'e}rrez-S{\'a}nchez}, {Guy}, {Hatzidimitriou}, {Hauser}, {Haywood},
  {Helmer}, {Helmi}, {Sarmiento}, {Hidalgo}, {Hilger}, {H{\l}adczuk}, {Hobbs},
  {Holland}, {Huckle}, {Jardine}, {Jasniewicz}, {Jean-Antoine Piccolo},
  {Jim{\'e}nez-Arranz}, {Jorissen}, {Juaristi Campillo}, {Julbe}, {Karbevska},
  {Kervella}, {Khanna}, {Kontizas}, {Kordopatis}, {Korn}, {K{\'o}sp{\'a}l},
  {Kostrzewa-Rutkowska}, {Kruszy{\'n}ska}, {Kun}, {Laizeau}, {Lambert},
  {Lanza}, {Lasne}, {Le Campion}, {Lebreton}, {Lebzelter}, {Leccia}, {Leclerc},
  {Lecoeur-Taibi}, {Liao}, {Licata}, {Lindstr{\o}m}, {Lister}, {Livanou},
  {Lobel}, {Lorca}, {Loup}, {Madrero Pardo}, {Magdaleno Romeo}, {Managau},
  {Mann}, {Manteiga}, {Marchant}, {Marconi}, {Marcos}, {Marcos Santos},
  {Mar{\'\i}n Pina}, {Marinoni}, {Marocco}, {Marshall}, {Polo},
  {Mart{\'\i}n-Fleitas}, {Marton}, {Mary}, {Masip}, {Massari},
  {Mastrobuono-Battisti}, {Mazeh}, {McMillan}, {Messina}, {Michalik}, {Millar},
  {Mints}, {Molina}, {Molinaro}, {Moln{\'a}r}, {Monari}, {Mongui{\'o}},
  {Montegriffo}, {Montero}, {Mor}, {Mora}, {Morbidelli}, {Morel}, {Morris},
  {Muraveva}, {Murphy}, {Musella}, {Nagy}, {Noval}, {Oca{\~n}a}, {Ogden},
  {Ordenovic}, {Osinde}, {Pagani}, {Pagano}, {Palaversa}, {Palicio},
  {Pallas-Quintela}, {Panahi}, {Payne-Wardenaar}, {Pe{\~n}alosa Esteller},
  {Penttil{\"a}}, {Pichon}, {Piersimoni}, {Pineau}, {Plachy}, {Plum}, {Poggio},
  {Pr{\v{s}}a}, {Pulone}, {Racero}, {Ragaini}, {Rainer}, {Raiteri}, {Rambaux},
  {Ramos}, {Ramos-Lerate}, {Re Fiorentin}, {Regibo}, {Richards}, {Rios Diaz},
  {Ripepi}, {Riva}, {Rix}, {Rixon}, {Robichon}, {Robin}, {Robin}, {Roelens},
  {Rogues}, {Rohrbasser}, {Romero-G{\'o}mez}, {Rowell}, {Royer}, {Ruz Mieres},
  {Rybicki}, {Sadowski}, {S{\'a}ez N{\'u}{\~n}ez}, {Sagrist{\`a} Sell{\'e}s},
  {Sahlmann}, {Salguero}, {Samaras}, {Sanchez Gimenez}, {Sanna},
  {Santove{\~n}a}, {Sarasso}, {Schultheis}, {Sciacca}, {Segol}, {Segovia},
  {S{\'e}gransan}, {Semeux}, {Shahaf}, {Siddiqui}, {Siebert}, {Siltala},
  {Silvelo}, {Slezak}, {Slezak}, {Smart}, {Snaith}, {Solano}, {Solitro},
  {Souami}, {Souchay}, {Spagna}, {Spina}, {Spoto}, {Steele},
  {Steidelm{\"u}ller}, {Stephenson}, {S{\"u}veges}, {Surdej}, {Szabados},
  {Szegedi-Elek}, {Taris}, {Taylo}, {Teixeira}, {Tolomei}, {Tonello}, {Torra},
  {Torra}, {Torralba Elipe}, {Trabucchi}, {Tsounis}, {Turon}, {Ulla}, {Unger},
  {Vaillant}, {van Dillen}, {van Reeven}, {Vanel}, {Vecchiato}, {Viala},
  {Vicente}, {Voutsinas}, {Weiler}, {Wevers}, {Wyrzykowski}, {Yoldas}, {Yvard},
  {Zhao}, {Zorec}, {Zucker}, \& {Zwitter}}]{GaiaCollaboration2022}
{Gaia Collaboration}, {Vallenari}, A., {Brown}, A.~G.~A., {et~al.} 2022, arXiv
  e-prints, arXiv:2208.00211.
\newblock \doarXiv{2208.00211}

\bibitem[{{Giavalisco} {et~al.}(2004){Giavalisco}, {Ferguson}, {Koekemoer},
  {Dickinson}, {Alexander}, {Bauer}, {Bergeron}, {Biagetti}, {Brandt},
  {Casertano}, {Cesarsky}, {Chatzichristou}, {Conselice}, {Cristiani}, {Da
  Costa}, {Dahlen}, {de Mello}, {Eisenhardt}, {Erben}, {Fall}, {Fassnacht},
  {Fosbury}, {Fruchter}, {Gardner}, {Grogin}, {Hook}, {Hornschemeier}, {Idzi},
  {Jogee}, {Kretchmer}, {Laidler}, {Lee}, {Livio}, {Lucas}, {Madau},
  {Mobasher}, {Moustakas}, {Nonino}, {Padovani}, {Papovich}, {Park},
  {Ravindranath}, {Renzini}, {Richardson}, {Riess}, {Rosati}, {Schirmer},
  {Schreier}, {Somerville}, {Spinrad}, {Stern}, {Stiavelli}, {Strolger},
  {Urry}, {Vandame}, {Williams}, \& {Wolf}}]{Giavalisco2004}
{Giavalisco}, M., {Ferguson}, H.~C., {Koekemoer}, A.~M., {et~al.} 2004, \apjl,
  600, L93, \dodoi{10.1086/379232}

\bibitem[{{Gonzaga} {et~al.}(2012){Gonzaga}, {Hack}, {Fruchter}, \&
  {Mack}}]{Gonzaga12}
{Gonzaga}, S., {Hack}, W., {Fruchter}, A., \& {Mack}, J. 2012, {The DrizzlePac
  Handbook}

\bibitem[{{Gordon} {et~al.}(2022){Gordon}, {Bohlin}, {Sloan}, {Rieke}, {Volk},
  {Boyer}, {Muzerolle}, {Schlawin}, {Deustua}, {Hines}, {Kraemer}, {Mullally},
  \& {Su}}]{Gordon2022}
{Gordon}, K.~D., {Bohlin}, R., {Sloan}, G.~C., {et~al.} 2022, \aj, 163, 267,
  \dodoi{10.3847/1538-3881/ac66dc}

\bibitem[{{Grogin} {et~al.}(2011){Grogin}, {Kocevski}, {Faber}, {Ferguson},
  {Koekemoer}, {Riess}, {Acquaviva}, {Alexander}, {Almaini}, {Ashby}, {Barden},
  {Bell}, {Bournaud}, {Brown}, {Caputi}, {Casertano}, {Cassata}, {Castellano},
  {Challis}, {Chary}, {Cheung}, {Cirasuolo}, {Conselice}, {Roshan Cooray},
  {Croton}, {Daddi}, {Dahlen}, {Dav{\'e}}, {de Mello}, {Dekel}, {Dickinson},
  {Dolch}, {Donley}, {Dunlop}, {Dutton}, {Elbaz}, {Fazio}, {Filippenko},
  {Finkelstein}, {Fontana}, {Gardner}, {Garnavich}, {Gawiser}, {Giavalisco},
  {Grazian}, {Guo}, {Hathi}, {H{\"a}ussler}, {Hopkins}, {Huang}, {Huang},
  {Jha}, {Kartaltepe}, {Kirshner}, {Koo}, {Lai}, {Lee}, {Li}, {Lotz}, {Lucas},
  {Madau}, {McCarthy}, {McGrath}, {McIntosh}, {McLure}, {Mobasher},
  {Moustakas}, {Mozena}, {Nandra}, {Newman}, {Niemi}, {Noeske}, {Papovich},
  {Pentericci}, {Pope}, {Primack}, {Rajan}, {Ravindranath}, {Reddy}, {Renzini},
  {Rix}, {Robaina}, {Rodney}, {Rosario}, {Rosati}, {Salimbeni}, {Scarlata},
  {Siana}, {Simard}, {Smidt}, {Somerville}, {Spinrad}, {Straughn}, {Strolger},
  {Telford}, {Teplitz}, {Trump}, {van der Wel}, {Villforth}, {Wechsler},
  {Weiner}, {Wiklind}, {Wild}, {Wilson}, {Wuyts}, {Yan}, \& {Yun}}]{Grogin2011}
{Grogin}, N.~A., {Kocevski}, D.~D., {Faber}, S.~M., {et~al.} 2011, \apjs, 197,
  35, \dodoi{10.1088/0067-0049/197/2/35}

\bibitem[{{Hoaglin} {et~al.}(1983){Hoaglin}, {Mosteller}, \&
  {Tukey}}]{hoaglin83_MAD}
{Hoaglin}, D.~C., {Mosteller}, F., \& {Tukey}, J.~W. 1983, {Understanding
  robust and exploratory data analysis}, Wiley Series in Probability and
  Mathematical Statistics, (John Wiley,)

\bibitem[{{Hsiao} {et~al.}(2022){Hsiao}, {Coe}, {Abdurro'uf}, {Whitler},
  {Jung}, {Khullar}, {Meena}, {Dayal}, {Barrow}, {Santos-Olmsted}, {Casselman},
  {Vanzella}, {Nonino}, {Jimenez-Teja}, {Oguri}, {Stark}, {Furtak}, {Zitrin},
  {Adamo}, {Brammer}, {Bradley}, {Diego}, {Zackrisson}, {Finkelstein},
  {Windhorst}, {Bhatawdekar}, {Hutchison}, {Broadhurst}, {Dimauro},
  {Andrade-Santos}, {Eldridge}, {Acebron}, {Avila}, {Bayliss}, {Benitez},
  {Binggeli}, {Bolan}, {Bradac}, {Carnall}, {Conselice}, {Donahue}, {Frye},
  {Fujimoto}, {Henry}, {James}, {Kassin}, {Kewley}, {Larson}, {Lauer}, {Law},
  {Mahler}, {Mainali}, {McCandliss}, {Nicholls}, {Pirzkal}, {Postman}, {Rigby},
  {Ryan}, {Senchyna}, {Sharon}, {Shimizu}, {Strait}, {Tang}, {Trenti},
  {Vikaeus}, \& {Welch}}]{Hsiao2022}
{Hsiao}, T. Y.-Y., {Coe}, D., {Abdurro'uf}, {et~al.} 2022, arXiv e-prints,
  arXiv:2210.14123.
\newblock \doarXiv{2210.14123}

\bibitem[{{Hunter}(2007)}]{matplotlib2007}
{Hunter}, J.~D. 2007, Computing in Science Engineering, 9, 90,
  \dodoi{10.1109/MCSE.2007.55}

\bibitem[{{Ilbert} {et~al.}(2006){Ilbert}, {Arnouts}, {McCracken},
  {Bolzonella}, {Bertin}, {Le F{\`e}vre}, {Mellier}, {Zamorani}, {Pell{\`o}},
  {Iovino}, {Tresse}, {Le Brun}, {Bottini}, {Garilli}, {Maccagni}, {Picat},
  {Scaramella}, {Scodeggio}, {Vettolani}, {Zanichelli}, {Adami}, {Bardelli},
  {Cappi}, {Charlot}, {Ciliegi}, {Contini}, {Cucciati}, {Foucaud}, {Franzetti},
  {Gavignaud}, {Guzzo}, {Marano}, {Marinoni}, {Mazure}, {Meneux}, {Merighi},
  {Paltani}, {Pollo}, {Pozzetti}, {Radovich}, {Zucca}, {Bondi}, {Bongiorno},
  {Busarello}, {de La Torre}, {Gregorini}, {Lamareille}, {Mathez}, {Merluzzi},
  {Ripepi}, {Rizzo}, \& {Vergani}}]{Ilbert2006}
{Ilbert}, O., {Arnouts}, S., {McCracken}, H.~J., {et~al.} 2006, \aap, 457, 841,
  \dodoi{10.1051/0004-6361:20065138}

\bibitem[{{Illingworth} {et~al.}(2016){Illingworth}, {Magee}, {Bouwens},
  {Oesch}, {Labbe}, {van Dokkum}, {Whitaker}, {Holden}, {Franx}, \&
  {Gonzalez}}]{Illingworth2016}
{Illingworth}, G., {Magee}, D., {Bouwens}, R., {et~al.} 2016, arXiv e-prints,
  arXiv:1606.00841.
\newblock \doarXiv{1606.00841}

\bibitem[{{Illingworth} {et~al.}(2013){Illingworth}, {Magee}, {Oesch},
  {Bouwens}, {Labb{\'e}}, {Stiavelli}, {van Dokkum}, {Franx}, {Trenti},
  {Carollo}, \& {Gonzalez}}]{Illingworth2013}
{Illingworth}, G.~D., {Magee}, D., {Oesch}, P.~A., {et~al.} 2013, \apjs, 209,
  6, \dodoi{10.1088/0067-0049/209/1/6}

\bibitem[{{Jarvis} {et~al.}(2013){Jarvis}, {Bonfield}, {Bruce}, {Geach},
  {McAlpine}, {McLure}, {Gonz{\'a}lez-Solares}, {Irwin}, {Lewis}, {Yoldas},
  {Andreon}, {Cross}, {Emerson}, {Dalton}, {Dunlop}, {Hodgkin}, {Le},
  {Karouzos}, {Meisenheimer}, {Oliver}, {Rawlings}, {Simpson}, {Smail},
  {Smith}, {Sullivan}, {Sutherland}, {White}, \& {Zwart}}]{Jarvis2013}
{Jarvis}, M.~J., {Bonfield}, D.~G., {Bruce}, V.~A., {et~al.} 2013, \mnras, 428,
  1281, \dodoi{10.1093/mnras/sts118}

\bibitem[{{Jauzac} {et~al.}(2015){Jauzac}, {Richard}, {Jullo}, {Cl{\'e}ment},
  {Limousin}, {Kneib}, {Ebeling}, {Natarajan}, {Rodney}, {Atek}, {Massey},
  {Eckert}, {Egami}, \& {Rexroth}}]{Jauzac2015}
{Jauzac}, M., {Richard}, J., {Jullo}, E., {et~al.} 2015, \mnras, 452, 1437,
  \dodoi{10.1093/mnras/stv1402}

\bibitem[{{Johnson} {et~al.}(2021){Johnson}, {Leja}, {Conroy}, \&
  {Speagle}}]{Johnson2021}
{Johnson}, B.~D., {Leja}, J., {Conroy}, C., \& {Speagle}, J.~S. 2021, \apjs,
  254, 22, \dodoi{10.3847/1538-4365/abef67}

\bibitem[{{Kikuchihara} {et~al.}(2020){Kikuchihara}, {Ouchi}, {Ono},
  {Mawatari}, {Chevallard}, {Harikane}, {Kojima}, {Oguri}, {Bruzual}, \&
  {Charlot}}]{Kikuchihara2020}
{Kikuchihara}, S., {Ouchi}, M., {Ono}, Y., {et~al.} 2020, \apj, 893, 60,
  \dodoi{10.3847/1538-4357/ab7dbe}

\bibitem[{{Koekemoer} {et~al.}(2011){Koekemoer}, {Faber}, {Ferguson}, {Grogin},
  {Kocevski}, {Koo}, {Lai}, {Lotz}, {Lucas}, {McGrath}, {Ogaz}, {Rajan},
  {Riess}, {Rodney}, {Strolger}, {Casertano}, {Castellano}, {Dahlen},
  {Dickinson}, {Dolch}, {Fontana}, {Giavalisco}, {Grazian}, {Guo}, {Hathi},
  {Huang}, {van der Wel}, {Yan}, {Acquaviva}, {Alexander}, {Almaini}, {Ashby},
  {Barden}, {Bell}, {Bournaud}, {Brown}, {Caputi}, {Cassata}, {Challis},
  {Chary}, {Cheung}, {Cirasuolo}, {Conselice}, {Roshan Cooray}, {Croton},
  {Daddi}, {Dav{\'e}}, {de Mello}, {de Ravel}, {Dekel}, {Donley}, {Dunlop},
  {Dutton}, {Elbaz}, {Fazio}, {Filippenko}, {Finkelstein}, {Frazer}, {Gardner},
  {Garnavich}, {Gawiser}, {Gruetzbauch}, {Hartley}, {H{\"a}ussler},
  {Herrington}, {Hopkins}, {Huang}, {Jha}, {Johnson}, {Kartaltepe},
  {Khostovan}, {Kirshner}, {Lani}, {Lee}, {Li}, {Madau}, {McCarthy},
  {McIntosh}, {McLure}, {McPartland}, {Mobasher}, {Moreira}, {Mortlock},
  {Moustakas}, {Mozena}, {Nandra}, {Newman}, {Nielsen}, {Niemi}, {Noeske},
  {Papovich}, {Pentericci}, {Pope}, {Primack}, {Ravindranath}, {Reddy},
  {Renzini}, {Rix}, {Robaina}, {Rosario}, {Rosati}, {Salimbeni}, {Scarlata},
  {Siana}, {Simard}, {Smidt}, {Snyder}, {Somerville}, {Spinrad}, {Straughn},
  {Telford}, {Teplitz}, {Trump}, {Vargas}, {Villforth}, {Wagner}, {Wandro},
  {Wechsler}, {Weiner}, {Wiklind}, {Wild}, {Wilson}, {Wuyts}, \&
  {Yun}}]{Koekemoer2011}
{Koekemoer}, A.~M., {Faber}, S.~M., {Ferguson}, H.~C., {et~al.} 2011, \apjs,
  197, 36, \dodoi{10.1088/0067-0049/197/2/36}

\bibitem[{{Kokorev} {et~al.}(2022){Kokorev}, {Brammer}, {Fujimoto}, {Kohno},
  {Magdis}, {Valentino}, {Toft}, {Oesch}, {Bauer}, {Coe}, {Egami}, {Oguri},
  {Ouchi}, {Postman}, {Richard}, {Jolly}, {Knudsen}, {Sun}, {Weaver}, {Ao},
  {Baker}, {Caputi}, {Espada}, {Hatsukade}, {Koekemoer}, {Mu{\~n}oz Arancibia},
  {Shimasaku}, {Umehata}, {Wang}, \& {Wang}}]{Kokorev2022}
{Kokorev}, V., {Brammer}, G., {Fujimoto}, S., {et~al.} 2022, arXiv e-prints,
  arXiv:2207.07125.
\newblock \doarXiv{2207.07125}

\bibitem[{{Kron}(1980)}]{Kron1980}
{Kron}, R.~G. 1980, \apjs, 43, 305, \dodoi{10.1086/190669}

\bibitem[{{Labb{\'e}} {et~al.}(2003){Labb{\'e}}, {Franx}, {Rudnick},
  {F{\"o}rster Schreiber}, {Rix}, {Moorwood}, {van Dokkum}, {van der Werf},
  {R{\"o}ttgering}, {van Starkenburg}, {van der Wel}, {Kuijken}, \&
  {Daddi}}]{Labbe2003}
{Labb{\'e}}, I., {Franx}, M., {Rudnick}, G., {et~al.} 2003, \aj, 125, 1107,
  \dodoi{10.1086/346140}

\bibitem[{{Livermore} {et~al.}(2017){Livermore}, {Finkelstein}, \&
  {Lotz}}]{Livermore2017}
{Livermore}, R.~C., {Finkelstein}, S.~L., \& {Lotz}, J.~M. 2017, \apj, 835,
  113, \dodoi{10.3847/1538-4357/835/2/113}

\bibitem[{{Lotz} {et~al.}(2017){Lotz}, {Koekemoer}, {Coe}, {Grogin}, {Capak},
  {Mack}, {Anderson}, {Avila}, {Barker}, {Borncamp}, {Brammer}, {Durbin},
  {Gunning}, {Hilbert}, {Jenkner}, {Khandrika}, {Levay}, {Lucas}, {MacKenty},
  {Ogaz}, {Porterfield}, {Reid}, {Robberto}, {Royle}, {Smith},
  {Storrie-Lombardi}, {Sunnquist}, {Surace}, {Taylor}, {Williams}, {Bullock},
  {Dickinson}, {Finkelstein}, {Natarajan}, {Richard}, {Robertson}, {Tumlinson},
  {Zitrin}, {Flanagan}, {Sembach}, {Soifer}, \& {Mountain}}]{Lotz2017}
{Lotz}, J.~M., {Koekemoer}, A., {Coe}, D., {et~al.} 2017, \apj, 837, 97,
  \dodoi{10.3847/1538-4357/837/1/97}

\bibitem[{{Mason} {et~al.}(2015){Mason}, {Trenti}, \& {Treu}}]{Mason2015}
{Mason}, C.~A., {Trenti}, M., \& {Treu}, T. 2015, \apj, 813, 21,
  \dodoi{10.1088/0004-637X/813/1/21}

\bibitem[{{Merlin} {et~al.}(2022){Merlin}, {Bonchi}, {Paris}, {Belfiori},
  {Fontana}, {Castellano}, {Nonino}, {Polenta}, {Santini}, {Yang},
  {Glazebrook}, {Treu}, {Roberts-Borsani}, {Trenti}, {Birrer}, {Brammer},
  {Grillo}, {Calabr{\`o}}, {Marchesini}, {Mason}, {Mercurio}, {Morishita},
  {Strait}, {Boyett}, {Leethochawalit}, {Nanayakkara}, {Vulcani}, {Bradac}, \&
  {Wang}}]{Merlin2022}
{Merlin}, E., {Bonchi}, A., {Paris}, D., {et~al.} 2022, \apjl, 938, L14,
  \dodoi{10.3847/2041-8213/ac8f93}

\bibitem[{{Merten} {et~al.}(2011){Merten}, {Coe}, {Dupke}, {Massey}, {Zitrin},
  {Cypriano}, {Okabe}, {Frye}, {Braglia}, {Jim{\'e}nez-Teja}, {Ben{\'\i}tez},
  {Broadhurst}, {Rhodes}, {Meneghetti}, {Moustakas}, {Sodr{\'e}}, {Krick}, \&
  {Bregman}}]{Merten2011}
{Merten}, J., {Coe}, D., {Dupke}, R., {et~al.} 2011, \mnras, 417, 333,
  \dodoi{10.1111/j.1365-2966.2011.19266.x}

\bibitem[{{Nardiello} {et~al.}(2022){Nardiello}, {Bedin}, {Burgasser},
  {Salaris}, {Cassisi}, {Griggio}, \& {Scalco}}]{Nardiello2022}
{Nardiello}, D., {Bedin}, L.~R., {Burgasser}, A., {et~al.} 2022, \mnras, 517,
  484, \dodoi{10.1093/mnras/stac2659}

\bibitem[{{Oke}(1974)}]{Oke1974}
{Oke}, J.~B. 1974, \apjs, 27, 21, \dodoi{10.1086/190287}

\bibitem[{{Pagul} {et~al.}(2021){Pagul}, {S{\'a}nchez}, {Davidzon}, \&
  {Mobasher}}]{Pagul2021}
{Pagul}, A., {S{\'a}nchez}, F.~J., {Davidzon}, I., \& {Mobasher}, B. 2021,
  \apjs, 256, 27, \dodoi{10.3847/1538-4365/abea9d}

\bibitem[{{Paris} {et~al.}(2023){Paris}, {Merlin}, {Fontana}, {Bonchi},
  {Brammer}, {Correnti}, {Treu}, {Boyett}, {Calabr{\`o}}, {Castellano}, {Chen},
  {Yang}, {Glazebrook}, {Kelly}, {Koekemoer}, {Leethochawalit}, {Mascia},
  {Mason}, {Morishita}, {Nonino}, {Pentericci}, {Polenta}, {Roberts-Borsani},
  {Santini}, {Trenti}, {Vanzella}, {Vulcani}, {Windhorst}, {Nanayakkara}, \&
  {Wang}}]{Paris2023}
{Paris}, D., {Merlin}, E., {Fontana}, A., {et~al.} 2023, \apj, 952, 20,
  \dodoi{10.3847/1538-4357/acda8a}

\bibitem[{{Perrin} {et~al.}(2014){Perrin}, {Sivaramakrishnan}, {Lajoie},
  {Elliott}, {Pueyo}, {Ravindranath}, \& {Albert}}]{Perrin2014}
{Perrin}, M.~D., {Sivaramakrishnan}, A., {Lajoie}, C.-P., {et~al.} 2014, in
  Society of Photo-Optical Instrumentation Engineers (SPIE) Conference Series,
  Vol. 9143, Space Telescopes and Instrumentation 2014: Optical, Infrared, and
  Millimeter Wave, ed. J.~{Oschmann}, Jacobus~M., M.~{Clampin}, G.~G. {Fazio},
  \& H.~A. {MacEwen}, 91433X, \dodoi{10.1117/12.2056689}

\bibitem[{{Perrin} {et~al.}(2012){Perrin}, {Soummer}, {Elliott}, {Lallo}, \&
  {Sivaramakrishnan}}]{Perrin2012}
{Perrin}, M.~D., {Soummer}, R., {Elliott}, E.~M., {Lallo}, M.~D., \&
  {Sivaramakrishnan}, A. 2012, in Society of Photo-Optical Instrumentation
  Engineers (SPIE) Conference Series, Vol. 8442, Space Telescopes and
  Instrumentation 2012: Optical, Infrared, and Millimeter Wave, ed. M.~C.
  {Clampin}, G.~G. {Fazio}, H.~A. {MacEwen}, \& J.~{Oschmann}, Jacobus~M.,
  84423D, \dodoi{10.1117/12.925230}

\bibitem[{{Richard} {et~al.}(2021){Richard}, {Claeyssens}, {Lagattuta},
  {Guaita}, {Bauer}, {Pello}, {Carton}, {Bacon}, {Soucail}, {Lyon}, {Kneib},
  {Mahler}, {Cl{\'e}ment}, {Mercier}, {Variu}, {Tamone}, {Ebeling}, {Schmidt},
  {Nanayakkara}, {Maseda}, {Weilbacher}, {Bouch{\'e}}, {Bouwens}, {Wisotzki},
  {de la Vieuville}, {Martinez}, \& {Patr{\'\i}cio}}]{Richard2021}
{Richard}, J., {Claeyssens}, A., {Lagattuta}, D., {et~al.} 2021, \aap, 646,
  A83, \dodoi{10.1051/0004-6361/202039462}

\bibitem[{{Rigby} {et~al.}(2022){Rigby}, {Perrin}, {McElwain}, {Kimble},
  {Friedman}, {Lallo}, {Doyon}, {Feinberg}, {Ferruit}, {Glasse}, {Rieke},
  {Rieke}, {Wright}, {Willott}, {Colon}, {Milam}, {Neff}, {Stark}, {Valenti},
  {Abell}, {Abney}, {Abul-Huda}, {Acton}, {Adams}, {Adler}, {Aguilar}, {Ahmed},
  {Albert}, {Alberts}, {Aldridge}, {Allen}, {Altenburg}, {Alvarez Marquez},
  {Alves de Oliveira}, {Andersen}, {Anderson}, {Anderson}, {Argyriou},
  {Armstrong}, {Arribas}, {Artigau}, {Arvai}, {Atkinson}, {Bacon}, {Bair},
  {Banks}, {Barrientes}, {Barringer}, {Bartosik}, {Bast}, {Baudoz}, {Beatty},
  {Bechtold}, {Beck}, {Bergeron}, {Bergkoetter}, {Bhatawdekar}, {Birkmann},
  {Blazek}, {Blome}, {Boccaletti}, {Boeker}, {Boia}, {Bonaventura}, {Bond},
  {Bosley}, {Boucarut}, {Bourque}, {Bouwman}, {Bower}, {Bowers}, {Boyer},
  {Bradley}, {Brady}, {Braun}, {Breda}, {Bresnahan}, {Bright}, {Britt},
  {Bromenschenkel}, {Brooks}, {Brooks}, {Brown}, {Brown}, {Brown}, {Bunker},
  {Burger}, {Bushouse}, {Cale}, {Cameron}, {Cameron}, {Canipe}, {Caplinger},
  {Caputo}, {Cara}, {Carey}, {Carniani}, {Carrasquilla}, {Carruthers}, {Case},
  {Catherine}, {Chance}, {Chapman}, {Charlot}, {Charlow}, {Chayer}, {Chen},
  {Cherinka}, {Chichester}, {Chilton}, {Chonis}, {Clampin}, {Clark}, {Clark},
  {Coe}, {Coleman}, {Comber}, {Comeau}, {Connolly}, {Cooper}, {Cooper},
  {Coppock}, {Correnti}, {Cossou}, {Coulais}, {Coyle}, {Cracraft}, {Curti},
  {Cuturic}, {Davis}, {Davis}, {Dean}, {DeLisa}, {deMeester}, {Dencheva},
  {Dencheva}, {DePasquale}, {Deschenes}, {Hunor Detre}, {Diaz}, {Dicken},
  {DiFelice}, {Dillman}, {Dixon}, {Doggett}, {Donaldson}, {Douglas}, {DuPrie},
  {Dupuis}, {Durning}, {Easmin}, {Eck}, {Edeani}, {Egami}, {Ehrenwinkler},
  {Eisenhamer}, {Eisenhower}, {Elie}, {Elliott}, {Elliott}, {Ellis},
  {Engesser}, {Espinoza}, {Etienne}, {Etxaluze}, {Falini}, {Feeney}, {Ferry},
  {Filippazzo}, {Fincham}, {Fix}, {Flagey}, {Florian}, {Flynn}, {Fontanella},
  {Ford}, {Forshay}, {Fox}, {Franz}, {Fu}, {Fullerton}, {Galkin}, {Galyer},
  {Garcia Marin}, {Gardner}, {Gardner}, {Garland}, {Garrett}, {Gasman},
  {Gaspar}, {Gaudreau}, {Gauthier}, {Geers}, {Geithner}, {Gennaro}, {Giardino},
  {Girard}, {Giuliano}, {Glassmire}, {Glauser}, {Glazer}, {Godfrey},
  {Golimowski}, {Gollnitz}, {Gong}, {Gonzaga}, {Gordon}, {Gordon},
  {Goudfrooij}, {Greene}, {Greenhouse}, {Grimaldi}, {Groebner}, {Grundy},
  {Guillard}, {Gutman}, {Ha}, {Haderlein}, {Hagedorn}, {Hainline}, {Haley},
  {Hami}, {Hamilton}, {Hammel}, {Hansen}, {Harkins}, {Harr}, {Hart}, {Hart},
  {Hartig}, {Hashimoto}, {Haskins}, {Hathaway}, {Havey}, {Hayden}, {Hecht},
  {Heller-Boyer}, {Henriques}, {Henry}, {Hermann}, {Hernandez}, {Hesman},
  {Hicks}, {Hilbert}, {Hines}, {Hoffman}, {Holfeltz}, {Holler}, {Hoppa},
  {Hott}, {Howard}, {Howard}, {Hunter}, {Hunter}, {Hurst}, {Husemann},
  {Hustak}, {Ilinca Ignat}, {Illingworth}, {Irish}, {Jackson}, {Jahromi},
  {Jakobsen}, {James}, {James}, {Januszewski}, {Jenkins}, {Jirdeh}, {Johnson},
  {Johnson}, {Jones}, {Jones}, {Jones}, {Jones}, {Jordan}, {Jordan}, {Jurczyk},
  {Jurling}, {Kaleida}, {Kalmanson}, {Kammerer}, {Kang}, {Kao}, {Karakla},
  {Kavanagh}, {Kelly}, {Kendrew}, {Kennedy}, {Kenny}, {Keski-kuha}, {Keyes},
  {Kidwell}, {Kinzel}, {Kirk}, {Kirkpatrick}, {Kirshenblat}, {Klaassen},
  {Knapp}, {Knight}, {Knollenberg}, {Koehler}, {Koekemoer}, {Kovacs}, {Kulp},
  {Kumari}, {Kyprianou}, {La Massa}, {Labador}, {Labiano Ortega}, {Lagage},
  {Lajoie}, {Lallo}, {Lam}, {Lamb}, {Lambros}, {Lampenfield}, {Langston},
  {Larson}, {Law}, {Lawrence}, {Lee}, {Leisenring}, {Lepo}, {Leveille},
  {Levenson}, {Levine}, {Levy}, {Lewis}, {Lewis}, {Libralato}, {Lightsey},
  {Link}, {Liu}, {Lo}, {Lockwood}, {Logue}, {Long}, {Long}, {Loomis},
  {Lopez-Caniego}, {Alvarez}, {Love-Pruitt}, {Lucy}, {Luetzgendorf}, {Maghami},
  {Maiolino}, {Major}, {Malla}, {Malumuth}, {Manjavacas}, {Mannfolk},
  {Marrione}, {Marston}, {Martel}, {Maschmann}, {Masci}, {Masciarelli},
  {Maszkiewicz}, {Mather}, {McKenzie}, {McLean}, {McMaster}, {Melbourne},
  {Mel{\'e}ndez}, {Menzel}, {Merz}, {Meyett}, {Meza}, {Miskey}, {Misselt},
  {Moller}, {Morrison}, {Morse}, {Moseley}, {Mosier}, {Mountain}, {Mueckay},
  {Mueller}, {Mullally}, {Murphy}, {Murray}, {Murray}, {Mustelier},
  {Muzerolle}, {Mycroft}, {Myers}, {Myrick}, {Nanavati}, {Nance}, {Nayak},
  {Naylor}, {Nelan}, {Nickson}, {Nielson}, {Nieto-Santisteban}, {Nikolov},
  {Noriega-Crespo}, {O'Shaughnessy}, {O'Sullivan}, {Ochs}, {Ogle}, {Oleszczuk},
  {Olmsted}, {Osborne}, {Ottens}, {Owens}, {Pacifici}, {Pagan}, {Page}, {Park},
  {Parrish}, {Patapis}, {Paul}, {Pauly}, {Pavlovsky}, {Pedder}, {Peek},
  {Pena-Guerrero}, {Pennanen}, {Perez}, {Perna}, {Perriello}, {Phillips},
  {Pietraszkiewicz}, {Pinaud}, {Pirzkal}, {Pitman}, {Piwowar}, {Platais},
  {Player}, {Plesha}, {Pollizi}, {Polster}, {Pontoppidan}, {Porterfield},
  {Proffitt}, {Pueyo}, {Pulliam}, {Quirt}, {Quispe Neira}, {Ramos Alarcon},
  {Ramsay}, {Rapp}, {Rapp}, {Rauscher}, {Ravindranath}, {Rawle}, {Regan},
  {Reichard}, {Reis}, {Ressler}, {Rest}, {Reynolds}, {Rhue}, {Richon},
  {Rickman}, {Ridgaway}, {Ritchie}, {Rix}, {Robberto}, {Robinson}, {Robinson},
  {Robinson}, {Rock}, {Rodriguez}, {Rodriguez Del Pino}, {Roellig}, {Rohrbach},
  {Roman}, {Romelfanger}, {Rose}, {Roteliuk}, {Roth}, {Rothwell}, {Rowlands},
  {Roy}, {Royer}, {Royle}, {Rui}, {Rumler}, {Runnels}, {Russ}, {Rustamkulov},
  {Ryden}, {Ryer}, {Sabata}, {Sabatke}, {Sabbi}, {Samuelson}, {Sapp},
  {Sappington}, {Sargent}, {Sauer}, {Scheithauer}, {Schlawin}, {Schlitz},
  {Schmitz}, {Schneider}, {Schreiber}, {Schulze}, {Schwab}, {Scott}, {Sembach},
  {Shanahan}, {Shaughnessy}, {Shaw}, {Shawger}, {Shay}, {Sheehan}, {Shen},
  {Sherman}, {Shiao}, {Shih}, {Shivaei}, {Sienkiewicz}, {Sing}, {Sirianni},
  {Sivaramakrishnan}, {Skipper}, {Sloan}, {Slocum}, {Slowinski}, {Smith},
  {Smith}, {Smith}, {Smith}, {Snyder}, {Soh}, {Sohn}, {Soto}, {Spencer},
  {Stallcup}, {Stansberry}, {Starr}, {Starr}, {Stewart}, {Stiavelli},
  {Straughn}, {Strickland}, {Stys}, {Summers}, {Sun}, {Sunnquist}, {Swade},
  {Swam}, {Swaters}, {Swoish}, {Taylor}, {Taylor}, {Te Plate}, {Tea}, {Teague},
  {Telfer}, {Temim}, {Thatte}, {Thompson}, {Thompson}, {Thomson}, {Tikkanen},
  {Tippet}, {Todd}, {Toolan}, {Tran}, {Trejo}, {Truong}, {Tsukamoto},
  {Tustain}, {Tyra}, {Ubeda}, {Underwood}, {Uzzo}, {Van Campen}, {Vandal},
  {Vandenbussche}, {Vila}, {Volk}, {Wahlgren}, {Waldman}, {Walker}, {Wander},
  {Warfield}, {Warner}, {Wasiak}, {Watkins}, {Weaver}, {Weilert}, {Weiser},
  {Weiss}, {Weissman}, {Welty}, {West}, {Wheate}, {Wheatley}, {Wheeler},
  {White}, {Whiteaker}, {Whitehouse}, {Whiteleather}, {Whitman}, {Williams},
  {Willmer}, {Willoughby}, {Wilson}, {Wirth}, {Wislowski}, {Wolf}, {Wolfe},
  {Wolff}, {Workman}, {Wright}, {Wu}, {Wu}, {Wymer}, {Yates}, {Yeager},
  {Yeates}, {Yerger}, {Yoon}, {Young}, {Yu}, {Zak}, {Zeidler}, {Zhou},
  {Zielinski}, {Zincke}, \& {Zonak}}]{Rigby2022}
{Rigby}, J., {Perrin}, M., {McElwain}, M., {et~al.} 2022, arXiv e-prints,
  arXiv:2207.05632.
\newblock \doarXiv{2207.05632}

\bibitem[{{Roberts-Borsani} {et~al.}(2022){Roberts-Borsani}, {Morishita},
  {Treu}, {Brammer}, {Strait}, {Wang}, {Bradac}, {Acebron}, {Bergamini},
  {Boyett}, {Calabr{\'o}}, {Castellano}, {Fontana}, {Glazebrook}, {Grillo},
  {Henry}, {Jones}, {Malkan}, {Marchesini}, {Mascia}, {Mason}, {Mercurio},
  {Merlin}, {Nanayakkara}, {Pentericci}, {Rosati}, {Santini}, {Scarlata},
  {Trenti}, {Vanzella}, {Vulcani}, \& {Willott}}]{Roberts-Borsani2022}
{Roberts-Borsani}, G., {Morishita}, T., {Treu}, T., {et~al.} 2022, \apjl, 938,
  L13, \dodoi{10.3847/2041-8213/ac8e6e}

\bibitem[{{Rowe} {et~al.}(2015){Rowe}, {Jarvis}, {Mandelbaum}, {Bernstein},
  {Bosch}, {Simet}, {Meyers}, {Kacprzak}, {Nakajima}, {Zuntz}, {Miyatake},
  {Dietrich}, {Armstrong}, {Melchior}, \& {Gill}}]{Rowe2015}
{Rowe}, B.~T.~P., {Jarvis}, M., {Mandelbaum}, R., {et~al.} 2015, Astronomy and
  Computing, 10, 121, \dodoi{10.1016/j.ascom.2015.02.002}

\bibitem[{{Salmon} {et~al.}(2020){Salmon}, {Coe}, {Bradley}, {Bouwens},
  {Brada{\v{c}}}, {Huang}, {Oesch}, {Stark}, {Sharon}, {Trenti}, {Avila},
  {Ogaz}, {Andrade-Santos}, {Carrasco}, {Cerny}, {Dawson}, {Frye}, {Hoag},
  {Johnson}, {Jones}, {Lam}, {Lovisari}, {Mainali}, {Past}, {Paterno-Mahler},
  {Peterson}, {Riess}, {Rodney}, {Ryan}, {Sendra-Server}, {Strait}, {Strolger},
  {Umetsu}, {Vulcani}, \& {Zitrin}}]{Salmon2020}
{Salmon}, B., {Coe}, D., {Bradley}, L., {et~al.} 2020, \apj, 889, 189,
  \dodoi{10.3847/1538-4357/ab5a8b}

\bibitem[{{Schlafly} \& {Finkbeiner}(2011)}]{Schlafly2011}
{Schlafly}, E.~F., \& {Finkbeiner}, D.~P. 2011, \apj, 737, 103,
  \dodoi{10.1088/0004-637X/737/2/103}

\bibitem[{{Schlegel} {et~al.}(1998){Schlegel}, {Finkbeiner}, \&
  {Davis}}]{Schlegel1998}
{Schlegel}, D.~J., {Finkbeiner}, D.~P., \& {Davis}, M. 1998, \apj, 500, 525,
  \dodoi{10.1086/305772}

\bibitem[{{Scoville} {et~al.}(2007){Scoville}, {Aussel}, {Brusa}, {Capak},
  {Carollo}, {Elvis}, {Giavalisco}, {Guzzo}, {Hasinger}, {Impey}, {Kneib},
  {LeFevre}, {Lilly}, {Mobasher}, {Renzini}, {Rich}, {Sanders}, {Schinnerer},
  {Schminovich}, {Shopbell}, {Taniguchi}, \& {Tyson}}]{Scoville2007}
{Scoville}, N., {Aussel}, H., {Brusa}, M., {et~al.} 2007, \apjs, 172, 1,
  \dodoi{10.1086/516585}

\bibitem[{{Sharon} {et~al.}(2020){Sharon}, {Bayliss}, {Dahle}, {Dunham},
  {Florian}, {Gladders}, {Johnson}, {Mahler}, {Paterno-Mahler}, {Rigby},
  {Whitaker}, {Akhshik}, {Koester}, {Murray}, {Remolina Gonz{\'a}lez}, \&
  {Wuyts}}]{Sharon2020}
{Sharon}, K., {Bayliss}, M.~B., {Dahle}, H., {et~al.} 2020, \apjs, 247, 12,
  \dodoi{10.3847/1538-4365/ab5f13}

\bibitem[{{Shipley} {et~al.}(2018){Shipley}, {Lange-Vagle}, {Marchesini},
  {Brammer}, {Ferrarese}, {Stefanon}, {Kado-Fong}, {Whitaker}, {Oesch},
  {Feinstein}, {Labb{\'e}}, {Lundgren}, {Martis}, {Muzzin}, {Nedkova},
  {Skelton}, \& {van der Wel}}]{shipley:18}
{Shipley}, H.~V., {Lange-Vagle}, D., {Marchesini}, D., {et~al.} 2018, \apjs,
  235, 14, \dodoi{10.3847/1538-4365/aaacce}

\bibitem[{{Skelton} {et~al.}(2014){Skelton}, {Whitaker}, {Momcheva}, {Brammer},
  {van Dokkum}, {Labb{\'e}}, {Franx}, {van der Wel}, {Bezanson}, {Da Cunha},
  {Fumagalli}, {F{\"o}rster Schreiber}, {Kriek}, {Leja}, {Lundgren}, {Magee},
  {Marchesini}, {Maseda}, {Nelson}, {Oesch}, {Pacifici}, {Patel}, {Price},
  {Rix}, {Tal}, {Wake}, \& {Wuyts}}]{Skelton2014}
{Skelton}, R.~E., {Whitaker}, K.~E., {Momcheva}, I.~G., {et~al.} 2014, \apjs,
  214, 24, \dodoi{10.1088/0067-0049/214/2/24}

\bibitem[{{Steinhardt} {et~al.}(2020){Steinhardt}, {Jauzac}, {Acebron}, {Atek},
  {Capak}, {Davidzon}, {Eckert}, {Harvey}, {Koekemoer}, {Lagos}, {Mahler},
  {Montes}, {Niemiec}, {Nonino}, {Oesch}, {Richard}, {Rodney}, {Schaller},
  {Sharon}, {Strolger}, {Allingham}, {Amara}, {Bah{\'e}}, {B{\oe}hm}, {Bose},
  {Bouwens}, {Bradley}, {Brammer}, {Broadhurst}, {Ca{\~n}as}, {Cen},
  {Cl{\'e}ment}, {Clowe}, {Coe}, {Connor}, {Darvish}, {Diego}, {Ebeling},
  {Edge}, {Egami}, {Ettori}, {Faisst}, {Frye}, {Furtak}, {G{\'o}mez-Guijarro},
  {Remolina Gonz{\'a}lez}, {Gonzalez}, {Graur}, {Gruen}, {Harvey}, {Hensley},
  {Hovis-Afflerbach}, {Jablonka}, {Jha}, {Jullo}, {Kneib}, {Kokorev},
  {Lagattuta}, {Limousin}, {von der Linden}, {Linzer}, {Lopez}, {Magdis},
  {Massey}, {Masters}, {Maturi}, {McCully}, {McGee}, {Meneghetti}, {Mobasher},
  {Moustakas}, {Murphy}, {Natarajan}, {Neyrinck}, {O'Connor}, {Oguri}, {Pagul},
  {Rhodes}, {Rich}, {Robertson}, {Sereno}, {Shan}, {Smith}, {Sneppen},
  {Squires}, {Tam}, {Tchernin}, {Toft}, {Umetsu}, {Weaver}, {van Weeren},
  {Williams}, {Wilson}, {Yan}, \& {Zitrin}}]{Steinhardt2020}
{Steinhardt}, C.~L., {Jauzac}, M., {Acebron}, A., {et~al.} 2020, \apjs, 247,
  64, \dodoi{10.3847/1538-4365/ab75ed}

\bibitem[{{Strait} {et~al.}(2021){Strait}, {Brada{\v{c}}}, {Coe}, {Lemaux},
  {Carnall}, {Bradley}, {Pelliccia}, {Sharon}, {Zitrin}, {Acebron}, {Neufeld},
  {Andrade-Santos}, {Avila}, {Frye}, {Mahler}, {Nonino}, {Ogaz}, {Oguri},
  {Ouchi}, {Paterno-Mahler}, {Stark}, {Mainali}, {Oesch}, {Trenti}, {Carrasco},
  {Dawson}, {Jones}, {Umetsu}, \& {Vulcani}}]{Strait2021}
{Strait}, V., {Brada{\v{c}}}, M., {Coe}, D., {et~al.} 2021, \apj, 910, 135,
  \dodoi{10.3847/1538-4357/abe533}

\bibitem[{{Treu} {et~al.}(2015){Treu}, {Schmidt}, {Brammer}, {Vulcani}, {Wang},
  {Brada{\v{c}}}, {Dijkstra}, {Dressler}, {Fontana}, {Gavazzi}, {Henry},
  {Hoag}, {Huang}, {Jones}, {Kelly}, {Malkan}, {Mason}, {Pentericci},
  {Poggianti}, {Stiavelli}, {Trenti}, \& {von der Linden}}]{Treu2015}
{Treu}, T., {Schmidt}, K.~B., {Brammer}, G.~B., {et~al.} 2015, \apj, 812, 114,
  \dodoi{10.1088/0004-637X/812/2/114}

\bibitem[{{Treu} {et~al.}(2022){Treu}, {Roberts-Borsani}, {Bradac}, {Brammer},
  {Fontana}, {Henry}, {Mason}, {Morishita}, {Pentericci}, {Wang}, {Acebron},
  {Bagley}, {Bergamini}, {Belfiori}, {Bonchi}, {Boyett}, {Boutsia},
  {Calabr{\'o}}, {Caminha}, {Castellano}, {Dressler}, {Glazebrook}, {Grillo},
  {Jacobs}, {Jones}, {Kelly}, {Leethochawalit}, {Malkan}, {Marchesini},
  {Mascia}, {Mercurio}, {Merlin}, {Nanayakkara}, {Nonino}, {Paris},
  {Poggianti}, {Rosati}, {Santini}, {Scarlata}, {Shipley}, {Strait}, {Trenti},
  {Tubthong}, {Vanzella}, {Vulcani}, \& {Yang}}]{Treu2022}
{Treu}, T., {Roberts-Borsani}, G., {Bradac}, M., {et~al.} 2022, \apj, 935, 110,
  \dodoi{10.3847/1538-4357/ac8158}

\bibitem[{{van der Walt} {et~al.}(2011){van der Walt}, {Colbert}, \&
  {Varoquaux}}]{numpy2011}
{van der Walt}, S., {Colbert}, S.~C., \& {Varoquaux}, G. 2011, Computing in
  Science Engineering, 13, 22, \dodoi{10.1109/MCSE.2011.37}

\bibitem[{{van der Wel} {et~al.}(2014){van der Wel}, {Franx}, {van Dokkum},
  {Skelton}, {Momcheva}, {Whitaker}, {Brammer}, {Bell}, {Rix}, {Wuyts},
  {Ferguson}, {Holden}, {Barro}, {Koekemoer}, {Chang}, {McGrath},
  {H{\"a}ussler}, {Dekel}, {Behroozi}, {Fumagalli}, {Leja}, {Lundgren},
  {Maseda}, {Nelson}, {Wake}, {Patel}, {Labb{\'e}}, {Faber}, {Grogin}, \&
  {Kocevski}}]{vanderWel2014}
{van der Wel}, A., {Franx}, M., {van Dokkum}, P.~G., {et~al.} 2014, \apj, 788,
  28, \dodoi{10.1088/0004-637X/788/1/28}

\bibitem[{{Wang} {et~al.}(2023){Wang}, {Leja}, {Bezanson}, {Johnson},
  {Khullar}, {Labb{\'e}}, {Price}, {Weaver}, \& {Whitaker}}]{Wang2023}
{Wang}, B., {Leja}, J., {Bezanson}, R., {et~al.} 2023, \apjl, 944, L58,
  \dodoi{10.3847/2041-8213/acba99}

\bibitem[{{Weaver} {et~al.}(2022){Weaver}, {Kauffmann}, {Ilbert}, {McCracken},
  {Moneti}, {Toft}, {Brammer}, {Shuntov}, {Davidzon}, {Hsieh}, {Laigle},
  {Anastasiou}, {Jespersen}, {Vinther}, {Capak}, {Casey}, {McPartland},
  {Milvang-Jensen}, {Mobasher}, {Sanders}, {Zalesky}, {Arnouts}, {Aussel},
  {Dunlop}, {Faisst}, {Franx}, {Furtak}, {Fynbo}, {Gould}, {Greve}, {Gwyn},
  {Kartaltepe}, {Kashino}, {Koekemoer}, {Kokorev}, {Le F{\`e}vre}, {Lilly},
  {Masters}, {Magdis}, {Mehta}, {Peng}, {Riechers}, {Salvato}, {Sawicki},
  {Scarlata}, {Scoville}, {Shirley}, {Silverman}, {Sneppen}, {Smolc̆i{\'c}},
  {Steinhardt}, {Stern}, {Tanaka}, {Taniguchi}, {Teplitz}, {Vaccari}, {Wang},
  \& {Zamorani}}]{Weaver2022}
{Weaver}, J.~R., {Kauffmann}, O.~B., {Ilbert}, O., {et~al.} 2022, \apjs, 258,
  11, \dodoi{10.3847/1538-4365/ac3078}

\bibitem[{{Welch} {et~al.}(2022){Welch}, {Coe}, {Zackrisson}, {de Mink},
  {Ravindranath}, {Anderson}, {Brammer}, {Bradley}, {Yoon}, {Kelly}, {Diego},
  {Windhorst}, {Zitrin}, {Dimauro}, {Jim{\'e}nez-Teja}, {Abdurro'uf}, {Nonino},
  {Acebron}, {Andrade-Santos}, {Avila}, {Bayliss}, {Ben{\'\i}tez},
  {Broadhurst}, {Bhatawdekar}, {Brada{\v{c}}}, {Caminha}, {Chen}, {Eldridge},
  {Farag}, {Florian}, {Frye}, {Fujimoto}, {Gomez}, {Henry}, {Hsiao},
  {Hutchison}, {James}, {Joyce}, {Jung}, {Khullar}, {Larson}, {Mahler},
  {Mandelker}, {McCandliss}, {Morishita}, {Newshore}, {Norman}, {O'Connor},
  {Oesch}, {Oguri}, {Ouchi}, {Postman}, {Rigby}, {Ryan}, {Sharma}, {Sharon},
  {Strait}, {Strolger}, {Timmes}, {Toft}, {Trenti}, {Vanzella}, \&
  {Vikaeus}}]{Welch2022}
{Welch}, B., {Coe}, D., {Zackrisson}, E., {et~al.} 2022, \apjl, 940, L1,
  \dodoi{10.3847/2041-8213/ac9d39}

\bibitem[{{Whitaker} {et~al.}(2019){Whitaker}, {Ashas}, {Illingworth}, {Magee},
  {Leja}, {Oesch}, {van Dokkum}, {Mowla}, {Bouwens}, {Franx}, {Holden},
  {Labb{\'e}}, {Rafelski}, {Teplitz}, \& {Gonzalez}}]{Whitaker2019}
{Whitaker}, K.~E., {Ashas}, M., {Illingworth}, G., {et~al.} 2019, \apjs, 244,
  16, \dodoi{10.3847/1538-4365/ab3853}

\bibitem[{Wiener(1949)}]{Wiener1949}
Wiener, N. 1949, Extrapolation, Interpolation, and Smoothing of Stationary Time
  Series: With Engineering Applications (The MIT Press),
  \dodoi{10.7551/mitpress/2946.001.0001}

\bibitem[{{Williams} {et~al.}(2022){Williams}, {Kelly}, {Chen}, {Brammer},
  {Zitrin}, {Treu}, {Scarlata}, {Koekemoer}, {Oguri}, {Lin}, {Diego}, {Nonino},
  {Hjorth}, {Broadhurst}, {Rogers}, {Perez-Fournon}, {Foley}, {Jha},
  {Filippenko}, {Strolger}, {Pierel}, \& {Poidevin}}]{Williams2022}
{Williams}, H., {Kelly}, P.~L., {Chen}, W., {et~al.} 2022, arXiv e-prints,
  arXiv:2210.15699.
\newblock \doarXiv{2210.15699}

\bibitem[{{Williams} {et~al.}(1996){Williams}, {Blacker}, {Dickinson}, {Dixon},
  {Ferguson}, {Fruchter}, {Giavalisco}, {Gilliland}, {Heyer}, {Katsanis},
  {Levay}, {Lucas}, {McElroy}, {Petro}, {Postman}, {Adorf}, \&
  {Hook}}]{Williams1996}
{Williams}, R.~E., {Blacker}, B., {Dickinson}, M., {et~al.} 1996, \aj, 112,
  1335, \dodoi{10.1086/118105}

\bibitem[{{Williams} {et~al.}(2009){Williams}, {Quadri}, {Franx}, {van Dokkum},
  \& {Labb{\'e}}}]{Williams2009}
{Williams}, R.~J., {Quadri}, R.~F., {Franx}, M., {van Dokkum}, P., \&
  {Labb{\'e}}, I. 2009, \apj, 691, 1879, \dodoi{10.1088/0004-637X/691/2/1879}

\bibitem[{{Willott} {et~al.}(2017){Willott}, {Abraham}, {Albert}, {Bradac},
  {Brammer}, {Chayer}, {Dixon}, {Doyon}, {Dupuis}, {Ferrarese}, {Goudfrooij},
  {Hutchings}, {Martel}, {Pacifici}, {Ravindranath}, \&
  {Sawicki}}]{Willott2017}
{Willott}, C.~J., {Abraham}, R.~G., {Albert}, L., {et~al.} 2017, {CANUCS: The
  CAnadian NIRISS Unbiased Cluster Survey}, JWST Proposal. Cycle 1, ID. \#1208

\bibitem[{{Windhorst} {et~al.}(2023){Windhorst}, {Cohen}, {Jansen}, {Summers},
  {Tompkins}, {Conselice}, {Driver}, {Yan}, {Coe}, {Frye}, {Grogin},
  {Koekemoer}, {Marshall}, {O'Brien}, {Pirzkal}, {Robotham}, {Ryan}, {Willmer},
  {Carleton}, {Diego}, {Keel}, {Porto}, {Redshaw}, {Scheller}, {Wilkins},
  {Willner}, {Zitrin}, {Adams}, {Austin}, {Arendt}, {Beacom}, {Bhatawdekar},
  {Bradley}, {Broadhurst}, {Cheng}, {Civano}, {Dai}, {Dole}, {D'Silva},
  {Duncan}, {Fazio}, {Ferrami}, {Ferreira}, {Finkelstein}, {Furtak}, {Gim},
  {Griffiths}, {Hammel}, {Harrington}, {Hathi}, {Holwerda}, {Honor}, {Huang},
  {Hyun}, {Im}, {Joshi}, {Kamieneski}, {Kelly}, {Larson}, {Li}, {Lim}, {Ma},
  {Maksym}, {Manzoni}, {Meena}, {Milam}, {Nonino}, {Pascale}, {Petric},
  {Pierel}, {del Carmen Polletta}, {R{\"o}ttgering}, {Rutkowski}, {Smail},
  {Straughn}, {Strolger}, {Swirbul}, {Trussler}, {Wang}, {Welch}, {B. Wyithe},
  {Yun}, {Zackrisson}, {Zhang}, \& {Zhao}}]{Windhorst2023}
{Windhorst}, R.~A., {Cohen}, S.~H., {Jansen}, R.~A., {et~al.} 2023, \aj, 165,
  13, \dodoi{10.3847/1538-3881/aca163}

\bibitem[{{Zheng} {et~al.}(2012){Zheng}, {Postman}, {Zitrin}, {Moustakas},
  {Shu}, {Jouvel}, {H{\o}st}, {Molino}, {Bradley}, {Coe}, {Moustakas},
  {Carrasco}, {Ford}, {Ben{\'\i}tez}, {Lauer}, {Seitz}, {Bouwens}, {Koekemoer},
  {Medezinski}, {Bartelmann}, {Broadhurst}, {Donahue}, {Grillo}, {Infante},
  {Jha}, {Kelson}, {Lahav}, {Lemze}, {Melchior}, {Meneghetti}, {Merten},
  {Nonino}, {Ogaz}, {Rosati}, {Umetsu}, \& {van der Wel}}]{Zheng2012}
{Zheng}, W., {Postman}, M., {Zitrin}, A., {et~al.} 2012, \nat, 489, 406,
  \dodoi{10.1038/nature11446}

\bibitem[{{Zhuang} \& {Shen}(2023)}]{Zhuang2023}
{Zhuang}, M.-Y., \& {Shen}, Y. 2023, arXiv e-prints, arXiv:2304.13776,
  \dodoi{10.48550/arXiv.2304.13776}

\bibitem[{{Zitrin} {et~al.}(2014){Zitrin}, {Zheng}, {Broadhurst}, {Moustakas},
  {Lam}, {Shu}, {Huang}, {Diego}, {Ford}, {Lim}, {Bauer}, {Infante}, {Kelson},
  \& {Molino}}]{Zitrin2014}
{Zitrin}, A., {Zheng}, W., {Broadhurst}, T., {et~al.} 2014, \apjl, 793, L12,
  \dodoi{10.1088/2041-8205/793/1/L12}

\end{thebibliography}
\bibliographystyle{aasjournal}

%% This command is needed to show the entire author+affiliation list when
%% the collaboration and author truncation commands are used.  It has to
%% go at the end of the manuscript.
% \allauthors

%% Include this line if you are using the \added, \replaced, \deleted
%% commands to see a summary list of all changes at the end of the article.
%\listofchanges

\end{document}